\documentclass[useAMS,usenatbib]{mn2e}

\usepackage[T1]{fontenc} 
\usepackage{aecompl}

\usepackage{amsmath}
\usepackage{amssymb}
\usepackage{epstopdf}
\usepackage{color}
\usepackage{graphicx}
\usepackage{float}
\usepackage{caption}
\usepackage{subcaption}
\usepackage{times}
\newcommand{\mo}{{\rm M}_\odot}
\newcommand{\lo}{{\rm L}_\odot}
\newcommand{\kpc}{{\rm kpc}}

\title{Baryonic and dark matter distribution in cosmological simulations of spiral galaxies} 

\author[Mollitor, Nezri, Teyssier]
 {Pol Mollitor$^1$\thanks{pol.mollitor@lam.fr (PM), emmanuel.nezri@lam.fr (EN), romain.teyssier@gmail.com (RT)}, 
 Emmanuel Nezri$^1$\footnotemark[1], 
 Romain Teyssier$^2$\footnotemark[1]\\ 
\normalsize $^1$Aix Marseille Universit\'e, CNRS, LAM (Laboratoire d'Astrophysique de Marseille) UMR 7326, 13388, Marseille, France \\
\normalsize $^2$Institute for Theoretical Physics, University of Zurich, Winterthurerstrasse 190, CH-8057, Zurich, Switzerland\\
 }

\date{Accepted Xxxx. Received Xxxx; in original form Xxxx}

\pagerange{\pageref{firstpage}--\pageref{lastpage}} \pubyear{2014}
\def\LaTeX{L\kern-.36em\raise.3ex\hbox{a}\kern-.15em
    T\kern-.1667em\lower.7ex\hbox{E}\kern-.125emX}

\begin{document}
\label{firstpage}

\maketitle 

\begin{abstract}

We study three cosmological hydrodynamical simulations of Milky Way(MW)-sized
halos including a comparison with the dark matter(DM)-only counterparts.
 We find one of our simulated galaxies with interesting MW-like features.
 Thanks to a consistently tuned star formation rate and
supernovae feedback we obtain an extended disk and a flat rotation curve with a satisfying
circular velocity and a reasonable DM density in the solar neighbourhood.
 Mimicking observational methods, we re-derive the stellar mass and obtain stellar-to-halo mass ratios reduced by
 more than 50\%. We show the interaction between the baryons and the dark matter which is first contracted by star formation and then cored by feedback processes. Indeed, we report an unprecedentedly observed effect in the DM density profile consisting of a central core combined with an adiabatic
contraction at larger galactic radii. The cores obtained are typically $\sim$ 5 kpc large. Moreover, this also impacts the DM density at the solar radius. In our simulation resembling most to the MW, the density is raised
from 0.23 GeV/cm$^3$ in the dark matter only run to 0.36 GeV/cm$^3$ (spherical shell)
or 0.54 GeV/cm$^3$ (circular ring) in the hydrodynamical run. Studying the subhalos,
 the dark matter within luminous satellites is also affected by baryonic processes and exihibits cored profiles whereas dark satellites are cuspy. We find a shift in mass compared to
DM-only simulations and obtain, for halos in the lower MW mass range, a
distribution of luminous satellites comparable to the MW spheroidal dwarf galaxies.

\end{abstract}


\section{Introduction}
If the $\Lambda$CDM structure formation scenario is successful at large
scales, the physics of baryons is dominant for small scale objects like galaxies.
Understanding galaxy formation in a cosmological context is a central
question of astrophysics. Starting from first principles,
hydrodynamical simulations are the most sophisticated and self-consistent
approach to address this question (see \citet{2014Natur.509..177V} for a recent achievement). A key issue is to
form realistic spiral galaxies like the Milky Way (MW).
Even if the formation of disks and spirals was observed in the first hydrodynamical
simulations \citep{1991ApJ...377..365K}, until recently the simulations
exhibited generic problems known as the overcooling and the angular
momentum problems \citep{1994MNRAS.267..401N}. Typically simulations
produced  objects that were too contracted with a star formation rate that was too high, 
especially at high redshifts, giving birth to massive bulges of old stars
as well as to disks that were too small. Consequently, they show some problems
with regard to observations like too peaked rotation curves or
stellar-to-halo mass ratios that were too high\citep{2012MNRAS.423.1726S}.
These features were encountered for all type of numerical treatments of
hydrodynamics (Lagrangian or Eulerian) \citep{2012MNRAS.423.1726S}.\\
In some way, the question of the baryon distribution in simulations can echo
the problem of the dark matter (DM) distribution in the
MW (see e.g \citep{1986ApJ...301...27B,2010Natur.463..203G,2009MNRAS.397...44R}. Actually both have a strong interplay
during the galaxy formation history and the
resulting dark matter configuration features, e.g. a cored/cuspy profile
in the center of (sub)halos, the velocity distribution shape in the solar neighborhood or a dark disk component, are critical issues for dark matter searches and identification.
Indeed, most of studies dedicated to DM detection assume DM distribution features inspired
from simulations results (see reviews like e.g
\citet{1996PhR...267..195J,2004PhR...405..279B,2012CRPhy..13..740L,2013arXiv1310.7039P}
and references therein.).
Thus, the results born of cosmological simulations and the acquired knowledge 
about the driving physical mechanisms intervening in the formation of MW-like spiral galaxies have crucial implications beyond the field of galaxy formation.\\
The numerical resolution was first thought to be a possible direction to cure the spiral
 simulated galaxy morphology (e.g \citet{2007MNRAS.374.1479G}), but the most popular way to address
this issue relies on how subgrid processes deal with
the physics of baryons (cooling, star formation, feedback modelling) are achieved.
Strictly speaking, star formation and stellar feedback are related in a complex and intricate dependency in the majority of works, as the
subgrid parameters depend on the resolution of the simulations \citep{2014MNRAS.437.1750M,2014arXiv1404.2613A}.\\
Currently, while stellar feedback processes allow to regulate star formation, a definite equilibrium with alternative schemes or new sources leading
to realistic star formation histories has still to be achieved.
Lately, a lot of efforts have been expanded on star formation and related feedback processes (e.g.
\citet{2006AAS...209.3802S,2013MNRAS.436.3031V,2013ApJ...770...25A,2013MNRAS.429.3068T,Roskar:2013pia,2013arXiv1311.2073H}).
Recently, a few interesting works have succeeded to form more realistic spiral
galaxies \citep{2011MNRAS.410.1391A,2011ApJ74276G,2013MNRAS.434.3142A,2013MNRAS.428..129S,2014MNRAS.437.1750M}
by tuning the parameters entering in those processes or by
introducing specific schemes, especially those involving the stellar
feedback.\\
Namely, the energy injected by stellar feedback processes may inhibit gas accretion. Different directions have been recently
investigated. \citet{2013MNRAS.428..129S}
implemented early feedback to prevent high redshift star
formation. 
Others \citep{2008MNRAS.389.1137S,2010Natur.463..203G,2010MNRAS.409.1541S} explored the effect of larger values of the injected amount of SN energy per explosion. Such approaches can also lead to the
destruction of the dark matter cusp \citep{2009MNRAS.395L..57P,2010Natur.463..203G,2014MNRAS.437..415D}. 
In \citet{2014MNRAS.437.1750M}, the 
introduction of an alternative
stellar feedback scheme that consists of kinetic winds that scale with the DM velocity dispersion and that
are decoupled from the hydrodynamical phase during their propagation time \citep{2013MNRAS.436.3031V} succeeds in some
cases to regulate the morphology of the disk (but preserves a cuspy
dark matter profile). The effect
of a supplementary source coming from stellar radiation has been
studied in \citet{Roskar:2013pia}, but such a scenario leads to the thickening of the
disk and the destruction of the spiral structure. \citet{2013arXiv1311.2073H} have proposed an explicit treatment of the multi-phase
interstellar medium and the stellar feedback, where the inputs are taken from stellar population models, and their results for the 
stellar-to-halo mass ratio agree up to $\sim 10^{12}\mo$, with predictions for the relation drawn from abundance matching works.\\
Concerning the parameters intervening in star formation in sub-grid models, 
a higher threshold \citep{2011ApJ74276G} can suppress star formation at high redshifts and, as a result,
avoid the formation of old low angular momentum stars. Additionally, a low star
formation efficiency seems necessary to form extended disks in \citet{2011MNRAS.410.1391A}. However, \citet{2013arXiv1311.2073H}
recently seem to reproduce realistic star formation rates without
adjusting the efficiency by hand thanks to a self-gravitating
criterium on the gas density and the aforementioned feedback prescription.\\
Even if the relevance of the adopted choices can be discussed, those
approaches on star formation and stellar feedback are interesting advances
in the field to address the remaining questions of
MW-like galaxy formation.\\
In the present article, we show our attempt to produce a MW-like disk galaxy in a
cosmological context. Three zoom-in simulations with RAMSES have been
performed and we study the
baryon and DM properties of the simulations. After a discussion of the properties of the baryonic galactic disks and a comparison with other simulations and 
recent observations,
we focus on the effects baryonic processes can induce into the distribution of DM, both in the case of the central parts of the halo
and in the gravitationally bound
satellites orbiting in the halo. In particular, we show that one of simulations is able to produce a galaxy with properties that resembles the Milky Way
and an environment which resembles that of the Milky Way surrounded by
a cored dark matter halo.
\\
The paper is organised as follows: in section \ref{sec:thesimulation},
we describe how we performed our simulations. Then, we analyse the
resulting galaxy properties at redshift 0 (section \ref{sec:galaxy0}). In
section \ref{sec:formHist} we discuss the formation history and
in section \ref{DMHaloproperties} we sudy the DM distribution features
(main original point of this work) in the
halo and its big satellites including a comparison between
the dark matter only runs and the hydrodynamical simulations.
Conclusions and
perspectives are given in section \ref{sec:summary}.

\section{the simulation}\label{sec:thesimulation}
In this section, we describe the methodology we applied to perform three high resolution "Zoom-in" simulations in a cosmological framework. 
We explain how we generated the initial conditions and ran the simulations using different physically motivated recipes in order to reproduce 
a large number of properties observed for redshift 0 spiral MW-like galaxies.\\
The analysis of the halos and their subhalos was done with the Amiga Halo Finder \citep{Knollmann:2009pb}.\\
In order to calculate densities and to extract the information of the snapshots, we used the unsio package\footnote{http://projets.lam.fr/projects/unsio}.\\
As a visualization tool, we used glnemo2\footnote{http://projets.lam.fr/projects/glnemo2}. The analysis and the assessment of the results and the physical properties
of our simulated objects was done with our own developed tools.

\subsection{Initial conditions}\label{sec:InitCond}
The "Zoom-in" technique we used is a powerful tool to simulate highly resolved structures within a cosmological context
while keeping the computational costs reasonably low. We used the MUSIC code \citep{Hahn:2011uy} to generate the primordial density fluctuations 
in a periodic 20 Mpc box in a Lambda-Cold-Dark Matter ($\Lambda$CDM) universe at redshift 50, with cosmological parameters
H$_{0}=70.3$km/s/Mpc the present-day Hubble constant, $\Omega_{\textrm{b},0}=0.045$ the baryonic matter density, 
$\Omega_{\textrm{m},0}=0.276$ the matter density, $\Omega_{\Lambda,0}=0.724$ the vacuum density.
First, we ran a DM low resolution simulation of $256^3$ particles to redshift 0 where we selected three MW candidate halos of different masses
that had a quiet merger history, i.e. no major merger after redshift 2, and no massive neighbour halo, i.e. 10 percent of the halo's virial mass,
closer than four times their virial radii. More specifically, 
we chose two halos in the lower mass range ($\lesssim10^{12}\mo$ inferred by stellar kinematics, see, e.g., 
\citet{2012ApJ...761...98K,2012ApJ...759..131B,2012MNRAS.425.2840D}) and
one halo in the higher mass range ($\sim 1-2\cdot10^{12}\mo $ favored by kinematics of satellite
galaxies or statistics of large cosmological DM simulations, see e.g. 
\citet{BoylanKolchin:2012xy,2011ApJ...743...40B}) to be resimulated. A discussion
on the MW-like virial mass range can be found in the introduction of \citet{2014A&A...562A..91P}.
For each halo, we repeated the following procedure: Selecting all the particles inside $3.5\cdot$R$_\textrm{97c}$ (see 
\citet{2014MNRAS.437.1894O} for a recent study on Lagrangian volumes in zoom simulations), 
we located their positions at the beginning of the simulation and increased locally in this volume the effective resolution to $1024^3$ particles, which corresponded 
to $10$ refinement levels of the mesh, a DM mass resolution of 230 812 $\mo$,
and a baryonic mass resolution of 44 963 $\mo$. The high resolution region is then successively enclosed by five cells of subsequent coarser refinement levels.
The mesh of the outer part of the box is kept at refinement level seven. Once the high resolution initial conditions were set up,
we re-ran the simulations until redshift 0. The primary properties of the three halos are listed 
in Table \ref{tab:properties}.\\
\begin{table*}
\centering
\begin{tabular}{lrrrrrrrrrrr}
\hline
Run & 
$R_{97}$  &
$M_{97,\rm tot}$  &
$M_{97,\rm gas}$  &
$M_\star$      &
$M_{97,\rm dm}$   &
$N_{\rm cells}$ &
$N_\star$      &
$N_{\rm dm}$  &
$R_{200}$  &
$M_{200,\rm tot}$ \\
 &
$(\kpc)$   & 
$(10^{10}\mo)$ & 
$(10^{10}\mo)$ & 
$(10^{10}\mo)$ & 
$(10^{10}\mo)$ & 
&
&
&
$(\kpc)$   & 
$(10^{10}\mo)$ 
\\
\hline
A & 344.9 & 227.52 & 23.96 &  18.23 & 185.32 &  8027923 & 6227518 & 11656318 &  253.69 &  186.68  \\
A-DM &  329.28 & 19.79 &  &   &  &   &   & 7178889  &  243.53  &  165.13 \\
B & 233.99 & 71.04 & 7.96 &  5.58 &  57.49 &  2491015 & 2153777 & 3910861 & 176.47 & 62.83  \\
B-DM & 220.85 & 59.73 &  &   &  &  &   & 2165979  & 162.90 & 49.42 \\
C & 244.60 & 81.15 & 9.58 & 5.50 & 6.60 & 2862113 & 2560388 & 4121429 & 181.83 & 68.73 \\
C-DM & 236.41 & 73.27 &  &  &  &  &   & 2657038  & 176.01  &  62.35 \\
\hline
\end{tabular}
\caption{Primary numerical parameters of the simulated halos at $z = 0$, A, B and C referring to the hydrodynamical versions
and *-DM to the corresponding dark matter only simulations. 
 We show the radius of the sphere whose mean density is equal to 97 (respectively 200) times the critical 
 density of the universe at redshift 0. The further columns give the
  total mass and DM mass
  inside $R_{97}$, the gas mass and stellar mass inside $R_{97}/10$ . The corresponding numbers of gaseous cells,
  star particles, and DM particles are given next.
    In all the runs, the spatial resolution reaches 150 parsec at $z=0$.}
\label{tab:properties}
\end{table*}
Traditionally, halo properties are given using a so-called "virial" radius, that defines a sphere in which the mean density 
$\frac{\rm M_{\rm Vir}}{4/3\cdot\pi\cdot\rm R_{\rm Vir}^{3}} $
is $\Delta_{\rm Vir}$ times denser 
than a certain mean reference density. Here, we take the critical density for a flat Universe  $\rho_{\rm crit} = 3 H^2(z) / (8\pi G)$.
As this definition is borrowed from structure growth theory, its use is not appropriate when one wants to define a physically
meaningful halo edge \citep{2008MNRAS.389..385C,2014ApJ...792..124Z}. 
In order to avoid ambiguity and to simplify comparisons we opted for giving our halo properties for two different $\Delta_{\rm vir}$ values,
namely 200 (commonly used value) and 97 (the value derived from the spherical top-hat collapse model for $\Lambda$CDM 
at z=0 for our cosmology \citep{1998ApJ...495...80B}), and put them in the subscript of the quantities we consider.\\
Halos B and C do not have any contamination of low resolution DM particles inside $R_{97}$. Halo A has a minor contamination between 85 kpc and
$R_{97}$  of 10 DM particles from lower resolution, but Halo A-DM does not.

\subsection{The simulation: dark matter and gas dynamics }
\label{subsec:Thesimulation:AMRandDM/gasdynamics}
We use the adaptive mesh refinement code RAMSES \citep{Teyssier:2001cp} to simulate the DM-only and the hydrodynamical simulations. 
Particles are modelled using a standard Particle Mesh method \citep{Teyssier:2001cp}.\\
Once the simulation started, we applied a "quasi-Lagrangian" strategy for triggering additional refinement. Attempting to maintain constant the mass per cell,
we refined the cell if the number of dark matter particles or baryonic (gas + stars) particles contained inside is bigger than eight. This refinement is 
limited to the high resolution zone as we use a passive scalar, which initially marked the high resolution cells. The scalar is advected passively by the flow and
allows refinement only if its value is above a certain threshold (here: 1\% \citep{Roskar:2013pia}). In order to avoid two-body relaxation effects, we first ran the DM-only 
simulation and we fixed the maximum refinement 
to the level this DM-only sibling reaches, i.e. 17 refinement levels corresponding to a spatial resolution of 150 pc.\\
The gas dynamics is modelled using a second-order unsplit Godunov scheme \citep{Teyssier:2006us}, based on the HLLC Riemann solver
and the MinMod slope limiter (see \citet{Fromang:2006aw} for technical details)). 
We assume a perfect gas equation of state, with $\gamma= 5/3$.\\
A particular concern of hydrodynamical simulations is to prevent numerical instabilities and artificial fragmentation: a gas medium that is stable 
in a theoretical framework is nevertheless prone
to be unstable in numerical models when the Jeans length is not well resolved within the numerical scheme
\citep{Truelove:1997dn}. In order to ensure this, we use the technique of the "polytropic pressure floor" 
\citep{Roskar:2013pia}. We define this pressure floor by imposing that the Jeans length is always resolved in at least
four resolution elements:
\begin{equation}
P=\frac{\rm G}{\pi\gamma}\rho(4\Delta x_{\rm min})^2
\end{equation}
When gas pressure hits this pressure floor, we make sure that the gas has reached a minimal temperature and its corresponding maximum density.   
Using standard cooling recipes, we approximated the gas 
temperature in equilibrium at solar metallicity by \citet{Bournaud:2010cf}
\begin{equation}
 T_{\rm eq}=10^4(\frac{n}{0.3})^{-1/2}\textrm{ Kelvin}.
\end{equation}
Equating the polytropic pressure floor with the gas pressure and having $\Delta x_{\rm min}=150\rm pc$, 
we set $\rm{T}_{\rm{min}}=3000\rm{  K}$ and $\rm{n}=2.7 \rm{ H}/\rm{cm}^3$.\\
Furthermore, we treat the gas thermodynamics using an optically thin cooling and heating function. Hydrogen and Helium chemical processes 
are computed with the assumption of photoionization equilibrium \citep{1991ApJ...377..365K}.
Metal cooling is also modelled at high and low temperatures, which included infrared hyperfine line cooling.
A uniform UV radiation background \citep{1996ApJ...461...20H} is switched on at reionization redshift chosen to be equal to 10.
We also use a simple model for self-shielding \citep{Roskar:2013pia}, which reduces the local UV radiation in high density gas regions and 
corrects the cooling and heating properties of the concerned gas.

\subsection{Star formation and feedback}
Star formation was modelled via a Schmidt law, where the formation rate was calculated as:
\begin{equation}
\dot \rho_\star = \epsilon_*\frac{\rho_{\rm gas}}{t_{\rm ff}}~~~{\rm for}~~~\rho_{\rm gas} > \rho_*{\rm .}
\end{equation}
\noindent
The star formation density threshold $\rho_*$ exactly corresponds to the maximum resolvable density defined in section \ref{subsec:Thesimulation:AMRandDM/gasdynamics}.
As suggested by observations \citep{Krumholz:2006jb}, the star formation efficiency parameter is set to a low value $\epsilon_{*}= 0.01$. 
This value is known to reproduce the Kennicutt-Schmidt relation quite well in comparable simulations \citep{2011MNRAS.410.1391A}, even though 
the physical star formation efficiency has more complex environmental dependencies \citep{2011ApJ...728...88G}. 
We use a stochastic model \citep{Rasera:2005gq} to create star particles of constant mass which are equal to the baryonic resolution of the simulation.
Once a cell is flagged for forming stars, it creates N stars where N is computed from a Poisson process with Poisson parameter 
$\lambda = \rho_* \Delta x^3 \Delta t/m_*$,  $\Delta t$ being the simulation time step.
According to the required rate, the star particles were then spawned in the cell. They received a velocity
equal to the local fluid velocity. The corresponding mass, momentum and internal energy
were removed from the gas of the parent cell.\\
Supernova (SN) feedback is modelled as thermal energy injection into the exploding star particle containing cell. After 10 Myr, each star particle releases 20\%
of its mass. This corresponds to a Chabrier Initial Mass Function \citep{Chabrier:2001dc} per single stellar population (SSP) undergoing 
a supernova. 10\% of the ejected mass is added to the metal content of the cell. Similar to the high resolution cell flag, the metal fraction is advected
passively by the hydrodynamical flow.\\
Furthermore, 
we use the implementation of \citet{2013MNRAS.429.3068T} and inject the SN energy in a non-thermal component.
The model tries to capture astrophysical non-thermal processes that are known to occur in supernova remnants, e.g. turbulance or cosmic rays, that cannot
be resolved within current simulations, but, on the other hand, they are able to reach energy densities that can significantly 
affect the dynamics of the propagating shock wave. These processes decrease with longer dynamical time scales compared to gas cooling, so that they 
gradually return energy back to the gas. The time evolution of this non-thermal energy can be described by 
\begin{equation}
 \rho \frac{\textrm{d}\epsilon_{\rm NT}}{\textrm{d}t}=\dot{E}_{\rm inj}-\frac{\rho \epsilon_{\rm NT}}{t_{\rm diss}}
\end{equation}
and is eventually driven by the non-thermal energy source $\dot{E}_{\rm inj}$ due to SN and the energy dissipation rate $t_{\rm diss}$ that we fixed 
to $20$ Myr. In practice, instead of modifying the hydrodynamical solver, the implementation adds the non-thermal pressure component directly to the total
gas pressure. As long as the non-thermal pressure in a gas cell is superior to the thermal component, cooling was neglected and was
re-activated when non-thermal becomes comparable to the thermal energy \citep{2013MNRAS.429.3068T}.\\
Because our spatial resolution is about of the order of magnitude of the galactic disk height, it
cannot capture the physics of individual molecular clouds. As a consequence, the star formation scheme tends to produce a spuriously
 homogenous distribution of young stars and, consequently, the associated feedback is similarly located. Therefore, 
 we chose to model stochastically exploding clouds \citep{Roskar:2013pia}: when a star particle reaches the required age,
 we draw a random number x between 0 and 1. If x is lower than the ratio of the star mass $m_{*}$ and the typical mass for GMC
 clouds in a galaxy $M_{\rm GMC}$ that we set to $2\cdot10^{6}$ M$_{\rm sun}$ \citep{2009ApJ...699..850K}, 
 a SN event is triggered. The released energy is then multiplied by the factor $M_{\rm GMC}/m_{*}$.
 In the end, these rare but more powerful explosion events conserve the total supernova energy and are closer
 to the energy released by individual giant molecular clouds in a disk galaxy.

\section{The galaxy at redshift 0}\label{sec:galaxy0}
This section analyses the morphology of the simulated galaxies at redshift 0. Stellar and gaseous disks are compared to observations. As we already stated, even
though we chose three MW-like DM halos for resimulation (Halo A is in the middle of MW-like galaxy mass range and Halo B and C are at the lower end. 
See section \ref{sec:InitCond} for the justification of our selection.), 
only one galaxy (Halo B) exhibits properties similar to the MW. Therefore, we will describe this galaxy more in detail without forgetting the 
results obtained for the other halos where we kept the same star formation and feedback recipes.
\subsection{Stellar disks}
\begin{figure*}
 \centering
 \begin{subfigure}[t]{0.45\linewidth}
   \includegraphics[width=1.\linewidth]{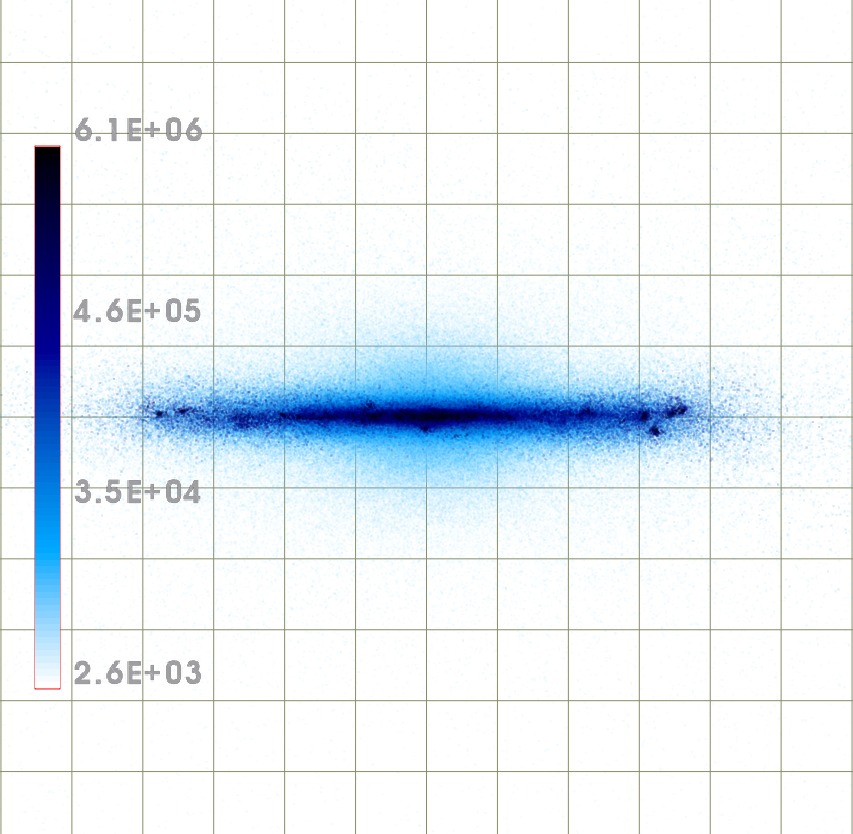}
   \includegraphics[width=1.\linewidth]{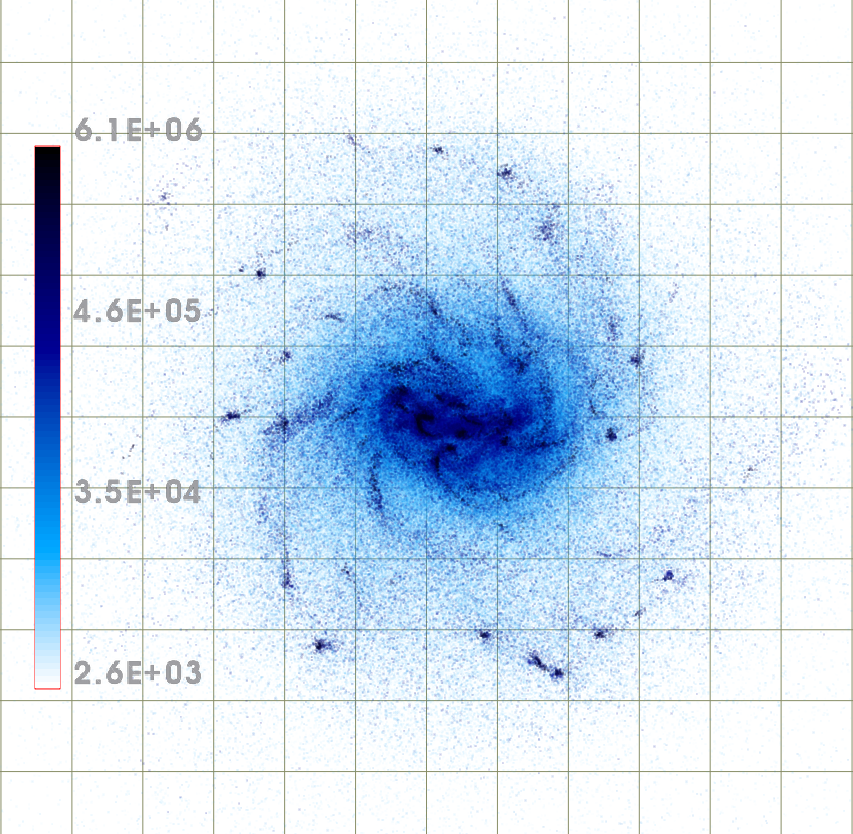}
  \caption{U-band luminosity $[\lo]$.}
  \end{subfigure}
  \begin{subfigure}[t]{0.45\linewidth}
     \includegraphics[width=1.0\linewidth]{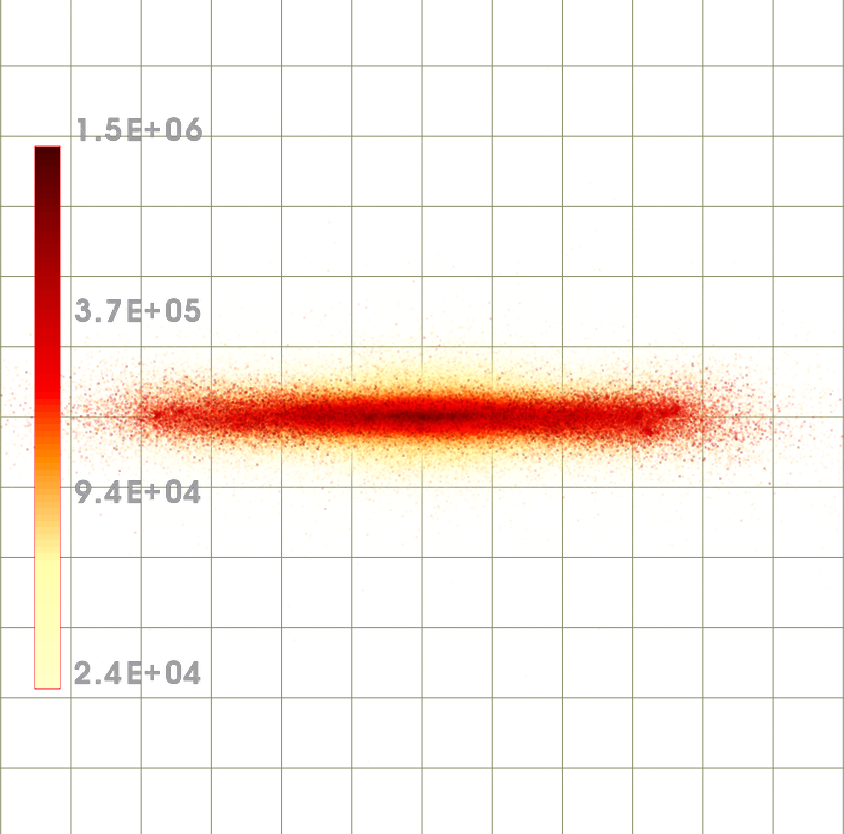}
     \includegraphics[width=1.0\linewidth]{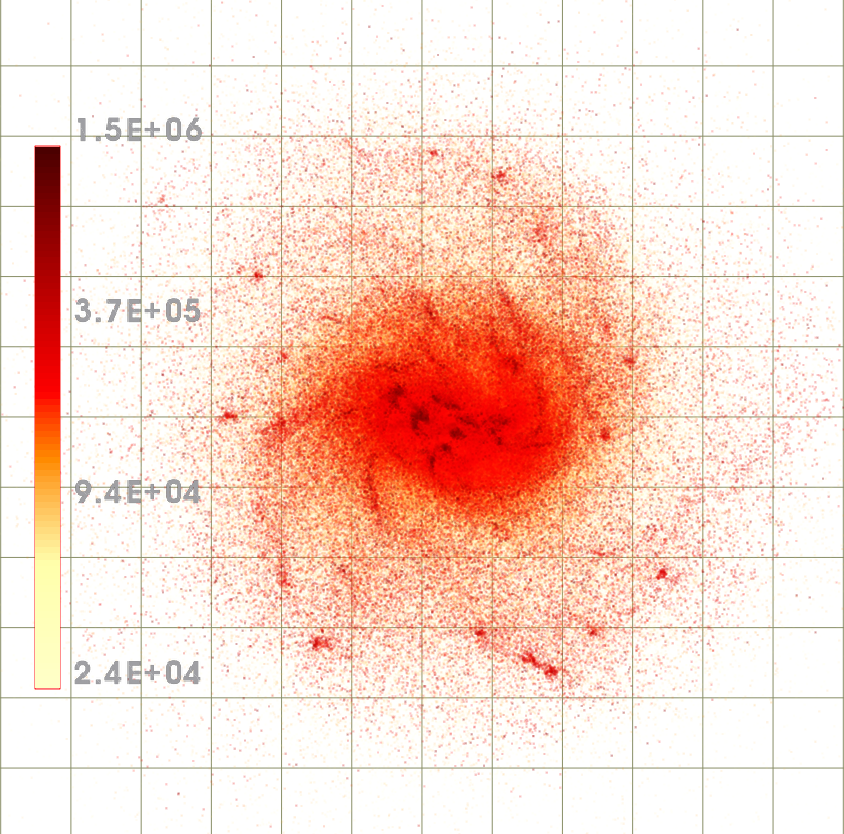}
       \caption{K-band luminosity $[\lo]$.}
  \end{subfigure}
  \caption{{\small Projected stellar luminosities  for Halo B at redshift 0. One grid cell is 5 kpc $\times$ 5 kpc. Luminosities (units are solar luminosities) 
  were computed from SSP
   integrated magnitudes calculated by the code CDM 2.5 \citep{Marigo:2007ns}, using the age and metallicity of the star particles.
   Visualization is done with glnemo2.}}
   \label{fig:starsmaps1}
\end{figure*}

\begin{figure*}
 \centering
 \begin{subfigure}[t]{0.45\linewidth}
   \includegraphics[width=1.\linewidth]{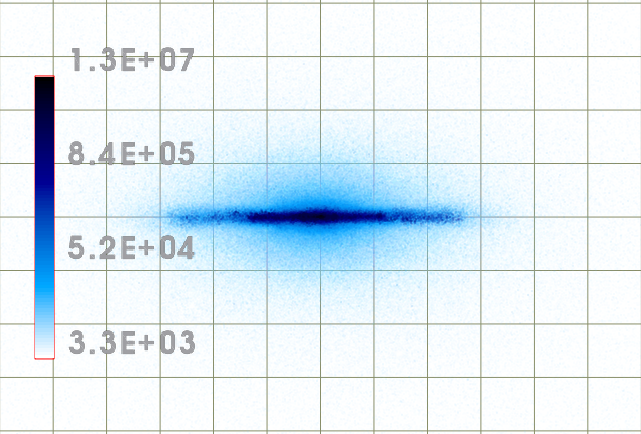}
   \includegraphics[width=1.\linewidth]{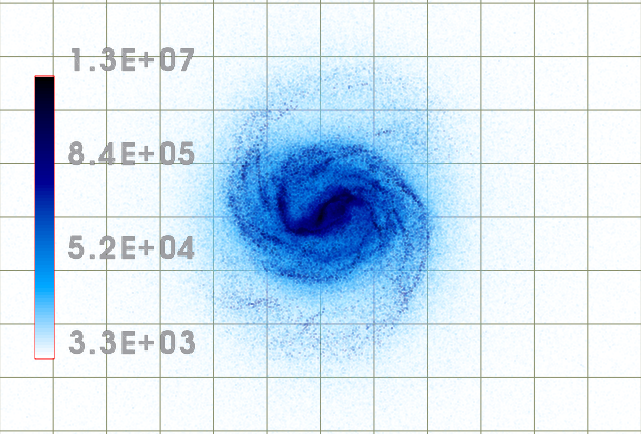}
  \caption{U-band luminosity $[\lo]$.}
  \end{subfigure}
  \begin{subfigure}[t]{0.45\linewidth}
     \includegraphics[width=1.0\linewidth]{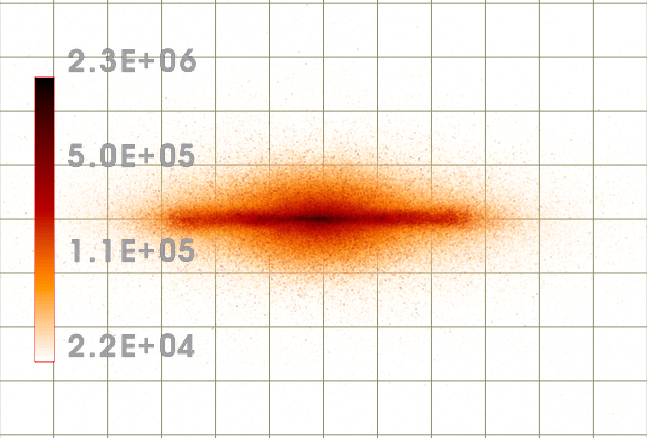}
     \includegraphics[width=1.0\linewidth]{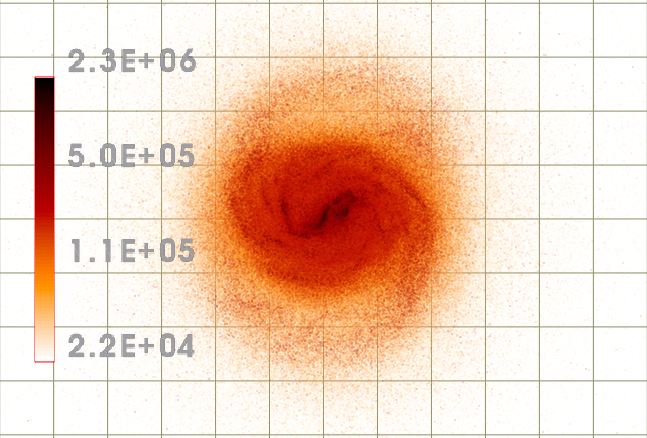}
       \caption{K-band luminosity  $[\lo]$.}
  \end{subfigure}
  \begin{subfigure}[t]{0.45\linewidth}
   \includegraphics[width=1.\linewidth]{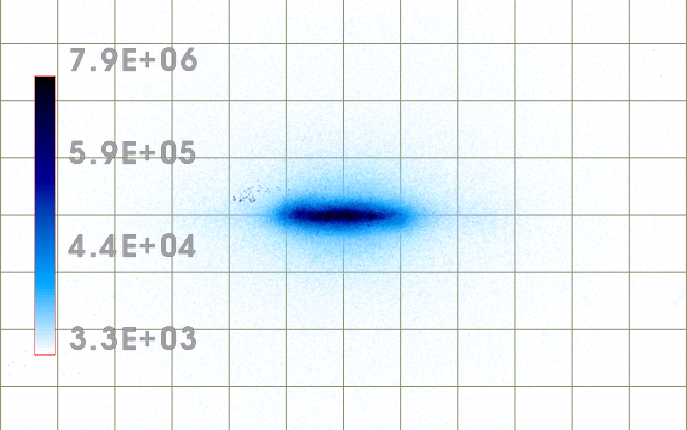}
   \includegraphics[width=1.\linewidth]{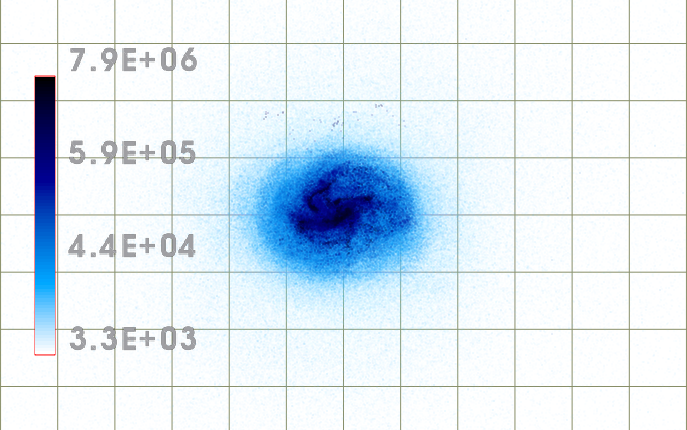}
  \caption{U-band luminosity  $[\lo]$.}
  \end{subfigure}
  \begin{subfigure}[t]{0.45\linewidth}
     \includegraphics[width=1.0\linewidth]{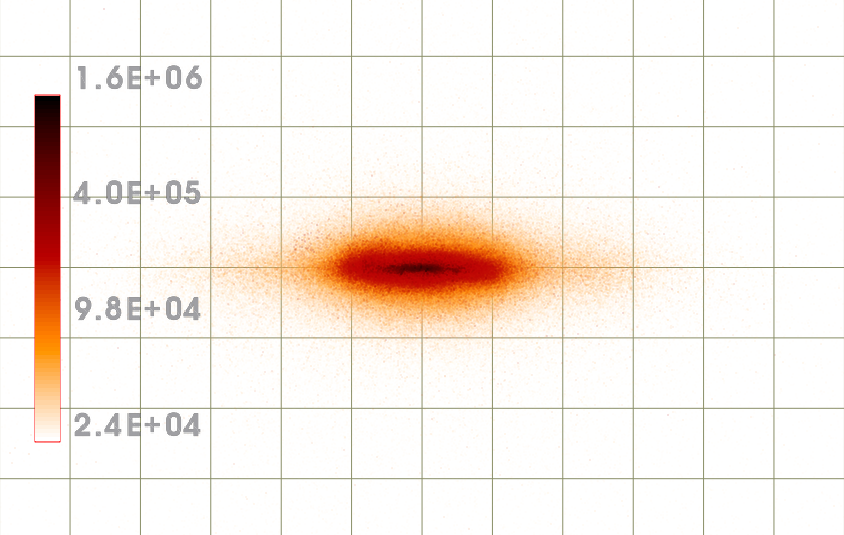}
     \includegraphics[width=1.0\linewidth]{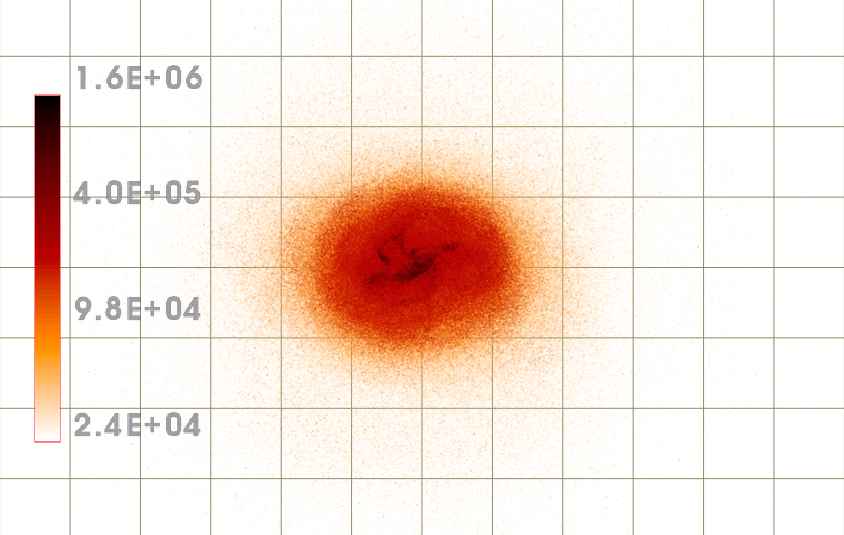}
       \caption{K-band luminosity  $[\lo]$.}
  \end{subfigure}
  \caption{Same figure than \ref{fig:starsmaps1} for the Halo A (above) and Halo C (below) at redshift 0.}
   \label{fig:starsmapsHaloA}
\end{figure*}

\begin{figure*}
 \centering
 \begin{subfigure}[t]{0.32\linewidth}
   \includegraphics[width=1.\linewidth]{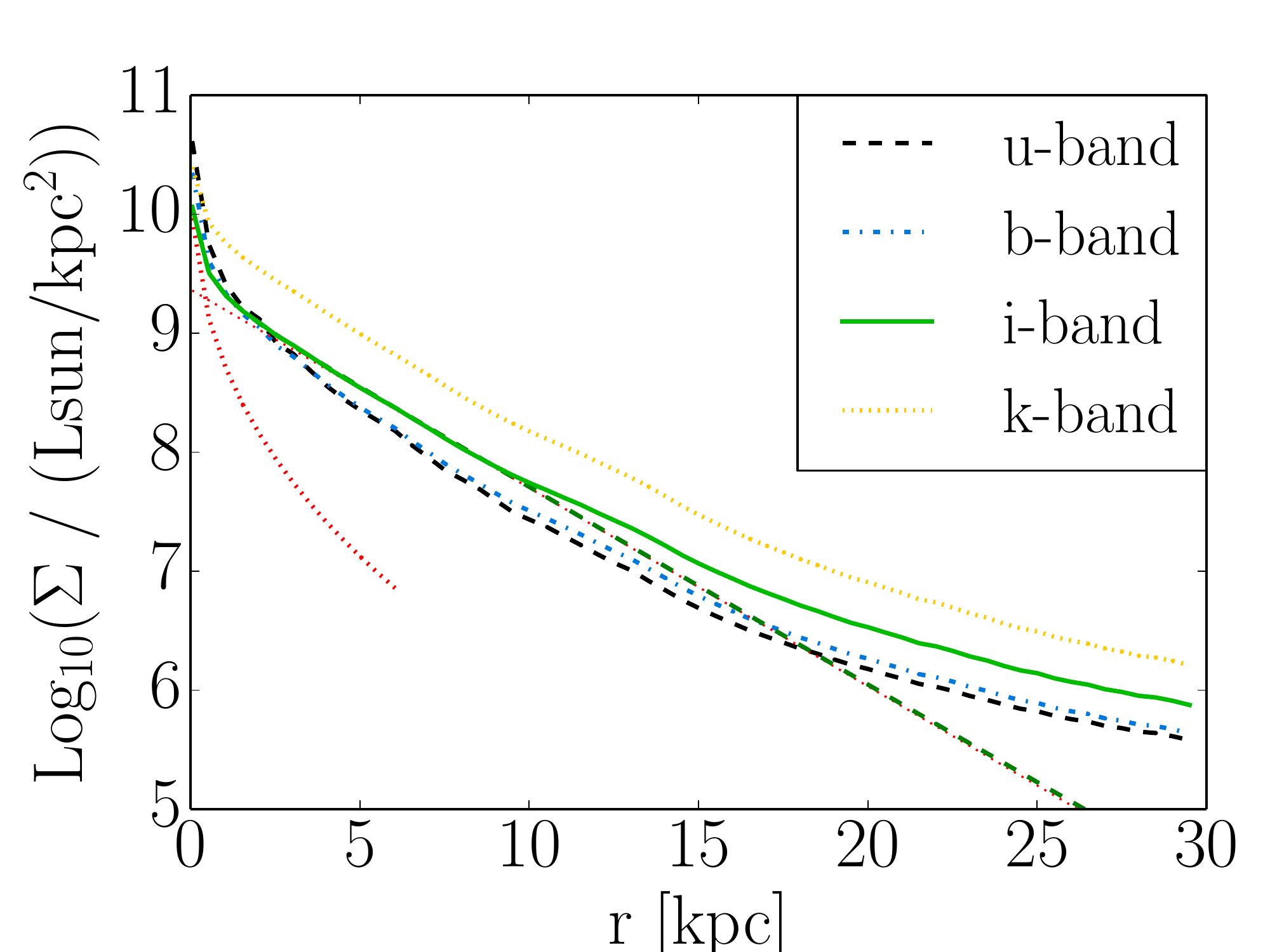}
  \caption{Halo A}
  \end{subfigure}
  \begin{subfigure}[t]{0.32\linewidth}
     \includegraphics[width=1.0\linewidth]{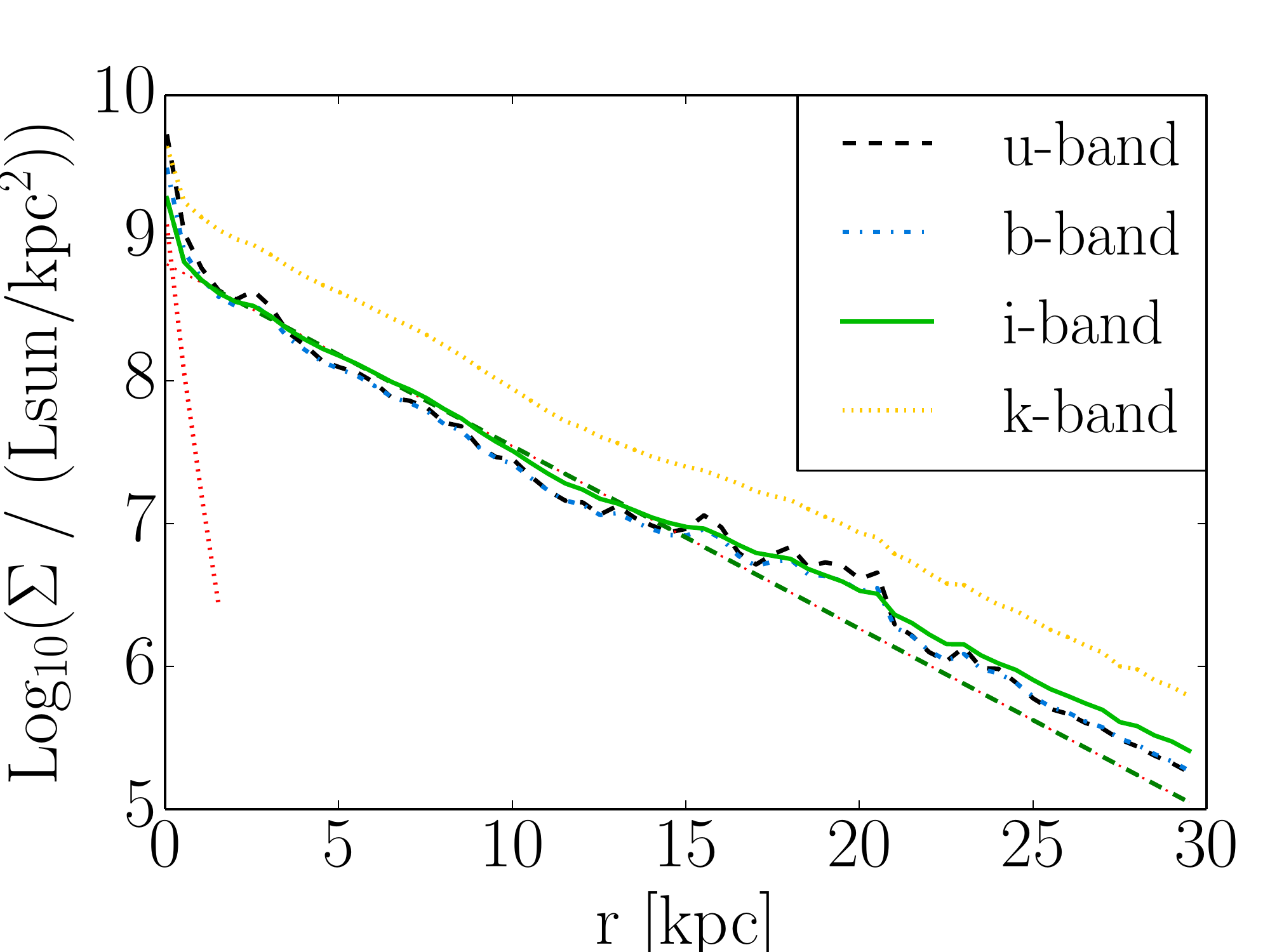}
       \caption{Halo B}
  \end{subfigure}
  \begin{subfigure}[t]{0.32\linewidth}
     \includegraphics[width=1.0\linewidth]{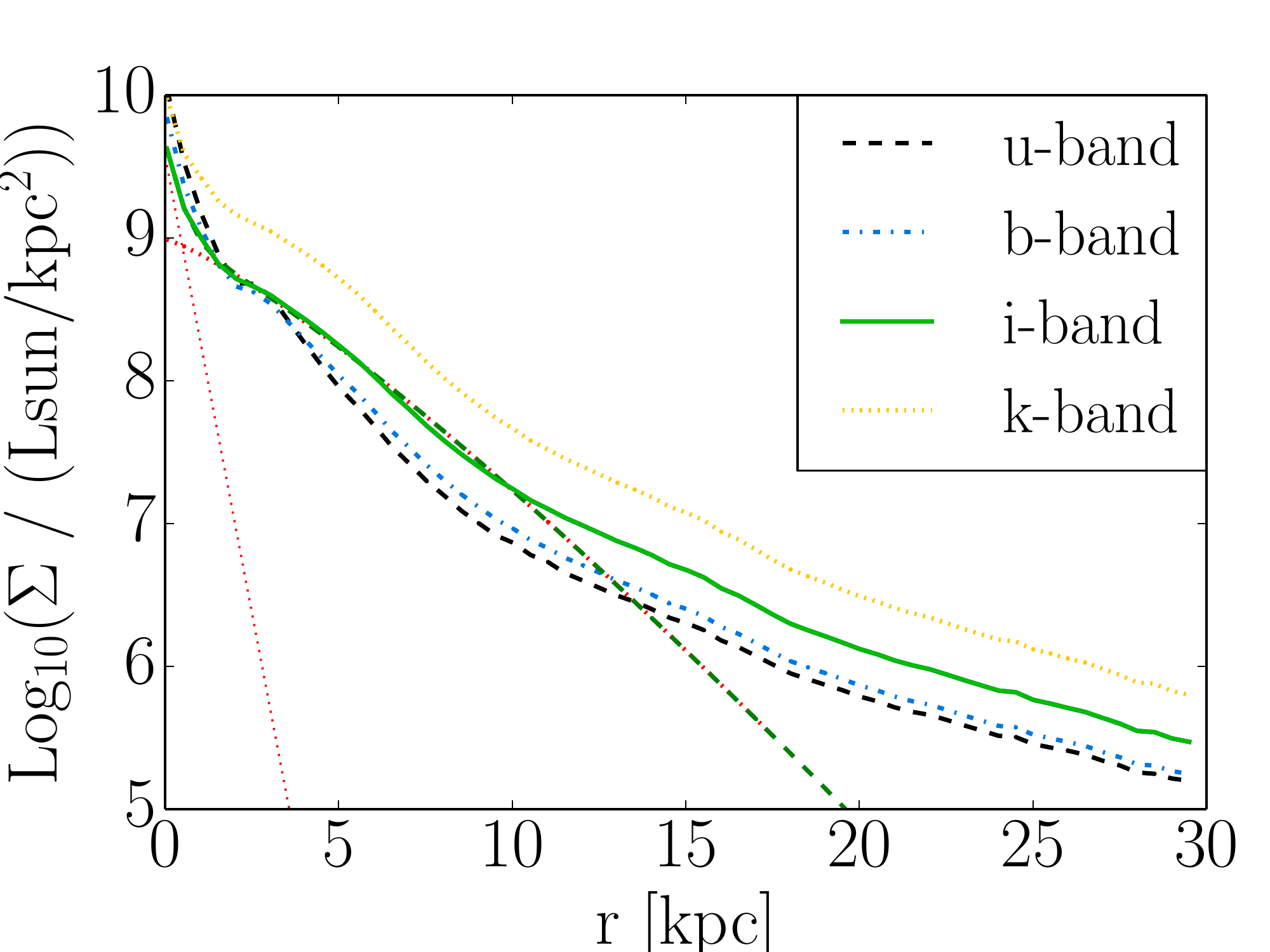}
       \caption{Halo C}
  \end{subfigure}
  \caption{Surface brightness profiles in x-y plane projection in 'UBIK' bands. The green dashed curve is the best fit for the I-band brightness profile. We fitted
  a two-component function consisting of an exponential disk function and a Sersic profile. The dotted red lines show the 2 components of the fit 
  whose values can be found in table \ref{tab:surfbright}.}
   \label{fig:surfacedensitystars}
\end{figure*}

In Figures \ref{fig:starsmaps1} and \ref{fig:starsmapsHaloA}, we show the side-on and face-on projection of the stellar luminosities 
of our simulated halos in the U \& K bands. 
The size of one grid cell is
5 kpc $\times$ 5 kpc. After extracting the halo 
from the simulation, we centered the particle positions on the highest star density. We diagonalised the position tensor of the star particles and rotated
the positions accordingly, with the result that the stellar disk lies in the x-y plane. The luminosities were computed 
using the code CDM 2.5\footnote{http://stev.oapd.inaf.it/cgi-bin/cmd}. More specifically, we  obtained the integrated magnitude for single stellar populations per 
solar mass, taking into account isochrones from \citet{Marigo:2007ns}, and assumed a Chabrier IMF and no dust absorption.
As a photometric system, we selected UBVRIJHK bands with Bessell filters. Then, we assigned the magnitudes to each stellar particle
considering its age and metal content and
converted the magnitudes to luminosities.\\
Halo A and Halo B show pronounced spiral disk features whereas Halo C has a much smaller stellar disk. The DM Halos B and C have almost the same mass whereas
Halo A has a mass three times higher. It is therefore not surprising that Halo A developed a more massive stellar disk. \\
In order to quantify the stellar disk properties and to compare our results with observations, we calculated the photometric surface brightness for different bands
shown in Figure \ref{fig:surfacedensitystars}. We projected the stars into the x-y plane, summed the luminosity in circular annuli and divided the total luminosity
per bin by the surface area to get the surface brightness. Moving away from the galactic center, it can be seen that the surface brightness decreases exponentially in all 
bands. At the center, in every simulated halo, we additionally find an excess on top of the disk component, generally called the galactic bulge.
Interestingly, Halo B presents an excess in the radius
range 15-20 kpc due to the presence of its massive elongated spiral arms. This feature is more pronounced in the shorter wavelengths (U \& B band),
because its spiral arms host massive star forming regions inside giant gas clouds (see Figure \ref{fig:starsmaps1}). 

As a next step, we performed a two-component fit of the brightness density profiles in the I band 
(green dashed line on Figure \ref{fig:surfacedensitystars}) where we fitted the central bulge with a S\'ersic profile and, at the same time, the disk component with an exponential profile (red dotted lines).
The best fit values are listed in Table \ref{tab:surfbright}. We verified that changing the fitting range did not alter too much the best fit values.\\
The central galaxy in Halo A has a disk scale length of about 2.45 kpc
as well as a massive bulge of old stars. The stellar disk reaches out to almost 15 kpc. In Figure \ref{fig:starsmapsHaloA},
the presence of a massive bar at the galactic center can be observed.\\
Halo B has formed a large spiral galaxy whose stellar disk reaches out to almost 20 kpc. The brightness profile
is dominated by an exponential disk of scale radius $\sim$ 3.4 kpc and a small bulge in the center. The Disk-to-Total (D/T) flux ratio in the I-band 
is remarkably high so that 
the Bulge-to-Disk (B/D) flux ratio is $\sim$ 14 \%, which is in agreement with expectations for Sbc-type galaxies (B/D = $0.138^{+0.164}_{-0.083}$) like the MW \citep{2008gady.book.....B} 
and Sc-type galaxies (B/D = $0.087^{+0.276}_{0.041} $) 
(median values$\pm$ 68/2 \% of the distribution on either side of the median, taken from \citet{2008MNRAS.388.1708G}, who 
included corrections for dust whereas we did not.).
Halo C's central galaxy is dominated by its central stars as is revealed by the dominance of the bulge fit. It has a shorter disk of less than 10 kpc.\\
\begin{table*}
\centering
\begin{tabular}{ccccccccccc}
\hline
   Run & $\log_{10}\Sigma_{\rm d}$ & $R_{\rm d}$ & $\log_{10}\Sigma_{\rm Sersic}$ & $r_{\rm Sersic}$ & $n$ & $B/D $ & $D/T$ & Fitting limit \\
       & $L_\odot {\rm kpc}^{-2}$ & $(\kpc)$ & $L_\odot {\rm kpc}^{-2}$ & $(\kpc)$ &  & & & kpc \\ 
\hline
Halo-A & 9.3673 & 2.6051 & 8.4864 & 1.3929 & 2.4559 & 0.509 & 0.663 & 17 \\ 
Halo-B & 8.8223 & 3.3943 & 8.3506 & 0.3899 & 1.2119 & 0.141 & 0.876 & 21 \\ 
Halo-C & 9.5747 & 0.3395 & 8.4205 & 3.9800 & 0.8205 & 2.289 & 0.303 & 10 \\ 
\hline
\end{tabular}
\caption{Parameters of the surface brightness profile decomposition in the I band. For
  each run the columns give (from left to right): the logarithm of the central brightness
  of the disk, the disk scale-length, the logarithm of the
  bulge brightness at the center,
  the bulge scale radius, the Sérsic index of the bulge, the B/D flux ratio and the D/T flux ratio.}
\label{tab:surfbright}
\end{table*}
\begin{figure}
 \centering
    \includegraphics[width=1.\linewidth]{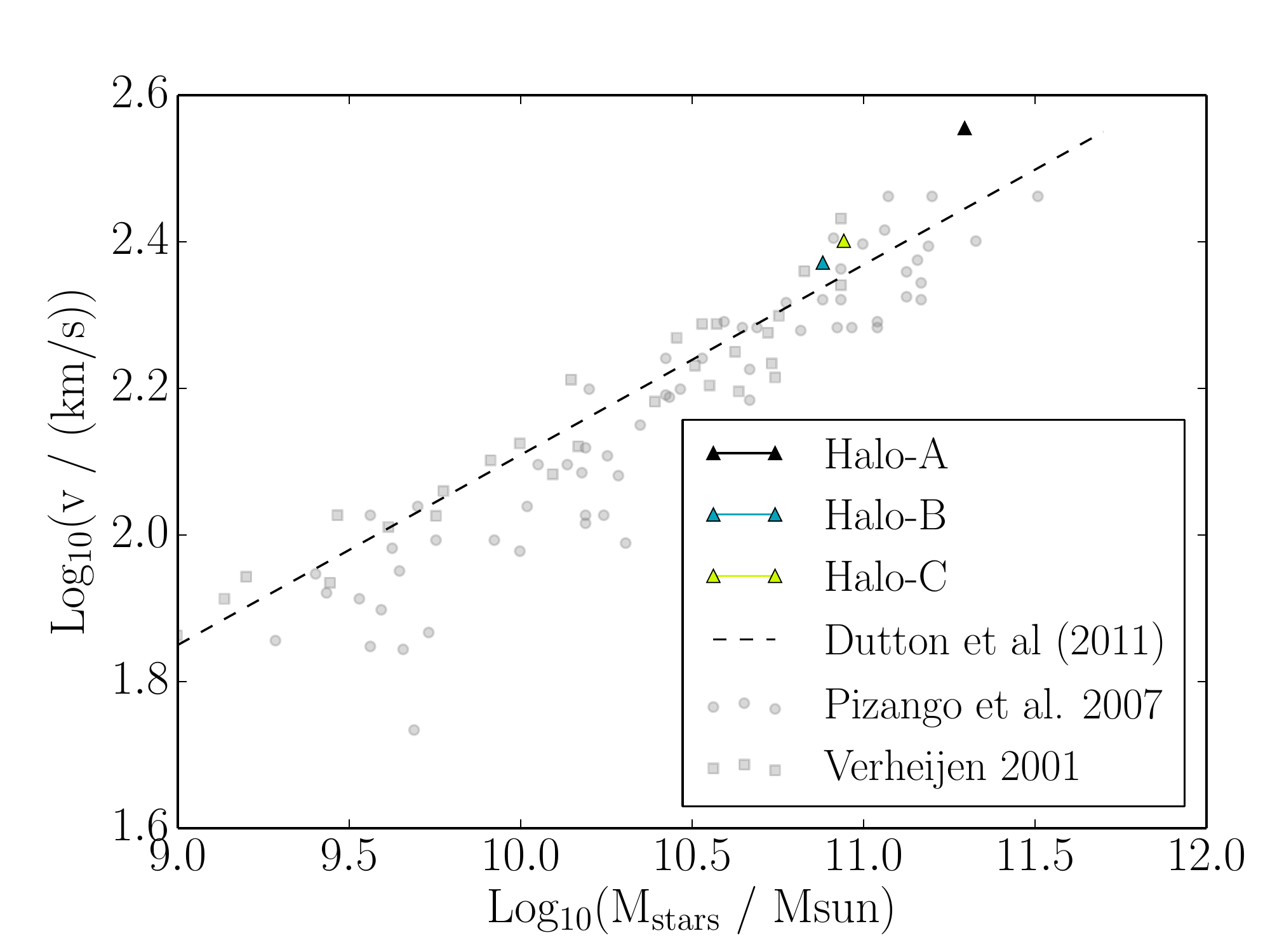}
 \caption{The Tully-Fisher relation for the 3 halos at redshift 0.}
 \label{fig:tullyfisher}
 \end{figure}
In figure \ref{fig:tullyfisher}, we show the Tully-Fisher relation \citep{1977A&A....54..661T} of the simulated galaxies at redshift 0.
In order to compute the total stellar mass belonging to the central galaxy, we considered the mass of all the star particles that lie between the galaxy center (defined as
the highest star density point) and 10\% of the radius R$_{97}$. The rotation velocity was calculated using the usual definition of the circular 
velocity, i.e. $v_{c}(r)=\sqrt{GM(<r)/r}$, where M($<$r) is the total mass inside radius r. We chose to calculate the value of $v_c$ at the radius that encloses
80\% of the stellar mass. At this point, the circular velocity has already reached its maximum. Because the obtained rotation curves are quite flat, they do not vary 
dramatically if one calculates the rotation velocity at a slightly shorter or longer radius. As grey symbols, we include a data set of observed galaxies 
\citep{2001ApJ...563..694V,2007AJ....134..945P}. \citet{2011MNRAS.410.1660D} already analysed their data and we show also their best fit to the Tully-Fisher relation. Halo B and C lie
within the observational trend. The galactic star mass in Halo A is a bit too high to lie within the observational data range but it follows the tendency of the fit and
is not further than some observed galaxies.\\
Gas fractions defined as $f_{\rm gas}=\frac{M_{\rm gas}}{M_{\rm gas}+M_{\rm stars}}$, where we take into account all baryons inside 10\% of R$_{97}$,
are equal to 0.12, 0.28 and 0.136 for Halo A, B and C respectively. These values allow an on-going star formation at redshift 0. Furthermore, having calculated
the R-band magnitudes of the stellar populations of the three galaxies (Halo A: -23.03; Halo B: -22.12; Halo C: -22.15), we conclude that our results are 
similar to the sample of simulations performed by \citet{2014MNRAS.437.1750M} (compare their figure 11) and
lie within the sample of the observational data of galaxies of HI and optical observations compiled
by \citet{1999AJ....117.1668H}.\\
In figure \ref{fig:vcirc}, we show the circular velocities (defined as above for the Tully-Fisher relation) of the three halos at redshift 0.
We calculated the mass contribution for each component
and plotted the corresponding curve for the stars in yellow, the DM in black, the gas in red.
The total circular velocity resulting from the sum of all the mass is shown in blue. We also add the observational Milky Way data from \citet{2009PASJ...61..227S}.
\begin{figure*}
 \centering
 \begin{subfigure}[t]{0.32\linewidth}
   \includegraphics[width=1.\linewidth]{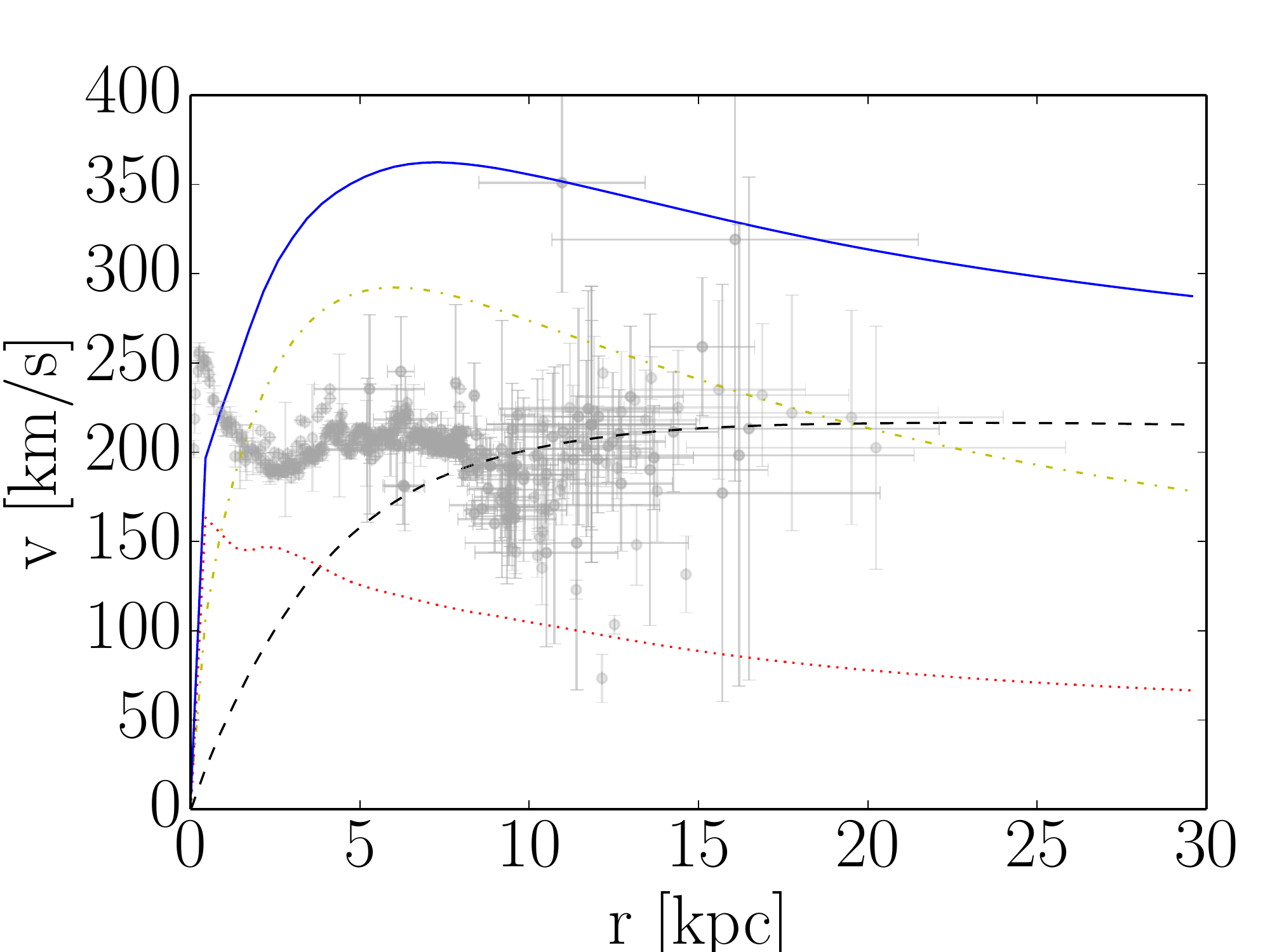}
  \caption{Halo A}
  \end{subfigure}
  \begin{subfigure}[t]{0.32\linewidth}
     \includegraphics[width=1.\linewidth]{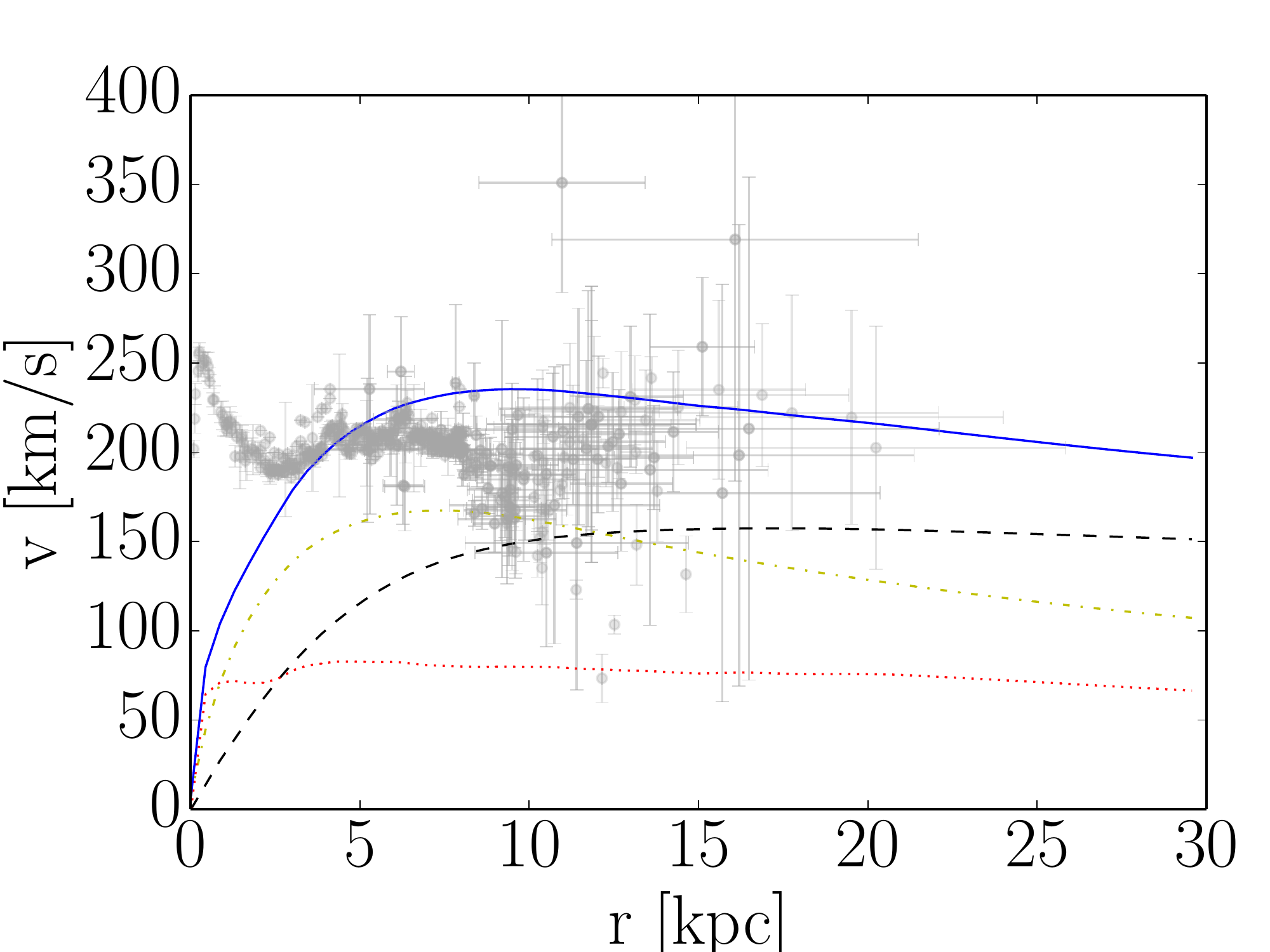}
       \caption{Halo B.}
  \end{subfigure}
  \begin{subfigure}[t]{0.35\linewidth}
     \includegraphics[width=1.\linewidth]{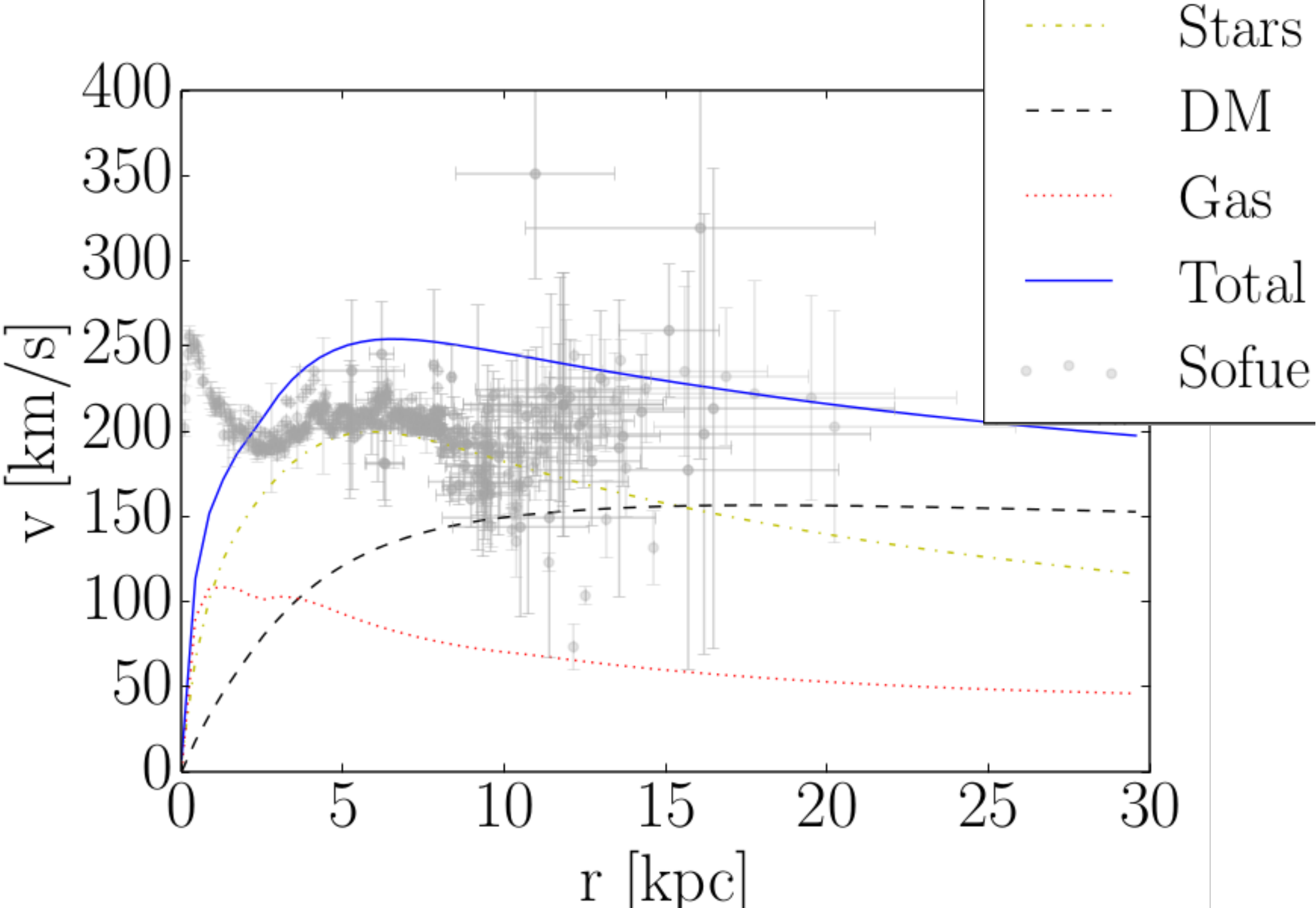}
       \caption{Halo C}
  \end{subfigure}
    \caption{Circular velocities $v_{c}(r)=\sqrt{GM(<r)/r}$ of the simulated halos at redshift 0. Shown are different mass components and the Milky Way observations
  from \citet{2009PASJ...61..227S} (see the plot legend).}
   \label{fig:vcirc}
\end{figure*}
Halo B and C show flat rotation curves and lie within the MW data. Halo A has a more peaked rotation curve because it locked a lot of mass into its stellar 
component. In the three cases, the gas component has a minor contribution at the outer part of the disk, but near the center it closely follows  
the stellar component and is superior to the contribution of the DM mass.\\
Regarding the rotation curve, Halo B formed the most MW resembling galaxy of our sample. The maximum circular velocity of 235 km/s is reached at 9.63 kpc.
Looking at the circular velocity at 8 kpc (that is about the distance separating the solar system from the MW center), it reaches 233 km/s and it lies well
within the observational data. Going out to 60 kpc, the rotational velocity is equal to 162 km/s, a value in the data range favored for the MW 
\citep{2008ApJ...684.1143X}.

\subsection{Gas}
\begin{figure*}
 \centering
 \begin{subfigure}[t]{0.45\linewidth}
   \includegraphics[width=1.\linewidth]{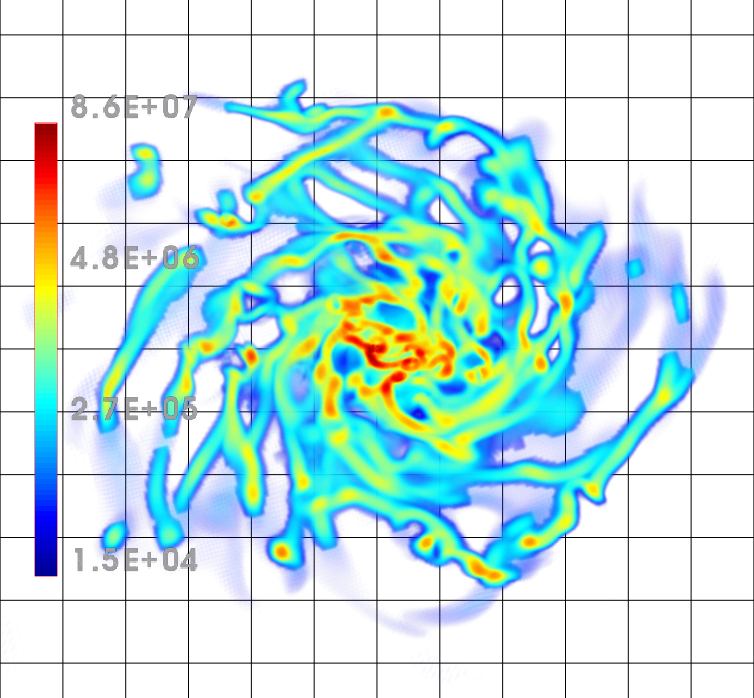}
  \caption{Halo B: Face-on view}
  \end{subfigure}
  \begin{subfigure}[t]{0.45\linewidth}
     \includegraphics[width=1.0\linewidth]{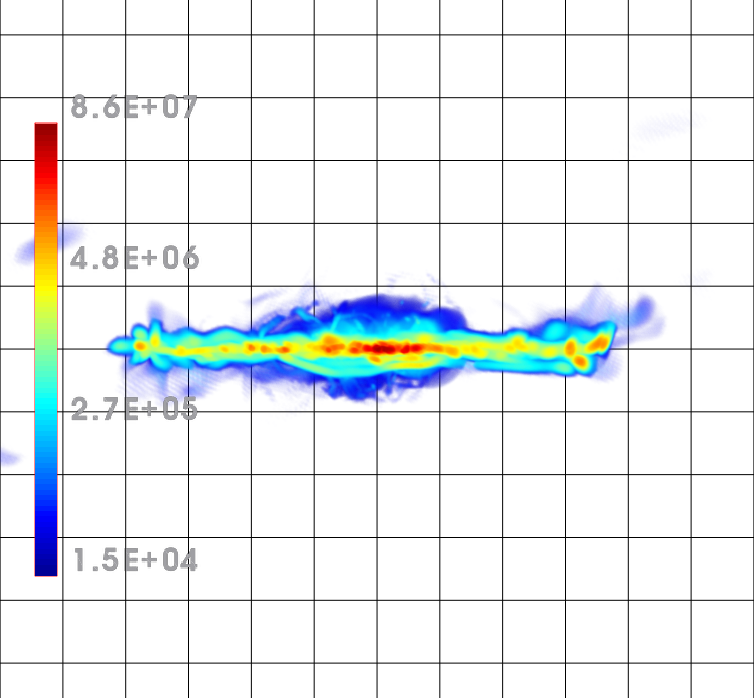}
       \caption{Halo B: Side-on view}
  \end{subfigure}
  \caption{The gas density [$\mo/\rm{kpc}^3$]. Grid cell size is 5 kpc$\times$5 kpc.}
   \label{fig:gasdensHaloB}
\end{figure*}

\begin{figure*}
 \centering
 \begin{subfigure}[t]{0.45\linewidth}
   \includegraphics[width=1.\linewidth]{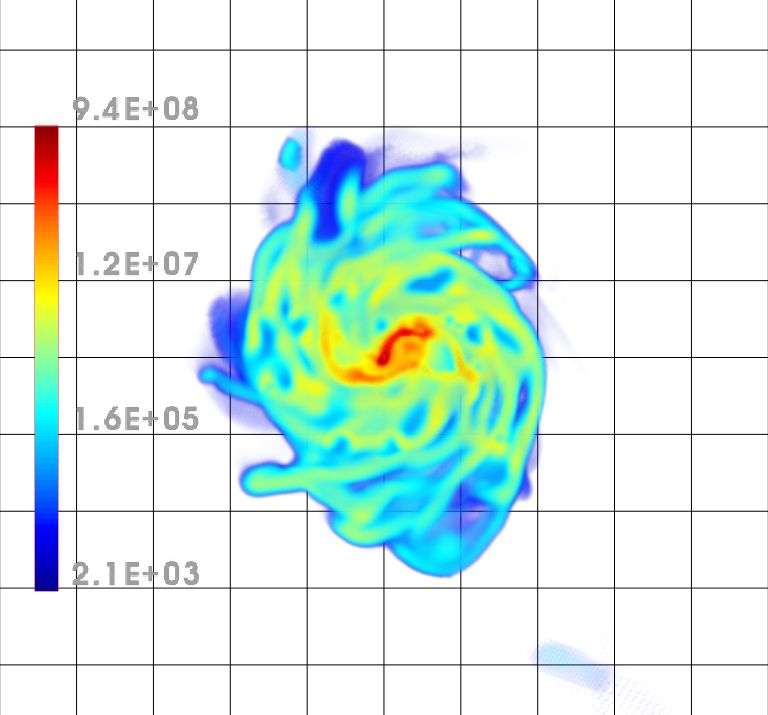}
  \caption{Halo A: Face-on view}
  \end{subfigure}
  \begin{subfigure}[t]{0.45\linewidth}
     \includegraphics[width=1.0\linewidth]{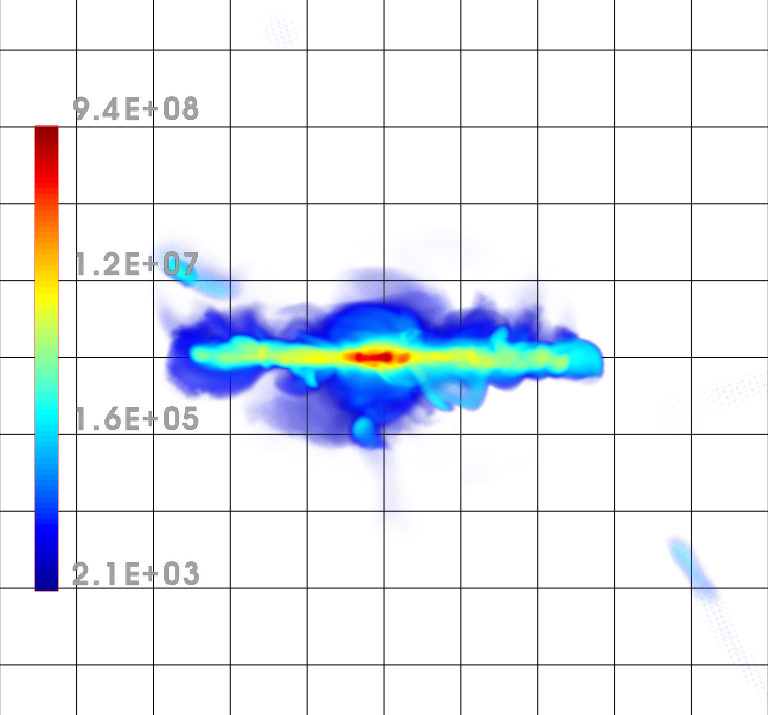}
       \caption{Halo A: Side-on view}
  \end{subfigure}
  
  \begin{subfigure}[t]{0.45\linewidth}
   \includegraphics[width=1.\linewidth]{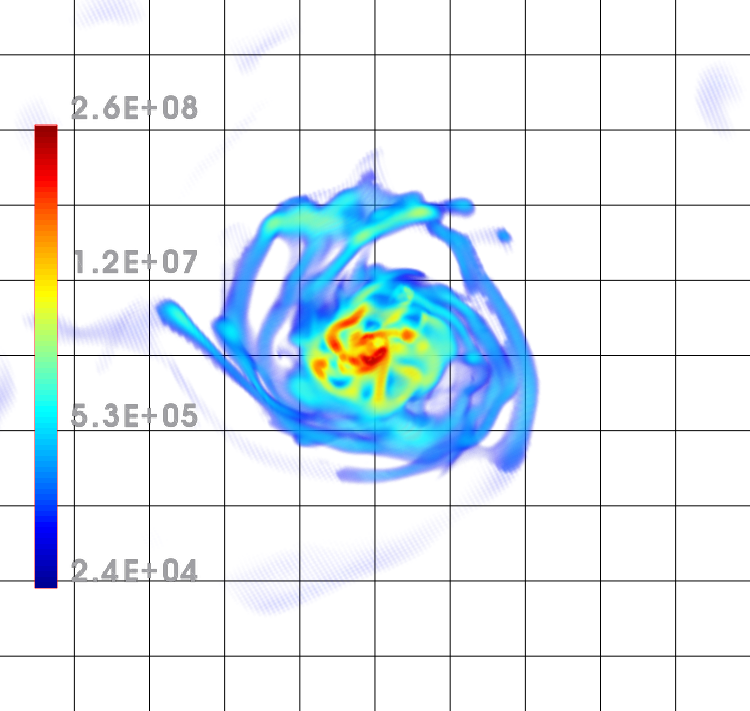}
  \caption{Halo C: Face-on view}
  \end{subfigure}
  \begin{subfigure}[t]{0.45\linewidth}
     \includegraphics[width=1.0\linewidth]{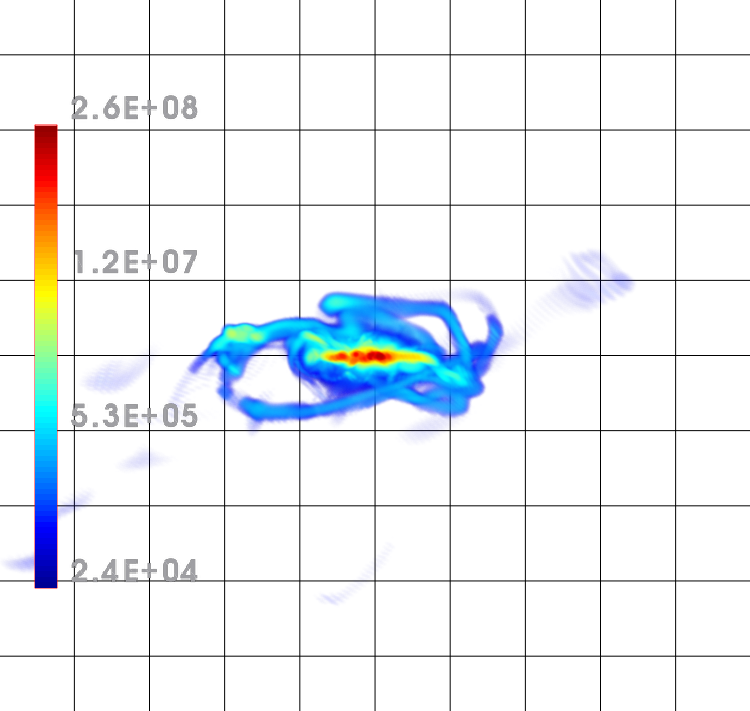}
       \caption{Halo C: Side-on view}
  \end{subfigure}
    \caption{The gas density [$\mo/\rm{kpc}^3$]  for Halo A (above) and Halo C (below). Grid cell size is 5 kpc$\times$5 kpc.}
    \label{fig:gasdensHaloAC}
\end{figure*}

In the Figures \ref{fig:gasdensHaloB} and \ref{fig:gasdensHaloAC}, we show the gas density of the three central galaxies in the same projection plane
as the stellar luminosity disk maps. The grid's cell size is again 5 kpc$\times$5 kpc. The gas disks are more "flocculent" in contrast with the smoother
gas disks produced by \citet{2014MNRAS.437.1750M}. The gas morphology of the disk is closer to the results obtained by 
\citet{2011ApJ74276G} and \citet{2013MNRAS.434.3142A}: SN feedback is at the origin of hot outflowing gas, creating large holes in the gas disk.\\
~
We verified that the densities that are reached when the gas is forming stars are realistic and that star formation rates follow a Kennicutt-Schmidt law 
$\Sigma_{\rm SFR}\propto\Sigma^{\rm n}_{\rm gas}$ with $n=1.4$ (see \citet{2012ARA&A..50..531K} for a recent review)
and lie
within observational data \citep{2008AJ....136.2846B} (figure \ref{fig:Kennicuttplot}). To do so, we projected the gas contained inside a sphere of a 20 kpc radius
to the x-y plane and built subsequent rings, each one being about 660 pc large. We included only the gas cells from the disk into the calculations, i.e., whose
vertical distance from the disk plane is less than 1 kpc. Only stars younger than 50 Myr were selected in the same corresponding rings. 
Our results concur with the low star formation efficiency model from \citet{2011MNRAS.410.1391A}.

\section{Formation History}\label{sec:formHist}
To calculate the star formation history (SFH), we selected only the stars at redshift 0 closer than 10\% of R$_{97}$ to the galactic center (defined as the 
highest star density) and trace the evolution of their formation in figure \ref{fig:SFH}. Halo B has the most non-perturbed SFH: it rises 
quickly until about redshift 2 
and then falls down steadily to redshift 0. Its star forming rate today, defined as the number of stars created in the last 50 Myr, is 4.54 $\mo$/year, a bit 
higher than estimations for the MW's rate (ranging from 0.9 to 2.2 $\mo/\rm yr$ for 
different IMF slopes \citep{2010ApJ...709..424M}).
In the case of Halo C, the star formation rate (SFR) stays high from redshift 3 to later than redshift 1, with a SFR of 3.62 $\mo$/yr at z$=0$.
Halo A has very high star formation rates during all its history. It increases rapidly up to 30 $\mo$/year at redshift 2, begins to diminish after redshift 1, and reaches the rate 
of 8.88 $\mo$/year at redshift 0. If one compares these SFH with predictions for MW-like halos derived 
from semi-analytical models combining stellar mass functions with merger histories of halos \citep{2013ApJ...770...57B}, the SFH of Halo
A is too high for halos with mass $\simeq 10^{12}\mo$. Halo B and C, though, show a star creation rate over time whose differences with the aforementioned models
are worth discussing. 
The peak of the SFR reaches a
rate of about 10 $\mo$/yr favored by these models, even though its position is shifted to higher redshifts with respect to these models 
(the peak of SFH predicted by \citet{2013ApJ...770...57B} for a similar halo mass range being between redshift 1 and 2). Furthermore, the 
SFR of Halo B and C stays at a too high level after redshift 1 and does not decrease sufficiently as can be seen on figure \ref{fig:SFH}. In fact, 
the SN feedback model that we used in the simulation is not able to address the challenges set by these semi-analytical models correctly.
Nevertheless, we will
argue in the next paragraph why we take these comparisons with caution.\\
\begin{figure*}
 \centering
 \begin{subfigure}[t]{0.45\linewidth}
  \includegraphics[width=1.\linewidth]{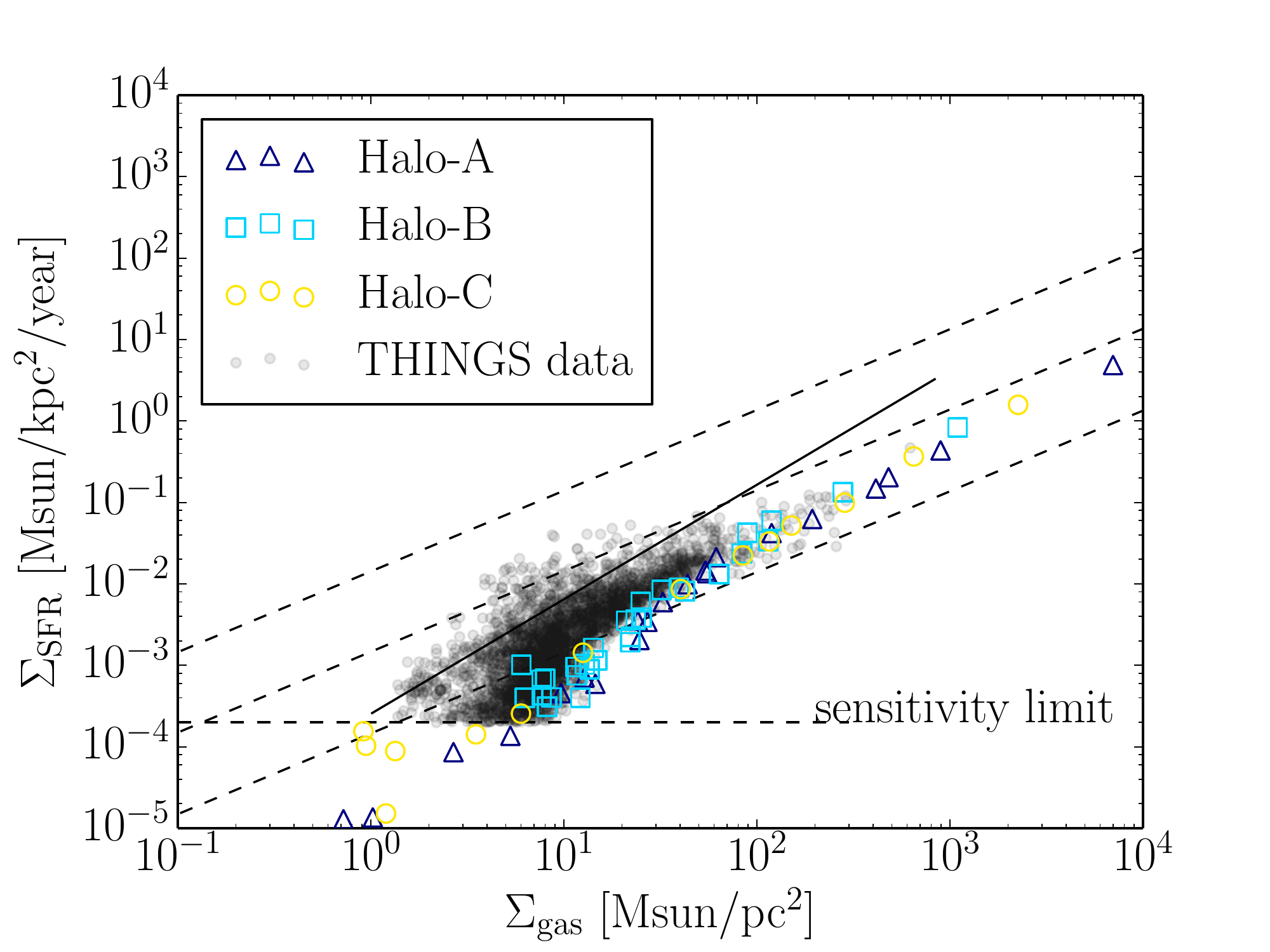}
 \end{subfigure}
 \begin{subfigure}[t]{0.45\linewidth}
   \includegraphics[width=1.\linewidth]{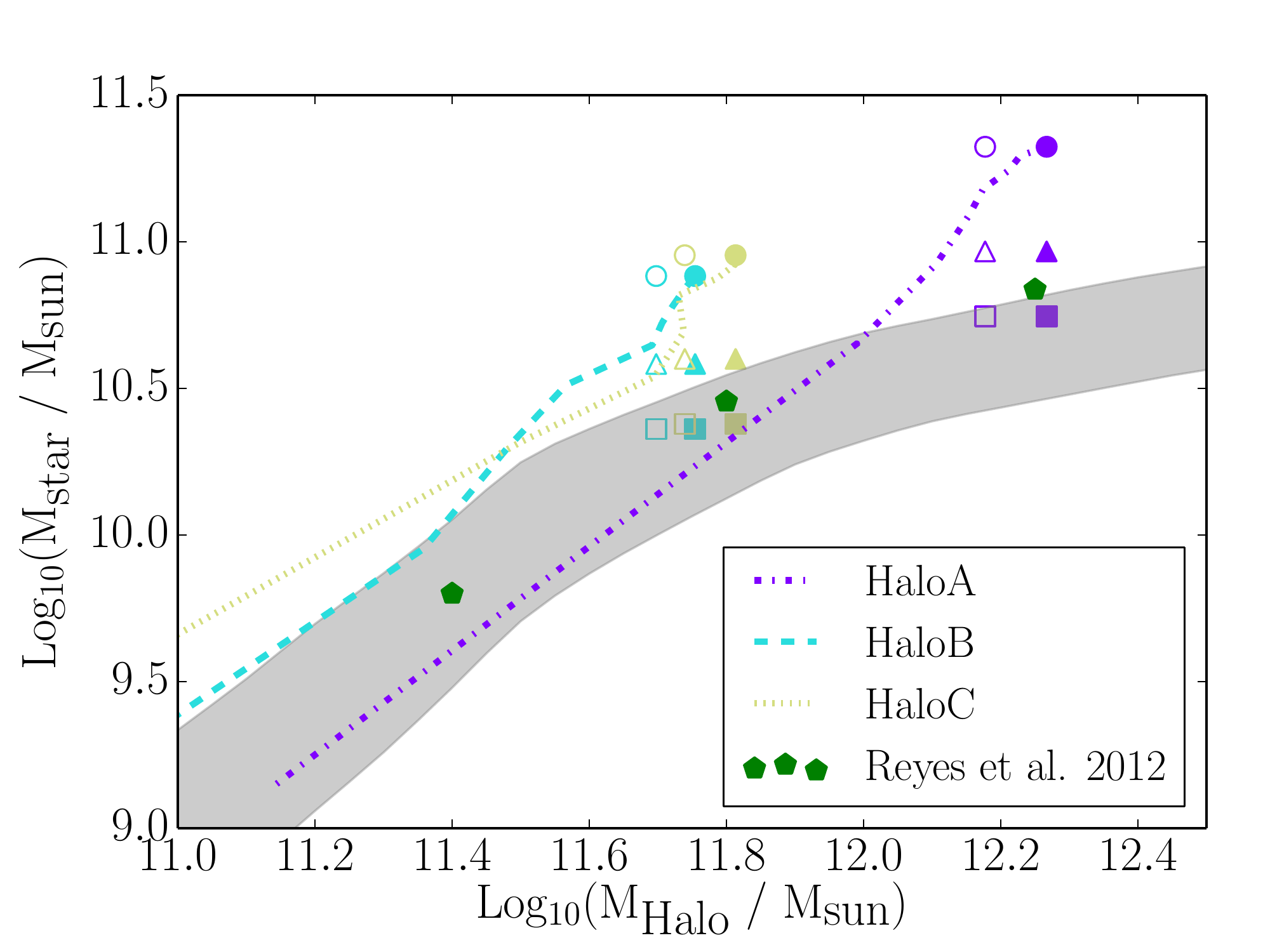}
 \end{subfigure}
  \caption{Left panel: star formation rate density for corresponding gas density at redshift 0. We considered all the stars younger than 50 Myr. Grey points are
 radial data from seven spiral galaxies taken from the THINGS survey \citep{2008AJ....136.2846B} where $\Sigma_{\rm gas}=1.36\cdot(\Sigma_{\rm HI}+\Sigma_{\rm H_2})$
 (with a factor of 1.36 accounting for helium). 
 Diagonal dotted lines show lines of constant SFE$=\Sigma_{\rm SFR}/\Sigma_{\rm gas}$ , indicating the level of $\Sigma_{\rm SFR}$ 
 needed to consume 1, 10 and 100 \% of the gas reservoir in 100 Myrs. 
 The solid black line is the original Kennicutt-Schmidt relation n=1.4 from \citet{1998ARA&A..36..189K}.\\
 Right Panel: the stellar mass evolution over time (from redshift 4 to 0) against the Halo Mass (DM inside R$_{97}$), compared to the prediction from abundance matching
 technique \citep{2013MNRAS.428.3121M} (the grey-shaded area indicates the 1$\sigma$ confidence level) and
 the results of a stacked weak lensing study of SDSS galaxies by \citet{2012MNRAS.425.2610R} (green pentagons). 
 Circles show the results at redshift 0 taking into account
 the sum of all the stellar mass inside 10\% of R$_{97}$. Triangles
 show the total stellar mass inside the Petrosian radii. Squares mark the stellar mass estimation based on the B-V color and
 V total magnitude of the central galaxies (see section \ref{sec:formHist} for further explanations). White edgecolored markers show the results for M$_{200}$.}
 \label{fig:Kennicuttplot}
 \end{figure*}
\begin{figure*}
 \centering
 \begin{subfigure}[t]{0.31\linewidth}
   \includegraphics[width=1.\linewidth]{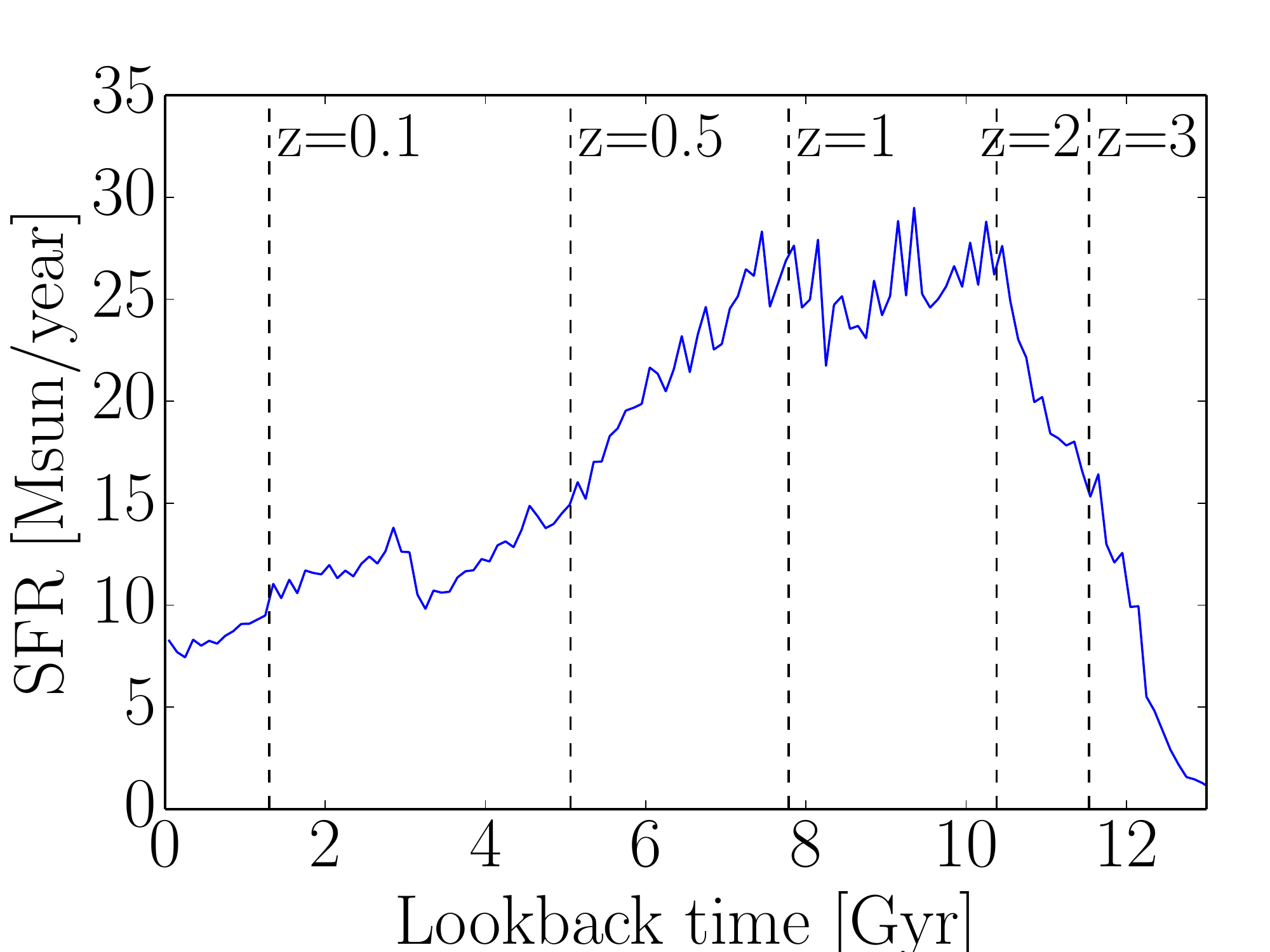}
  \caption{Halo A}
  \end{subfigure}
  \begin{subfigure}[t]{0.31\linewidth}
     \includegraphics[width=1.0\linewidth]{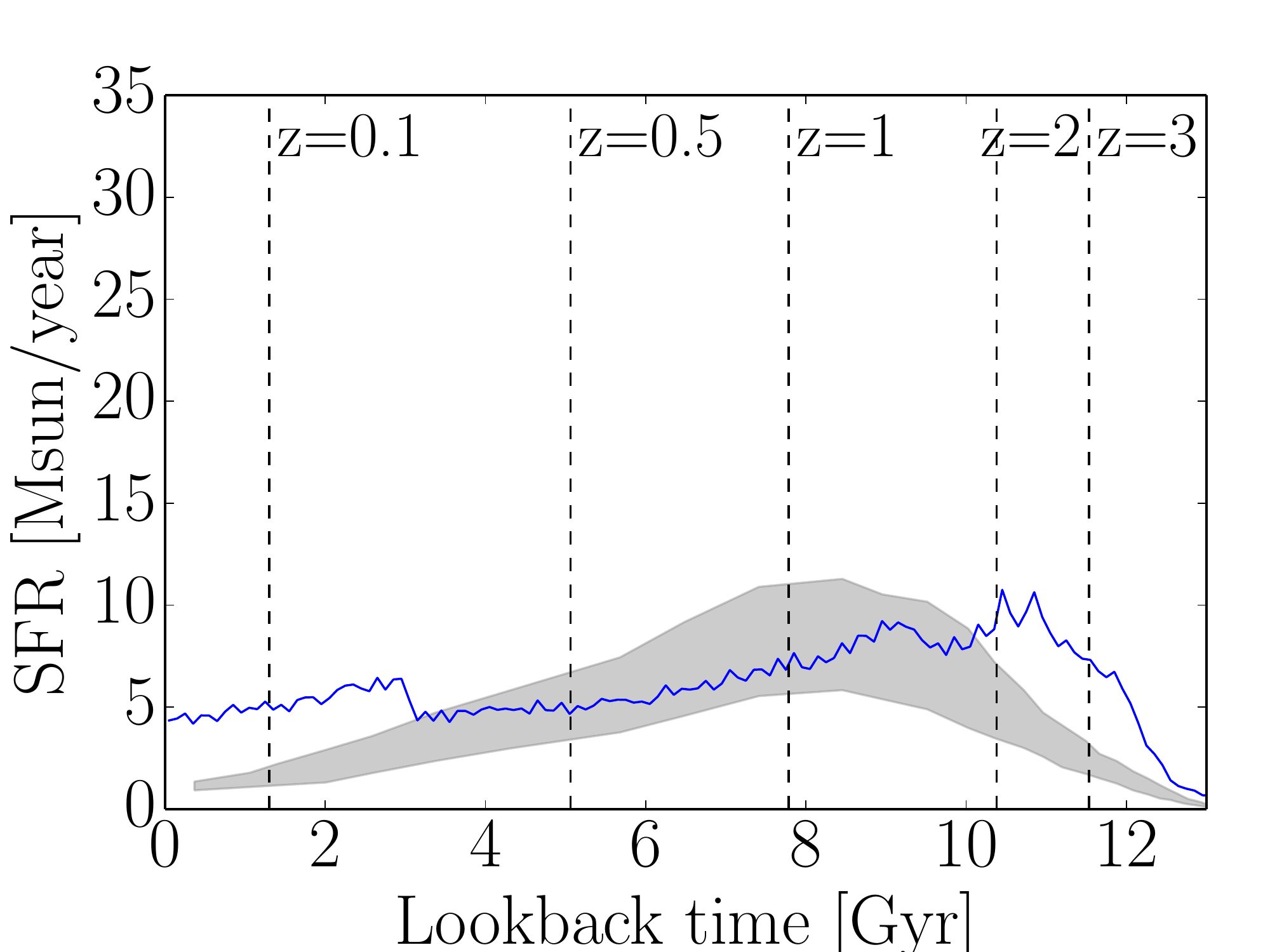}
       \caption{Halo B.}
  \end{subfigure}
  \begin{subfigure}[t]{0.31\linewidth}
     \includegraphics[width=1.0\linewidth]{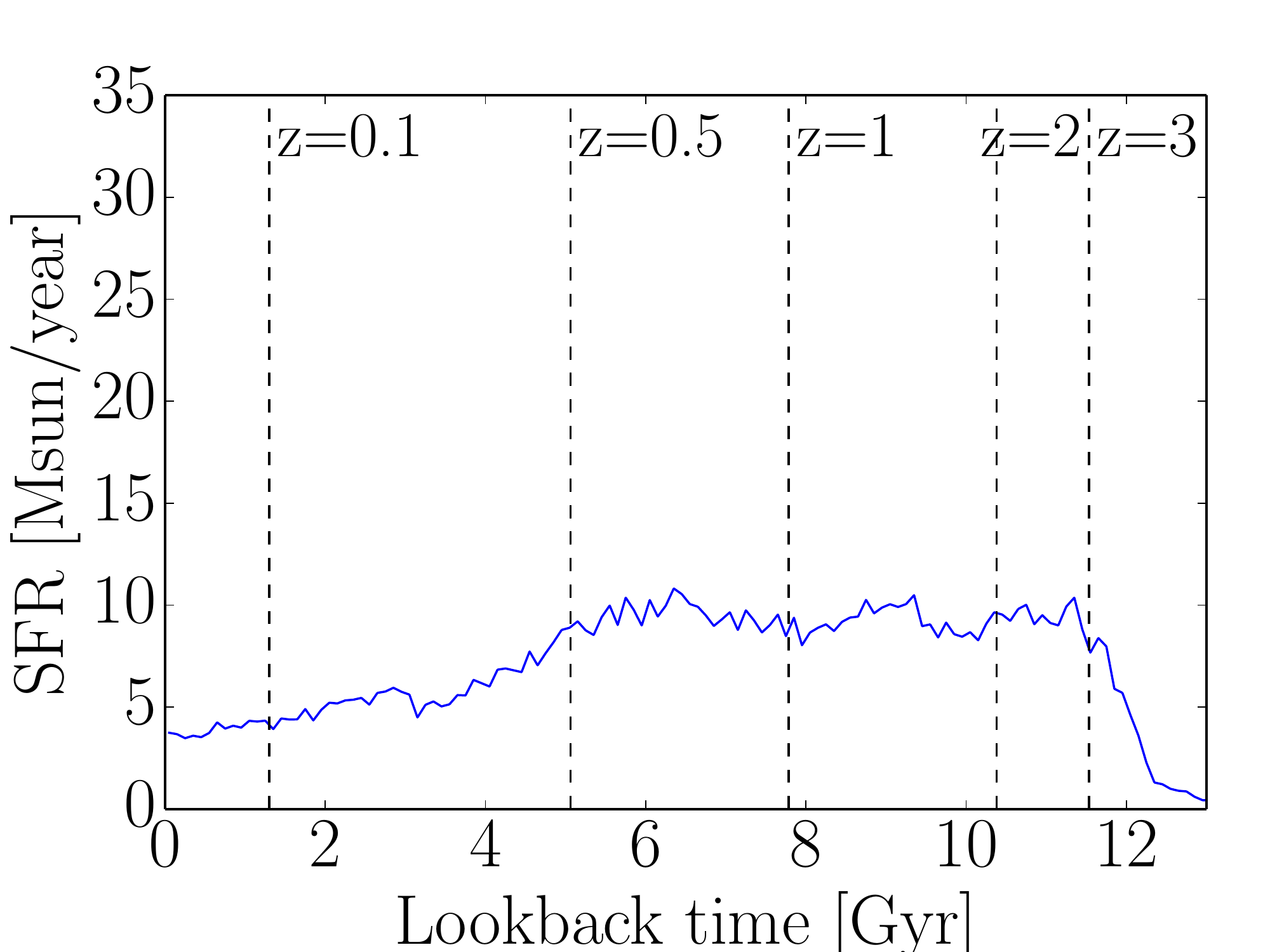}
       \caption{Halo C}
  \end{subfigure}
  \caption{Star formation histories as a function of lookback time (Gyr) and redshift for the three halos simulated for this study. Only stars inside a distance of 10\% of R$_{97}$ from
  the halo center were considered. For Halo B, we added the SFH prediction for halos with virial mass M$_{97}=10^{12}\mo$ from 
  \citet{2013ApJ...770...57B}.}
   \label{fig:SFH}
\end{figure*}
Seen from a different angle, the caveat exposed above has a direct impact on
the expected stellar mass contained within a halo at redshift 0. Moreover, several techniques 
(abundance mass matching (e.g. \citet{2013MNRAS.428.3121M,2013ApJ...770...57B}) and weak lensing 
\citep{2012MNRAS.425.2610R}) seem to reproduce the same results (see figure 16 in \citet{2012ApJ...759..138P}).
In figure \ref{fig:Kennicuttplot}, we show the star mass evolution of the three halos from redshift 4 to 0 against the DM halo mass (DM mass inside 
R$_{97}$ !). First, we explicitly warn the reader that the DM mass growth in the plot is overestimated as we fixed M$_{97}$ when the mean density
inside the sphere defined by radius R$_{97}$ drops below $97\times \rho_c(\rm z)$. As \citet{2014ApJ...792..124Z} shows, this criterion ill-defines the
physical halo border and artificially overestimates DM halo growth. Physically, the DM halo mass is presumably in place much earlier so that 
the curve should be much steeper. Secondly, we plot in the shaded area the prediction for the stellar-to-halo mass relation for redshift 0 galaxies
from \citet{2013MNRAS.428.3121M}.
This prediction is valid for M$_{200}$, so that the curves and points shown here are slightly right-shifted with respect to the observations. For comparison,
we add as white-filled edge-colored points the results taking into account the DM mass inside R$_{200}$.
The curves on Figure \ref{fig:Kennicuttplot} show the evolution over time with the ending point capturing the final state of the halos at redshift 0,
which considers the total stellar mass inside 10\% of R$_{97}$ from the halo center. Observational surveys, on the other hand, in order to derive stellar masses, 
base their estimates on photometric data. Several works in the literature have pointed out that "re-observing" the stellar masses in simulations can
reduce the existing gap between results from simulations and observations \citep{2010MNRAS.407L..41S,2011ApJ74276G}.
Attempting to mimic the observational procedure, we decided to follow \citet{2013ApJ...766...56M} in the following.
As described by \citet{2001AJ....121.2358B} and \citet{2001AJ....122.1104Y}, 
we define first the Petrosian ratio at a radius r from the center of an object to be the ratio of the local 
surface brightness in an annulus at r to the mean surface brightness within r 
\begin{equation}
 R_{\rm P}=\frac{\int_{0.8r}^{1.25r}dr'2\pi r'I(r')/(\pi(1.25^2-0.8^2)r^2)}{\int_{0}^{r}dr'2\pi I(r')/(\pi r^2)}
\end{equation}
where I(r) is the azimuthally averaged surface brightness profile in the R-band. The Petrosian radius is defined as the radius at which R$_{\rm P}$(r) equals 
a reference value, here we fixed it to 0.2 (like in the SDSS survey). We then sum up the star mass inside the Petrosian radius for the three halos (triangles in figure
\ref{fig:Kennicuttplot}). The final stellar mass estimation is based on the B-V color and V total magnitude using the fitting values of Bell \&
de Jong (2001), namely $L_V=10^{-(V-4.8)/2.5}$ and $M_{\rm star}=L_V\times10^{-0.734+1.404\cdot(B-V)}$. We correct the value for a Chabrier IMF from a Salpeter
IMF (lowering the value in logarithmical scale by 0.26 dex \citep{2009ApJ...690.1236I}) and plot the derived stellar mass as squares. With this
procedure we obtain a stellar mass estimation that is about 55 - 65 \% lower than the simulation data. 
The stellar mass of Halo A, B and C then lies largely within the abundance mass matching relation. 
Of course, we do not claim that this procedure solves the problems current galaxy formation models encounter when confronted to stringent observational data,
but at least
it shows how much uncertainty can be related to observational procedures that are used
to derive stellar masses. 

\section{DM halo properties}
\label{DMHaloproperties}
In the following section, we analyse the DM halo properties. In particular, we focus on comparing the results obtained for the hydrodynamical
simulations with the corresponding DM-only counterparts.

\subsection{Baryon's impact on radial density profiles}\label{sec:BaryonImpactonDM}
\begin{figure*}
 \centering
 \begin{subfigure}[t]{0.32\linewidth}
   \includegraphics[width=1.\linewidth]{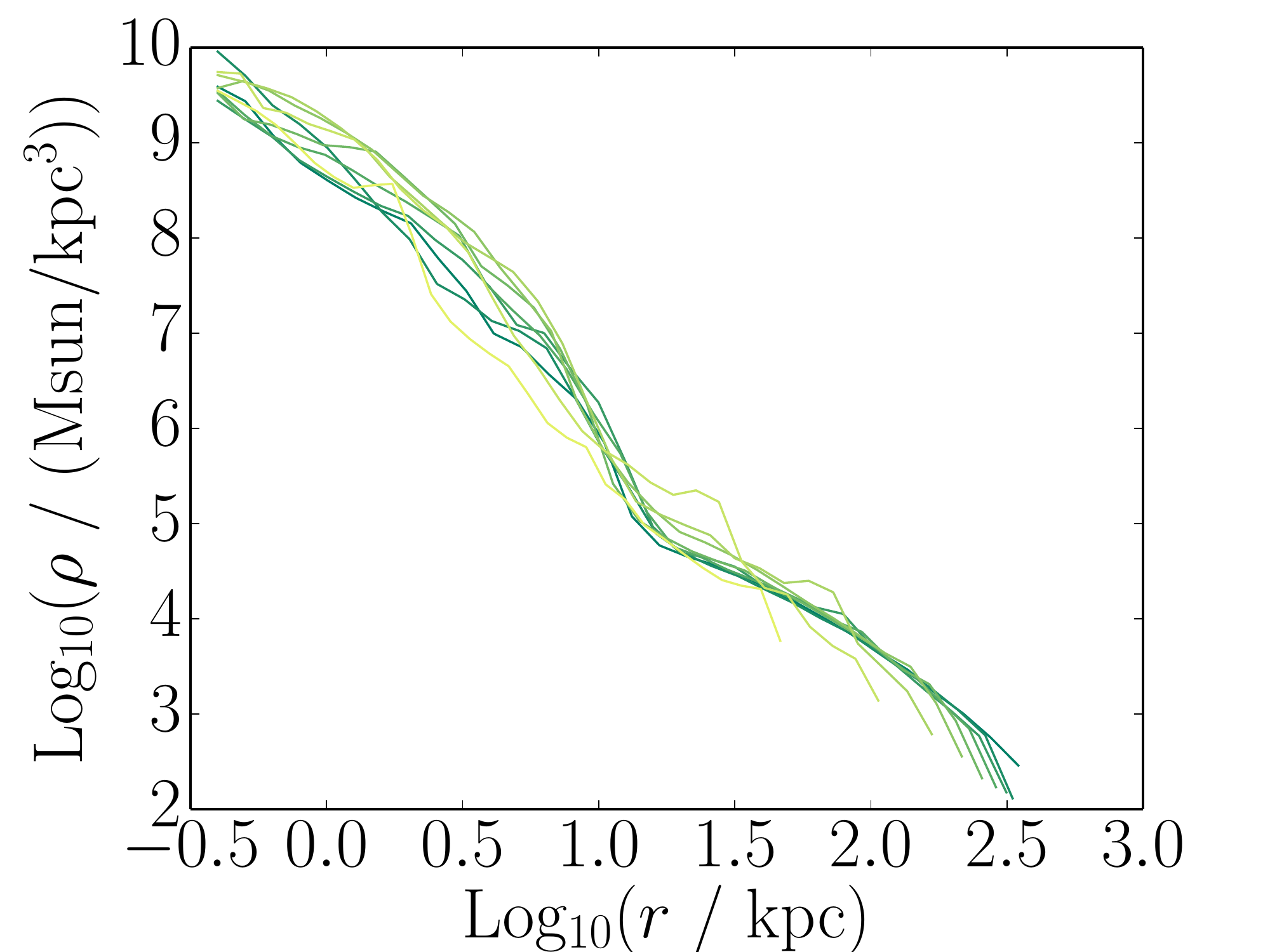}
  \end{subfigure}
  \begin{subfigure}[t]{0.32\linewidth}
  \includegraphics[width=1.\linewidth]{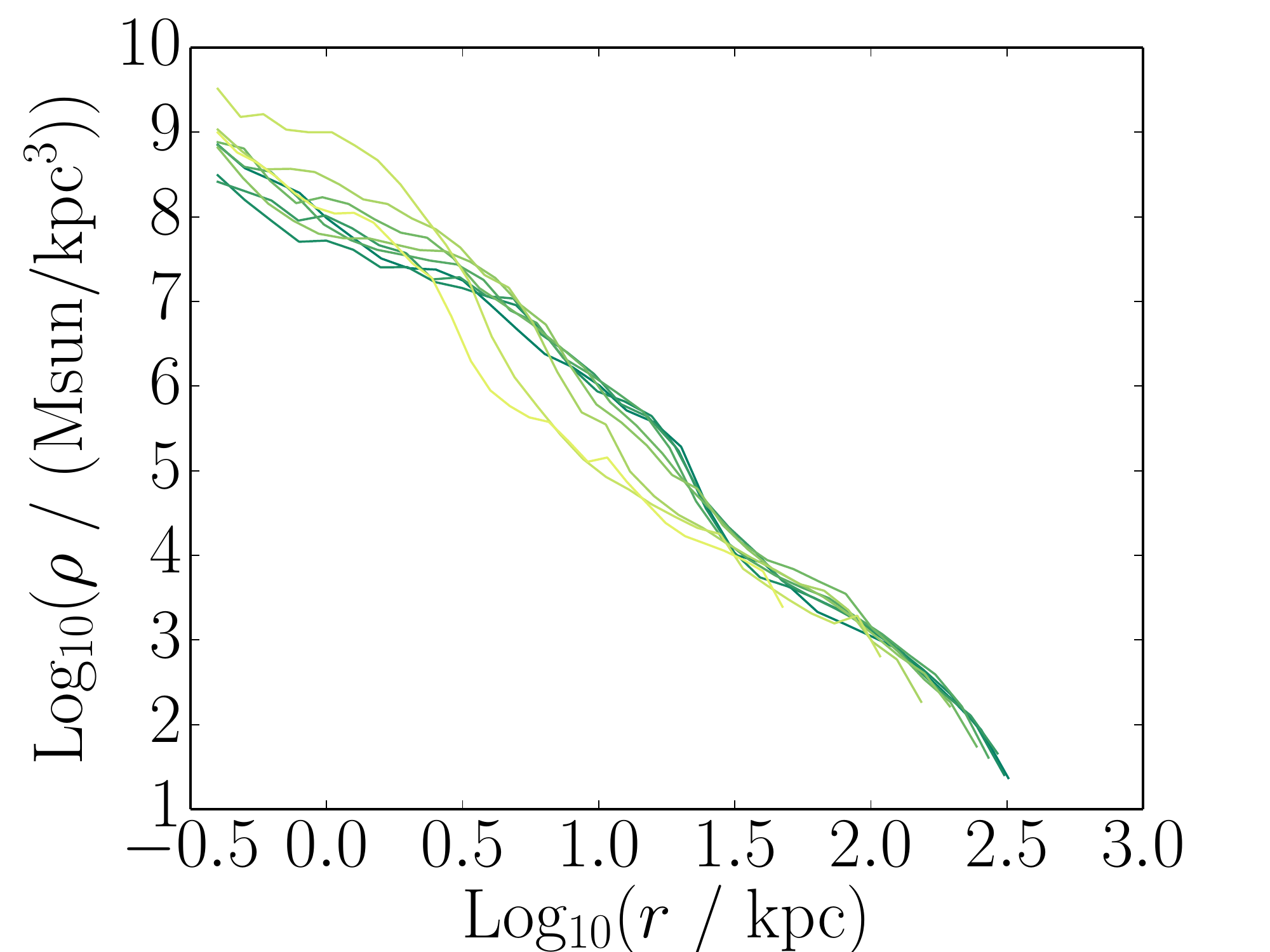}
  \end{subfigure}
  \begin{subfigure}[t]{0.35\linewidth}
  \includegraphics[width=1.\linewidth]{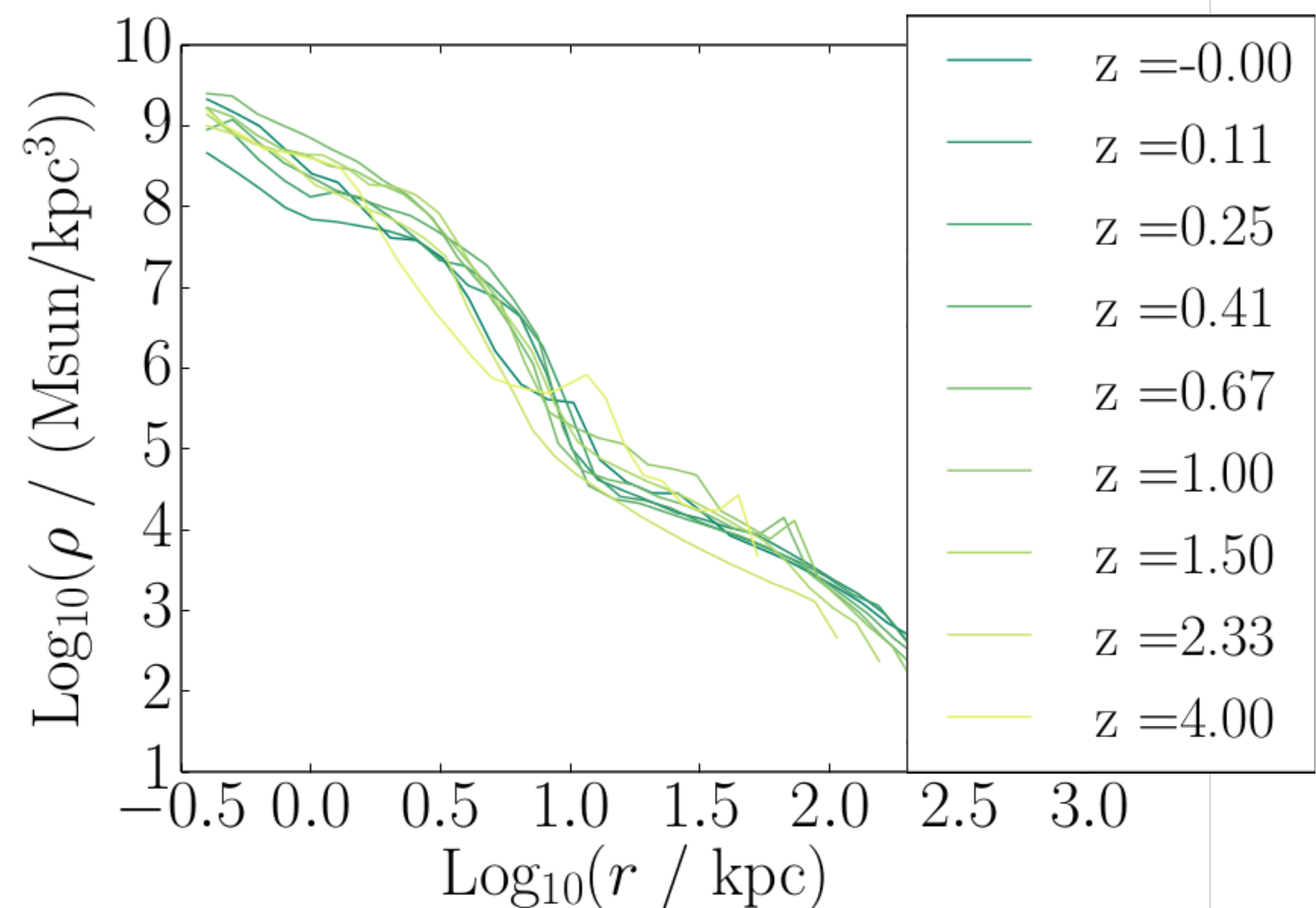}
  \end{subfigure}
  
  a) Gas density distribution for Halo A, B and C (from left to right).
    
  \begin{subfigure}[t]{0.32\linewidth}
   \includegraphics[width=1.\linewidth]{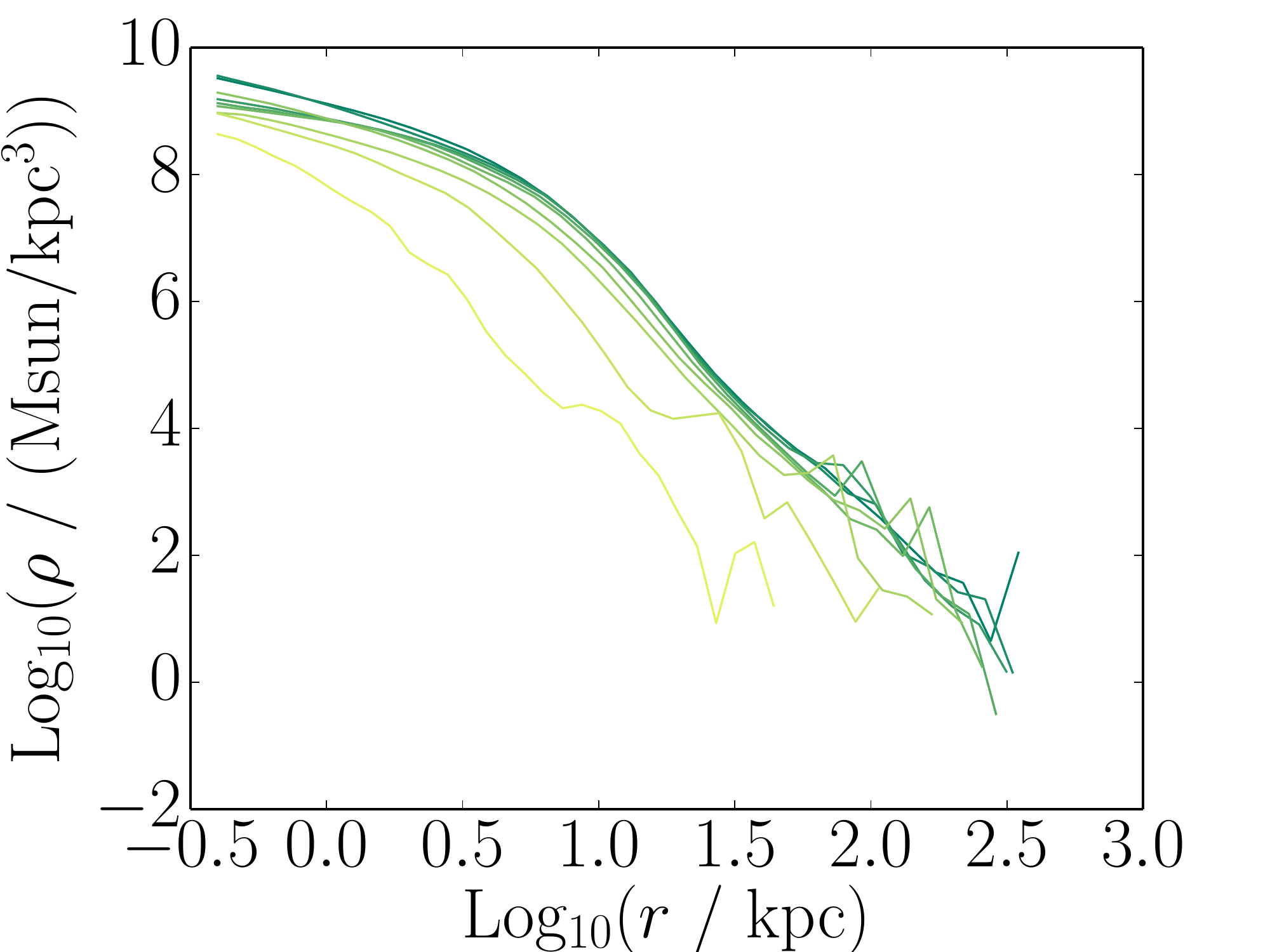}
  \end{subfigure}
  \begin{subfigure}[t]{0.32\linewidth}
  \includegraphics[width=1.\linewidth]{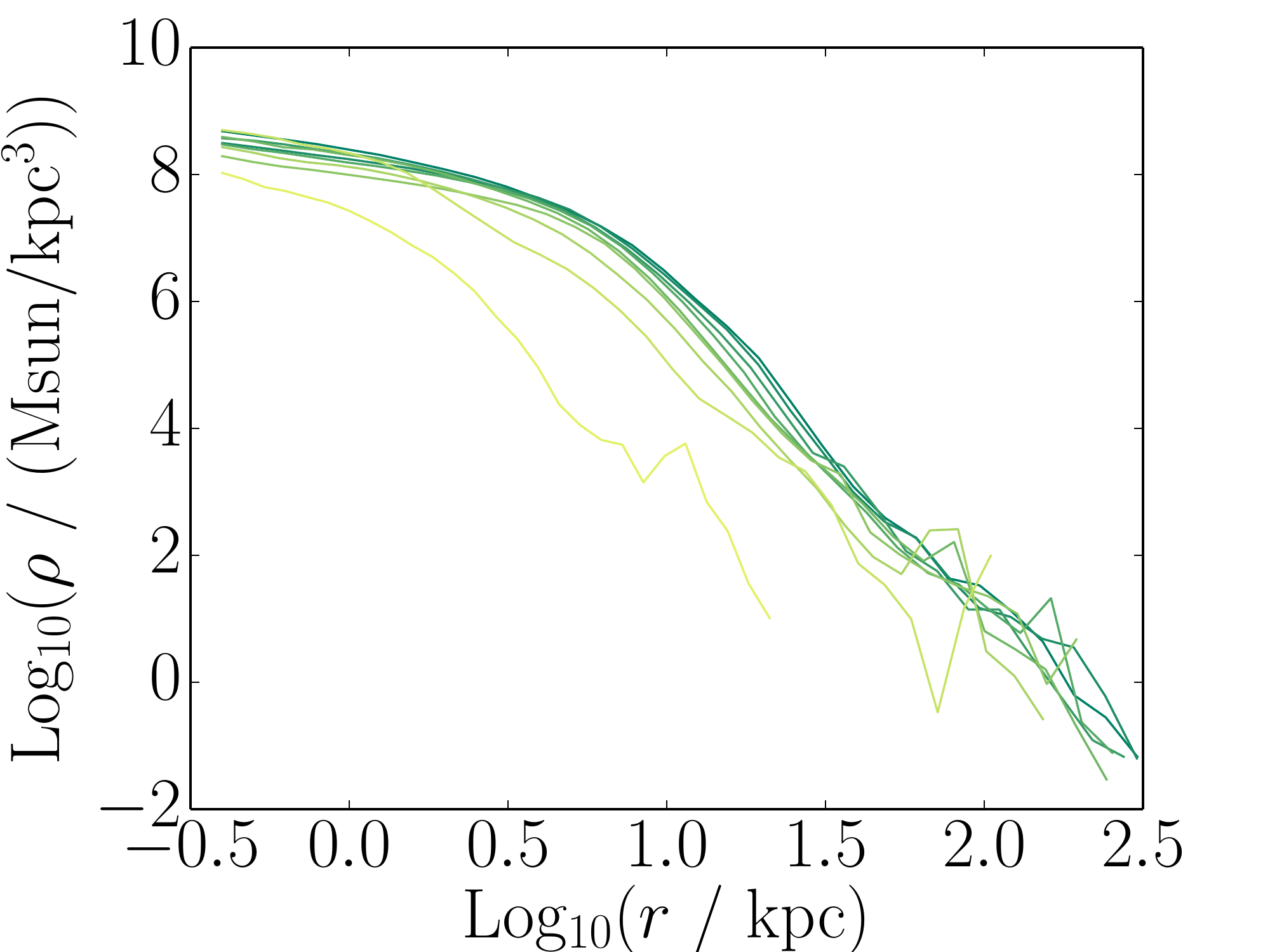}
  \end{subfigure}
  \begin{subfigure}[t]{0.35\linewidth}
     \includegraphics[width=1.\linewidth]{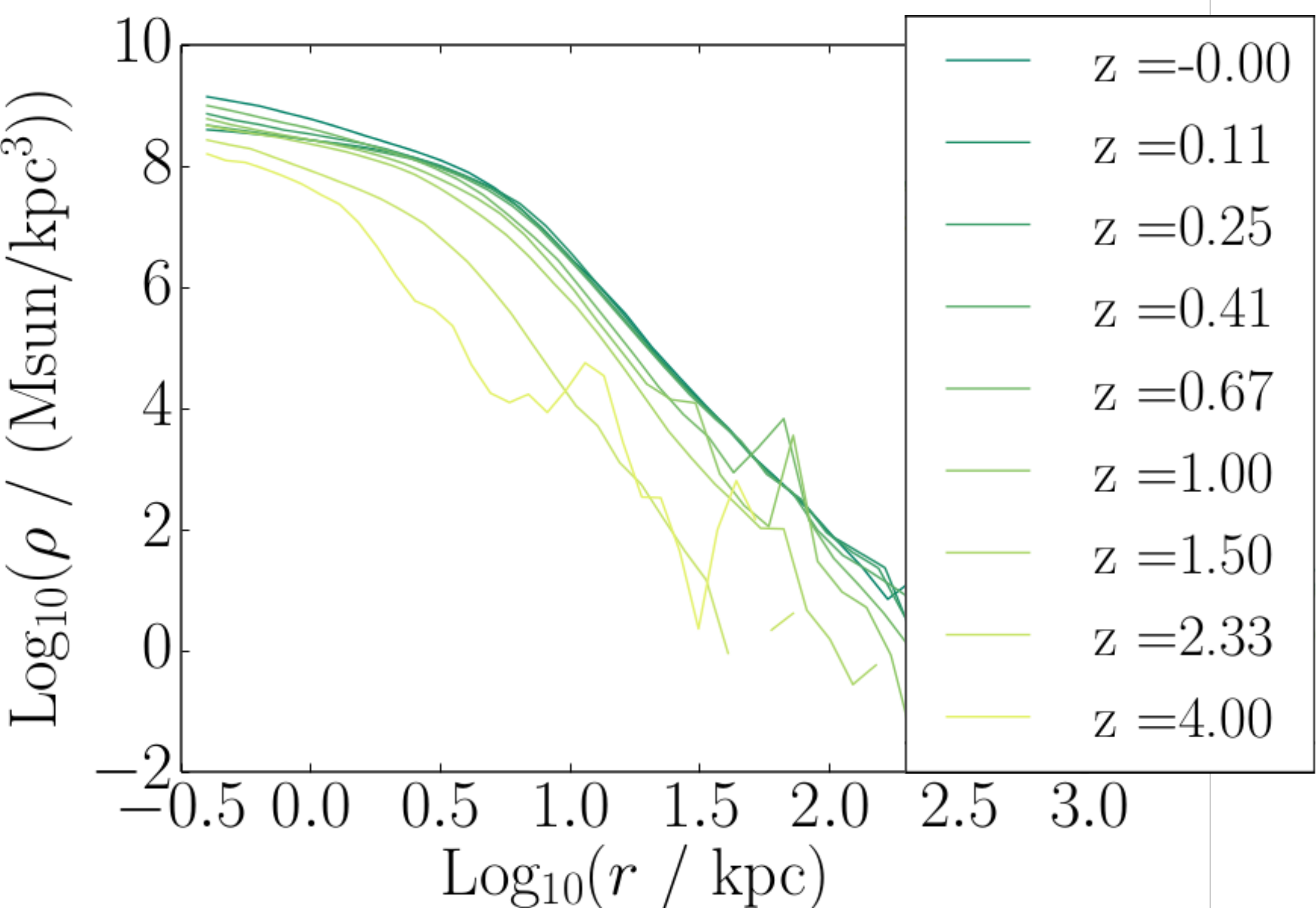}
  \end{subfigure}
  
   b) Star density distribution for Halo A, B and C (from left to right).
    
  \begin{subfigure}[t]{0.32\linewidth}
   \includegraphics[width=1.\linewidth]{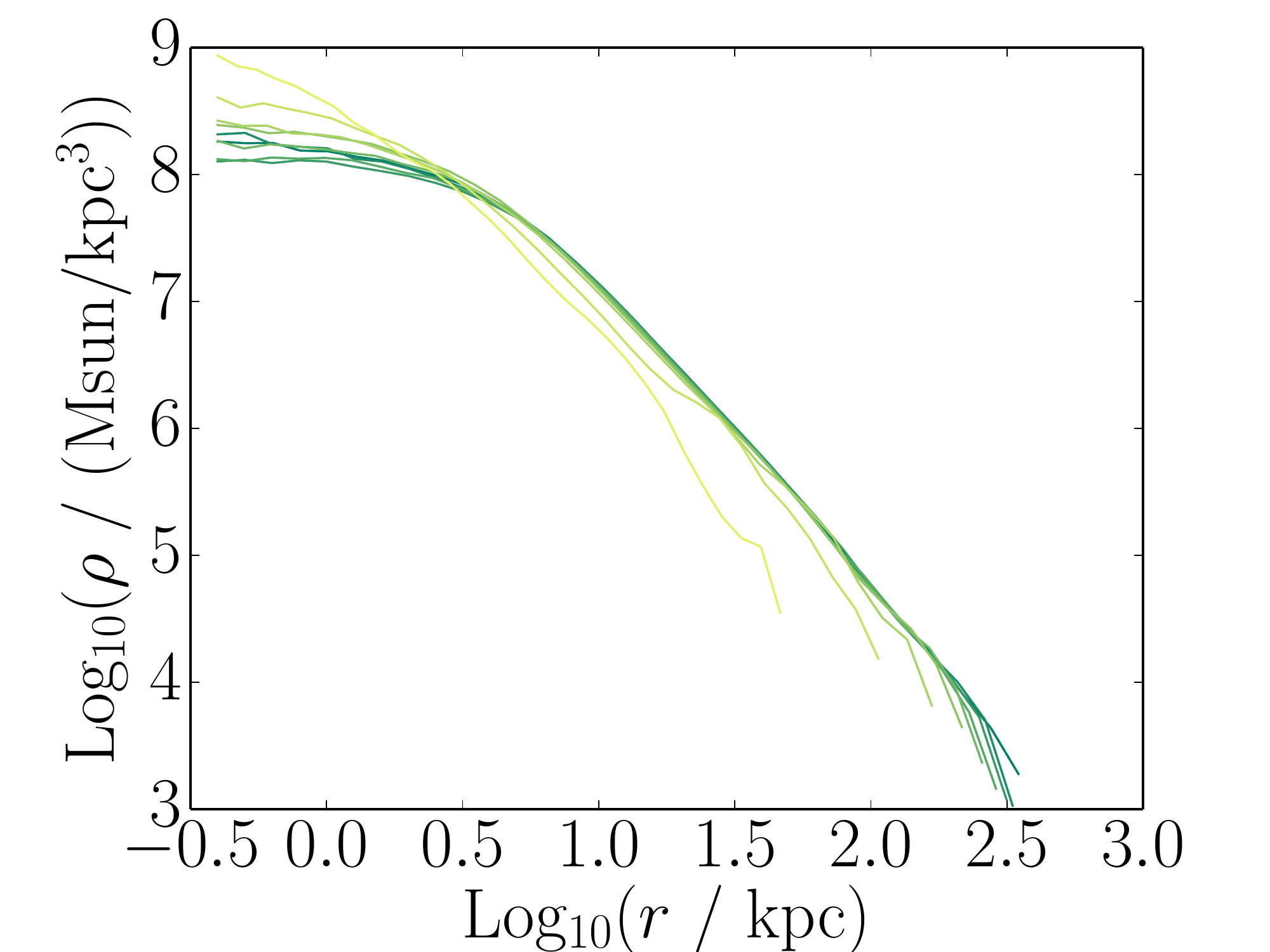}
  \end{subfigure}
  \begin{subfigure}[t]{0.32\linewidth}
  \includegraphics[width=1.\linewidth]{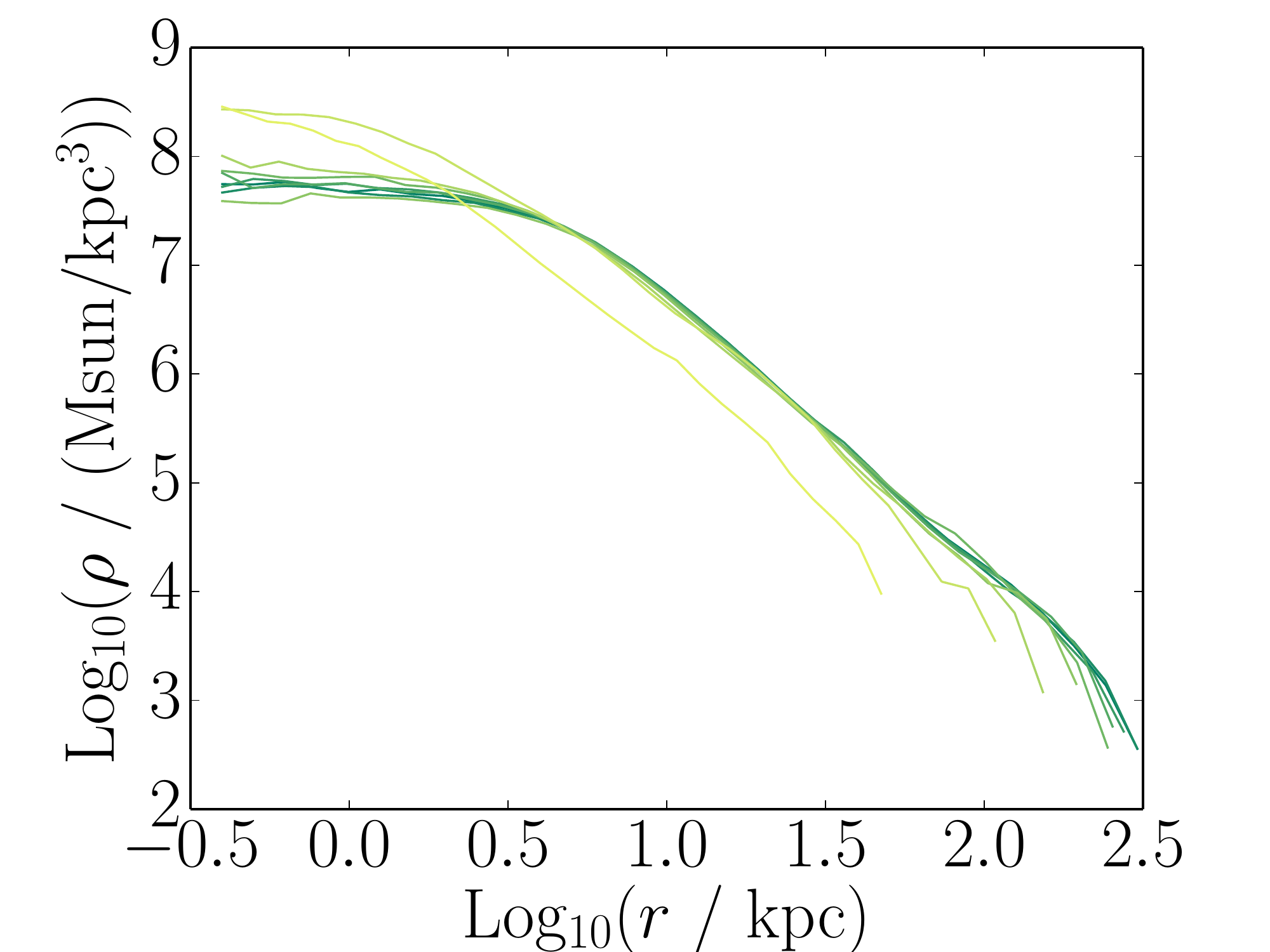}
  \end{subfigure}
  \begin{subfigure}[t]{0.35\linewidth}
     \includegraphics[width=1.\linewidth]{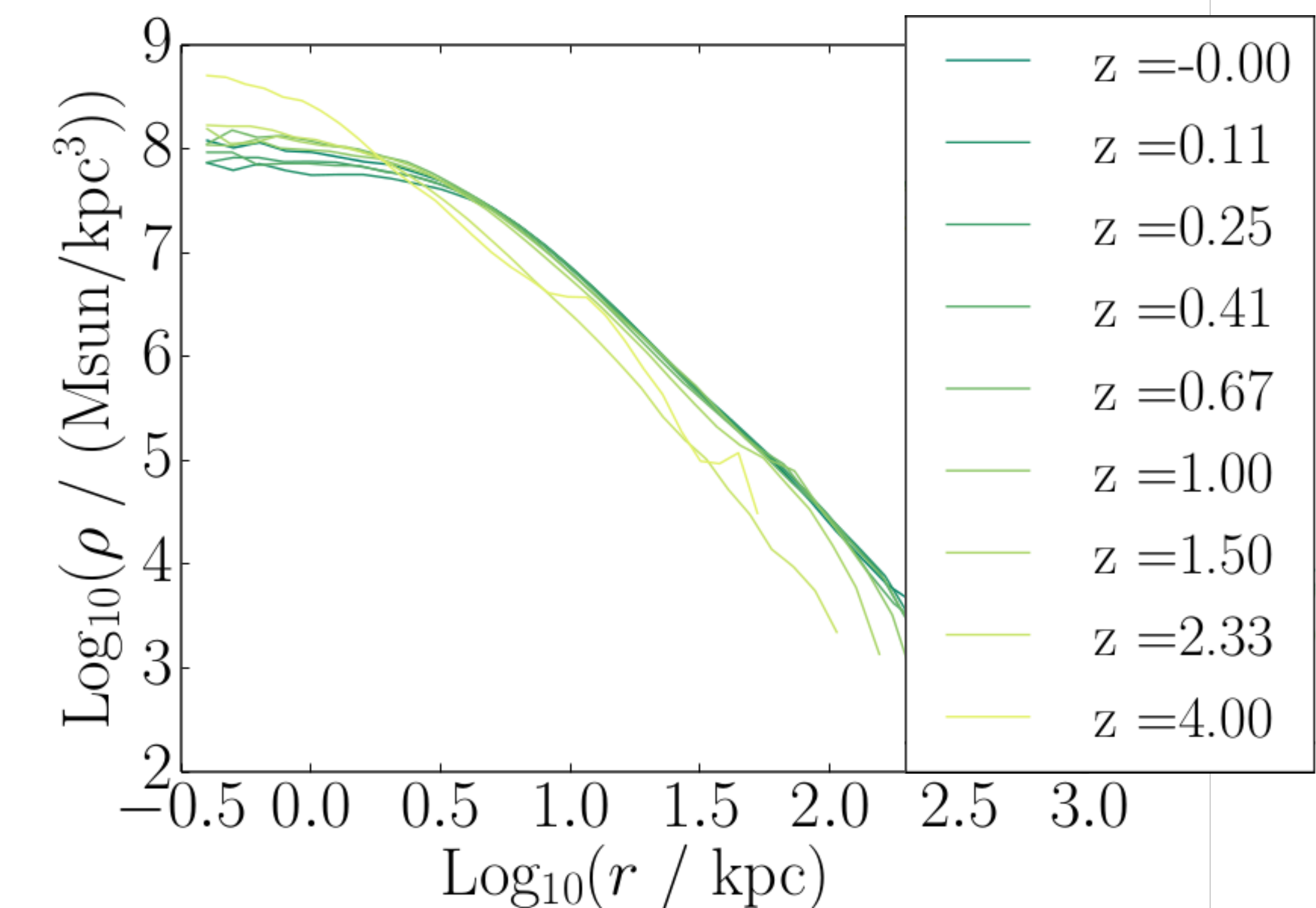}
  \end{subfigure}
  
  c) DM density distribution for Halo A, B and C (from left to right).
    
  \caption{Spherical averaged density profiles of the gas, stars and DM for Halo A, B and C (from left to right) over redshift range.}
   \label{fig:DensityProfiles}
\end{figure*}

\begin{figure*}
 \centering
 \begin{subfigure}[t]{0.32\linewidth}
   \includegraphics[width=1.\linewidth]{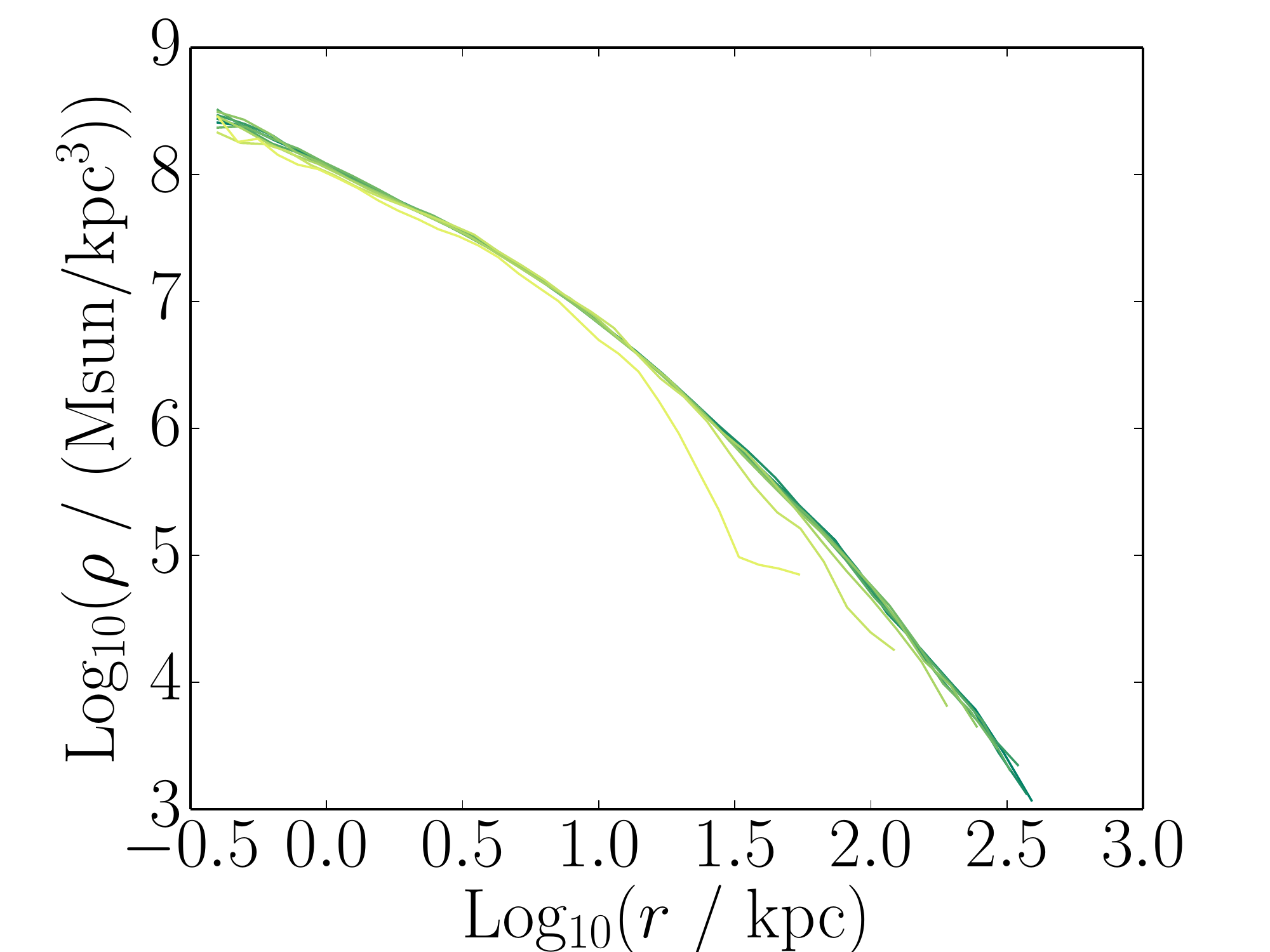}
  \end{subfigure}
  \begin{subfigure}[t]{0.32\linewidth}
     \includegraphics[width=1.\linewidth]{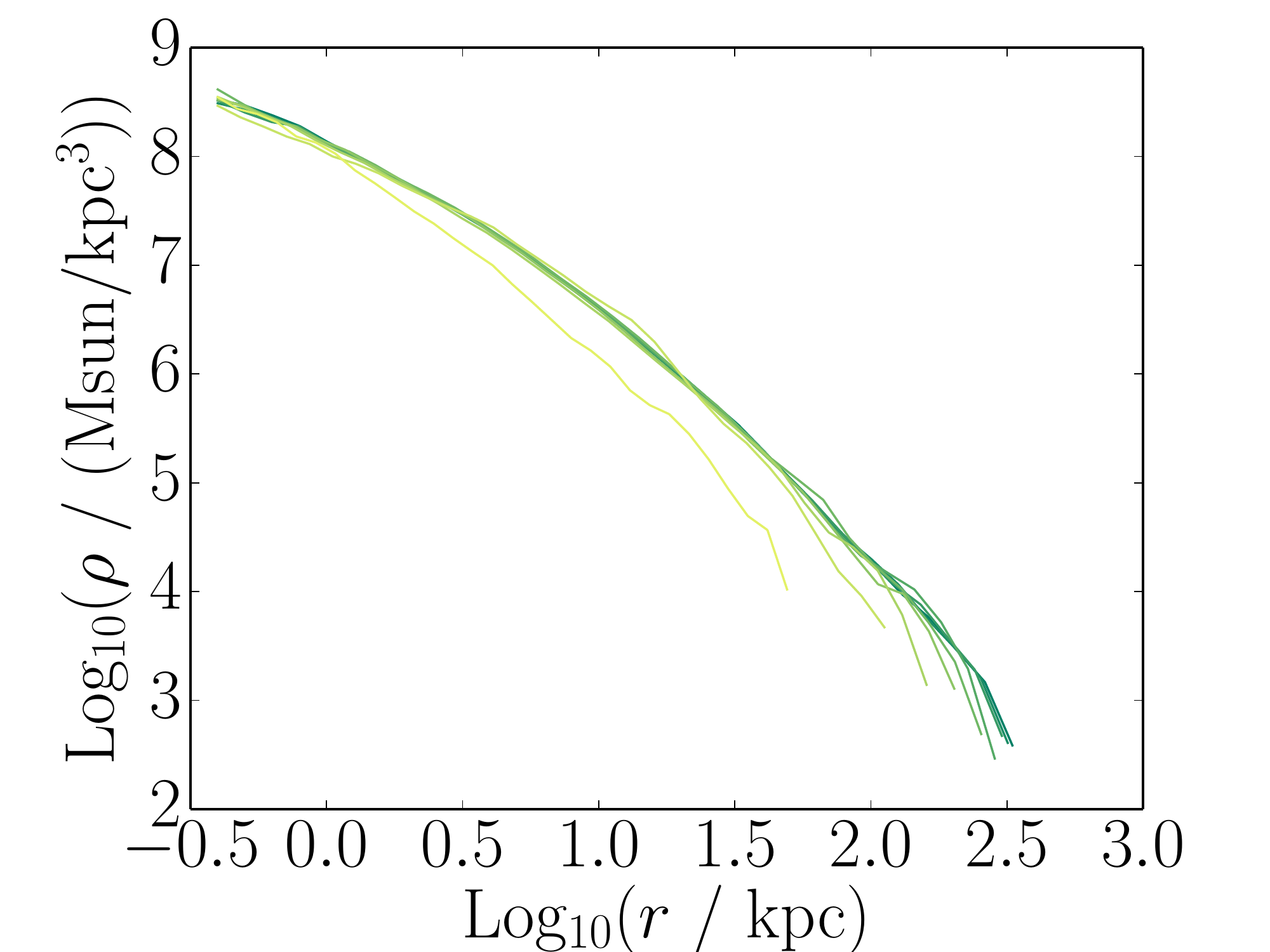}
  \end{subfigure}
  \begin{subfigure}[t]{0.35\linewidth}
     \includegraphics[width=1.\linewidth]{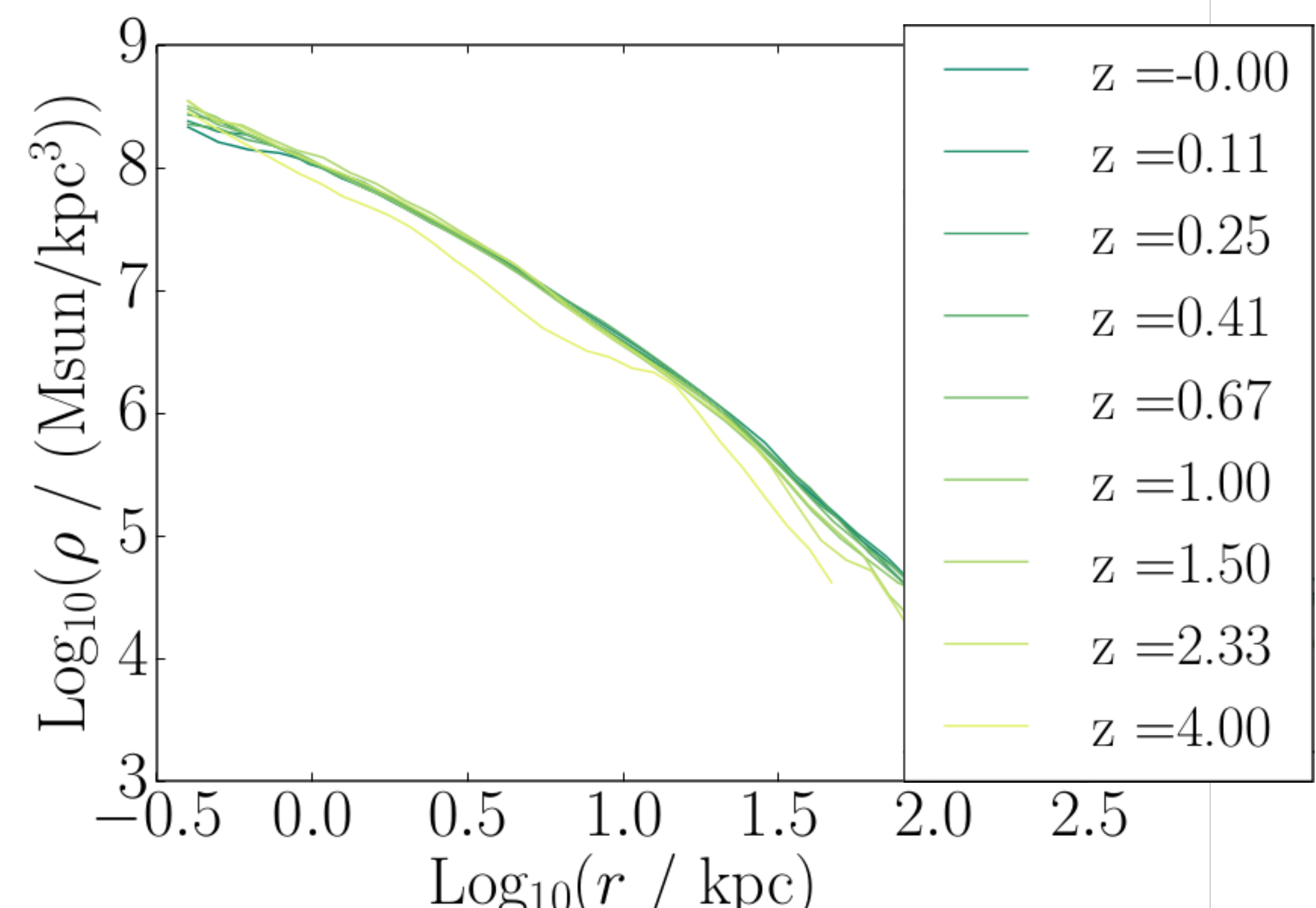}
  \end{subfigure}
  \caption{From the DM-only siblings: DM density profiles for Halo A, B and C over redshift range.}
   \label{fig:DensityProfilesDMonly}
\end{figure*}

Although cosmological N-Body simulations provide realistic environments and mass accretion histories for
galaxies, it is not clear which effects on the DM density are induced by modelling the baryon physics on small scales.
The DM density profiles provided by pure N-Body models favor "cuspy" slopes in the inner regions
of DM halos \citep{1997ApJ...490..493N}  or are well-fitted by an Einasto profile \citep{2008MNRAS.391.1685S}. 
Recently, it has been demonstrated that star formation and
its associated feedback schemes for dwarf-sized/MW-sized halos and AGN feedback for cluster-sized halos
could cause a flattening of the slope of DM density profiles, which has been dubbed the "cusp-core" transformation 
\citep{2012ApJ...744L...9M,2012ASPC..453..365M,2013MNRAS.429.3068T,2013MNRAS.433.3297D}. 
The inclusion of these processes could reconcile predictions with observations that favor flat inner density profiles
\citep{2009MNRAS.397.1169D}, in particular also for the MW \citep{2013JCAP...07..016N}.
\citet{2012MNRAS.421.3464P} provided an analytical model demonstrating that impulsive gas motions are responsible for 
cusp-core transformations. In agreement with simulation results of \citet{2012MNRAS.422.1231G}, 
\citet{2013MNRAS.429.3068T} showed that the core formation mechanisms imply a bursty star formation history
and "hot" stellar velocity distributions, predictions that are in agreement with observations \citep{2012ApJ...744...44W}. 
In the simulations presented in this paper, we used similar feedback mechanisms
and, in the following section, we give the results when they are applied to more massive halos.\\
In Figure \ref{fig:DensityProfiles} we show the spherical averaged density profiles for stars, DM, and gas components 
evolving over time, i.e., from redshift 4 to 0. We centered the halos on the highest star density. In general, 
there is a small offset between this point and the highest DM density point, but as its order of magnitude is $\sim$0(10pc), it is not noticable on the density figures in 
\ref{fig:DensityProfiles}.\\ 
If we analyse the star density profile for the three halos, we see that star formation acts first in the galactic center and then raises 
the density constantly in the outer parts of the galaxy, which extends out to 10 kpc, or even further in the case of Halo B. 
Additionally, Halo A and C show a recent increase in their central density, revealing the formation of a massive central bulge.\\ 
The DM density profile  for the three halos in the hydrodynamical run converges over time in the outer part of the halos and
presents a cored central profile. Starting from redshift 4, the initially steep DM profiles of Halo A and C are immediately flattened
due to intense star formation and its associated feedback. Besides, interesting to notice is the small increase of the 
profile in the snapshots of the last redshifts shown: the increased central mass coming from the bulge formation contracts the DM profile. The profile
of Halo B presents a special feature. From redshift 4 to 2.3, the steep DM density profile is adiabatically contracted as feedback
triggered by insetting star formation was not powerful enough to start a core formation. Then, between redshift 2.3 and 1.5, a redshift range that
coincides with the peak of
star formation (c.f. figure \ref{fig:SFH}), the central DM density is efficiently flattened.\\
As the time steps of the snaphots shown here are not sufficiently small to capture single starburst events, we do not see the "breathing" effect of 
the gas densities that make the central gravitational potential oscillate and induce the flattening of the DM density profile \citep{2013MNRAS.429.3068T}. 
On the figure, it is difficult to disentangle one specific physical process due to the complex nature of gas physics on large scales captured
in the simulation, which incorporates, e.g., the cosmological gas infall, the aforementioned galactic gas outblowing provocated 
by stellar feedback, gas cooling or the transformation from gas mass into star particles.
However, for the three halos, one clearly identifies the transition
between the galactic disk gas and the halo gas indicated by the slope change in the profile. 
The central gas density profile of Halo A and C is observed to fluctuate considerably over time. Sudden increases of the gas reservoir in the disk alternate with periods
where the density diminishes as gas is locked into stars. Halo B evolves less abruptly, the central gas density tends to go down steadily while the gas
density in the extended galactic disk increases and is stabilized with time.\\
The DM profile of the hydrodynamical runs are in striking difference with the results from the DM-only simulations, which is plotted in Figure \ref{fig:DensityProfilesDMonly}.
For them, we obtain a cuspy central profile whose slope is close to 1. We use the function
\begin{equation}
 \rho(r,\rho_s,r_s,\alpha,\beta,\gamma)=\frac{\rho_s}{(\frac{r}{r_s})^{\gamma}(1+(\frac{r}{r_s})^{\alpha})^{(\beta-\gamma)/\alpha}}
 \label{eq:NFW}
\end{equation}
to fit the DM profile from Log$_{10}$r=-0.6 to R$_{97}$ of each halo. For the DM-only runs we imposed $\alpha=1$ for the fit so that $\gamma$ 
immediately reflects the inner slope and $\beta-\gamma$ the outer slope of the profile. The results can be found in Table \ref{tab:DMprofileFit}.
Enticing is the comparison between the simulation cores ($\sim 2-5$ kpc) with the core sizes inferred by the fitted profiles to observational data of the MW, e.g., isothermal profile
(Core size in \citet{2009PASJ...61..227S}: 5.5 kpc) or Burkert profile (Core size in 
\citet{2013JCAP...07..016N} $\sim10$ kpc). 
 
\begin{figure*}
 \centering
 \begin{subfigure}[t]{0.45\linewidth}
    \includegraphics[width=1.\linewidth]{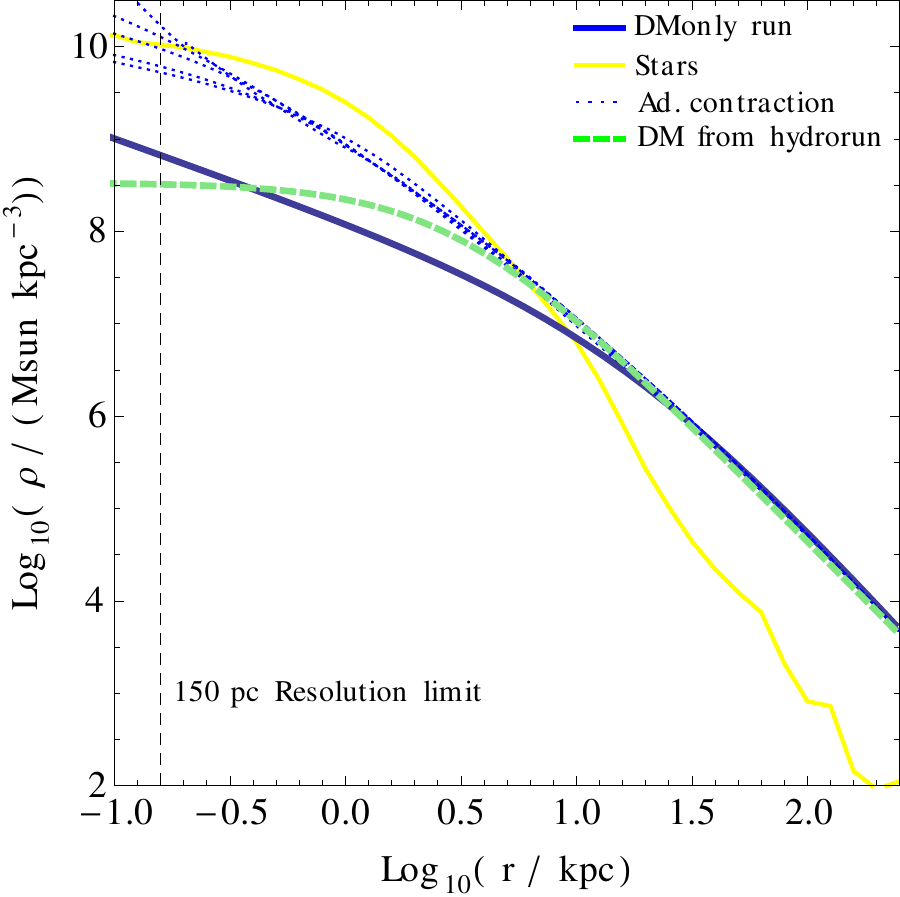}
  \end{subfigure} 
  \begin{subfigure}[t]{0.45\linewidth}
    \includegraphics[width=1.\linewidth]{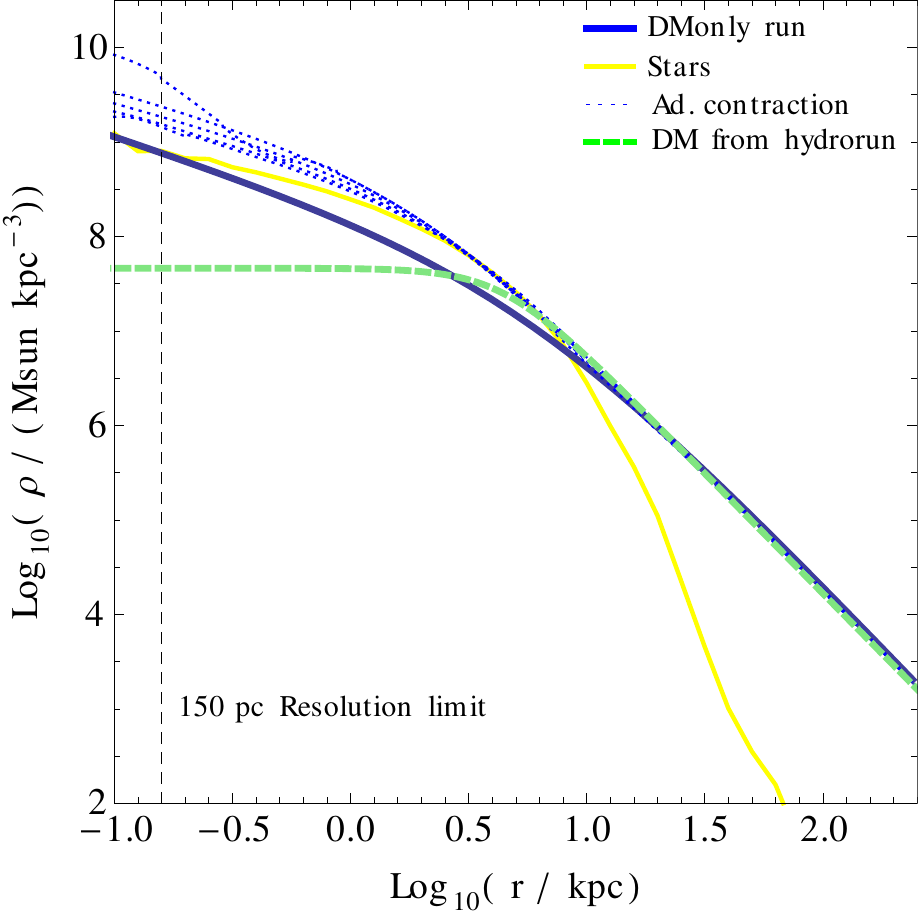}
  \end{subfigure} 
 \caption{Comparison between the DM density profiles from the DM-only and hydrodynamical runs for Halos A (left) and B (right). The DM profile from the DM-only run 
 is adiabatically contracted via the stellar density profile from the hydro run. See section \ref{sec:BaryonImpactonDM} for further explanations.}
 \label{fig:Adiabaticcompression}
 \end{figure*}

\begin{table}
\centering
\begin{tabular}{lrrrrr}
\hline
Run & $ \textrm{Log}_{10}\rho_{s}$ & $r_s$ & $\alpha$ & $\beta$ & $\gamma$  \\
    & \small [$\mo$/kpc$^3$] &[ kpc]& \normalsize & 	    &		\\
\hline
 Halo A & 8.005  & 4.39 & 1.879 & 2.469 & 0.126\\
 Halo A-DM & 7.232 & 13.026 & 1 & 2.707 & 0.794 \\
 Halo B & 7.663 & 4.425 & 2.895 &  2.541 & 8*$10^{-9}$\\
 Halo B-DM & 7.639 & 5.552  & 1 & 2.636 & 0.819 \\
 Halo C & 7.678 & 4.317 & 2.451 & 2.477 & 0.268  \\
 Halo C-DM & 6.992 & 13.148 & 1 & 2.871 & 0.927 \\
\hline
\end{tabular}
\caption{Best fitting values for the spherical averaged density profiles fitted with equation \ref{eq:NFW}. For the DM-only simulations,
we fixed $\alpha=1$. The fit was performed for $r\in[250 \rm{pc},R_{97}]$.}
\label{tab:DMprofileFit}
\end{table}
\noindent We checked that the cored DM profiles are not introduced by a lack of resolution ($\sim$ 150 pc, see Figure \ref{fig:Adiabaticcompression}).\\

Our result is in contradiction with \citet{2014MNRAS.437..415D} (see also \citet{2014MNRAS.441.2986D}) who analysed the DM profiles of the MaGiCC simulations
performed by \citet{2013MNRAS.428..129S}.
They found that the stellar feedback models used in the simulations were able to turn the DM cusp into a core, only for a certain 
galactic stellar mass range that lies beneath the stellar mass of our simulated galaxies.\\

In order to emphasize the flattening of the DM profile, we compared our results with a case where the DM-only simulation halos were adiabatically
contracted by the baryonic distribution obtained in the corresponding hydro run. 
The standard adiabatic contraction model systematically overestimates its effect in the inner region \citep{1986ApJ...301...27B},
because the assumptions of spherical symmetry and homologuous contraction are not fulfilled in the violent hierarchical structure formation processes.
Therefore, we employ a modified adiabatic contraction model \citep{2004ApJ...616...16G} calibrated on cosmological simulations. The formalism of this 
model attempts to account for orbital eccentricities of the particles. In Figure \ref{fig:Adiabaticcompression}, we plot first the DM density 
profile (green-dashed line) and the star density (yellow line) from the hydrodynamical run. Then, taking into account the realistic stellar mass from the hydrodynamical
run,
we contract via the aforementioned model the DM density profile from the DM-only run (blue-thick) and show the predicted contracted density profile as blue-dotted curves.
In Figure \ref{fig:Adiabaticcompression}, we see that the stellar feedback is able to transform the inner cuspy DM profile into a cored profile and to counteract against the adiabatic contraction.
Nevertheless, moving further away from the galactic center, around 10 kpc, we do see a contraction. The DM from the hydro-run departs from the DM-only profile, and
exactly follows the predicted contracted profiles. As the stellar mass of Halo A is bigger and reaches higher densities, the DM is drained further inside the galaxy
of Halo A than for that of Halo B and the effect is more visible. In Figure \ref{fig:DensityProfiles}, it can be seen that the DM profile from Halo C has the same behaviour than Halo B,
but is less interesting because the scale radius where the slope of DM density profile changes is shorter, which is due to the smallness of its galactic disk.\\
This newly observed effect, combining core formation at the center and adiabatic contraction at larger galactic radii, is the result of two different regimes: at the 
center, stellar feedback can compensate the additional gravitational attraction coming from the stars, whereas the feedback released by the stars in the outer parts of the galaxy
is not strong enough to prevent the contraction.

The described contraction has an important effect on the DM density at the solar neighbourhood radius. In the case of Halo B, comparing the local mean DM density in 
a spherical shell at 8 kpc, we notice that it is increased from 0.23 GeV/cm$^3$ in the DM-only run to 0.36 GeV/cm$^3$
in the hydro run.
Calculating the mean DM density in the ring at 8 kpc situated in the disk plane, we obtain a mean DM density of 0.50 GeV/cm$^3$,
a result that is in agreement with estimations for the local MW DM density
(see \citet{2014JPhG...41f3101R} for a recent review). A detailed study of the solar neighbourhood DM distribution in our simulation
will be the subject of a forthcoming paper.
 
\subsection{Satellites}\label{sec:satellites}
\begin{figure*}
 \centering
 \begin{subfigure}[t]{0.32\linewidth}
  \includegraphics[width=1.\linewidth]{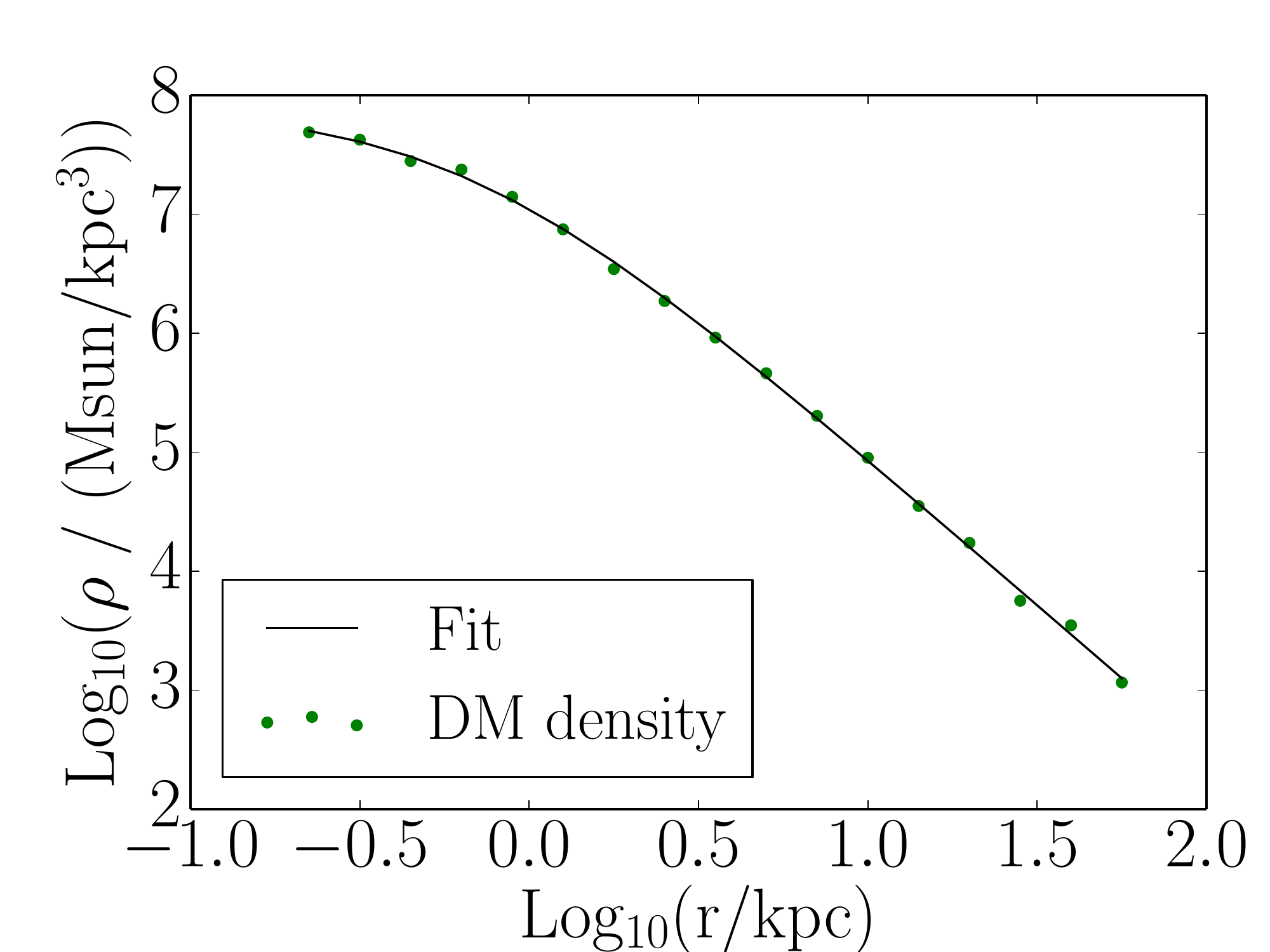}
 \end{subfigure}
 \begin{subfigure}[t]{0.32\linewidth}
  \includegraphics[width=1.\linewidth]{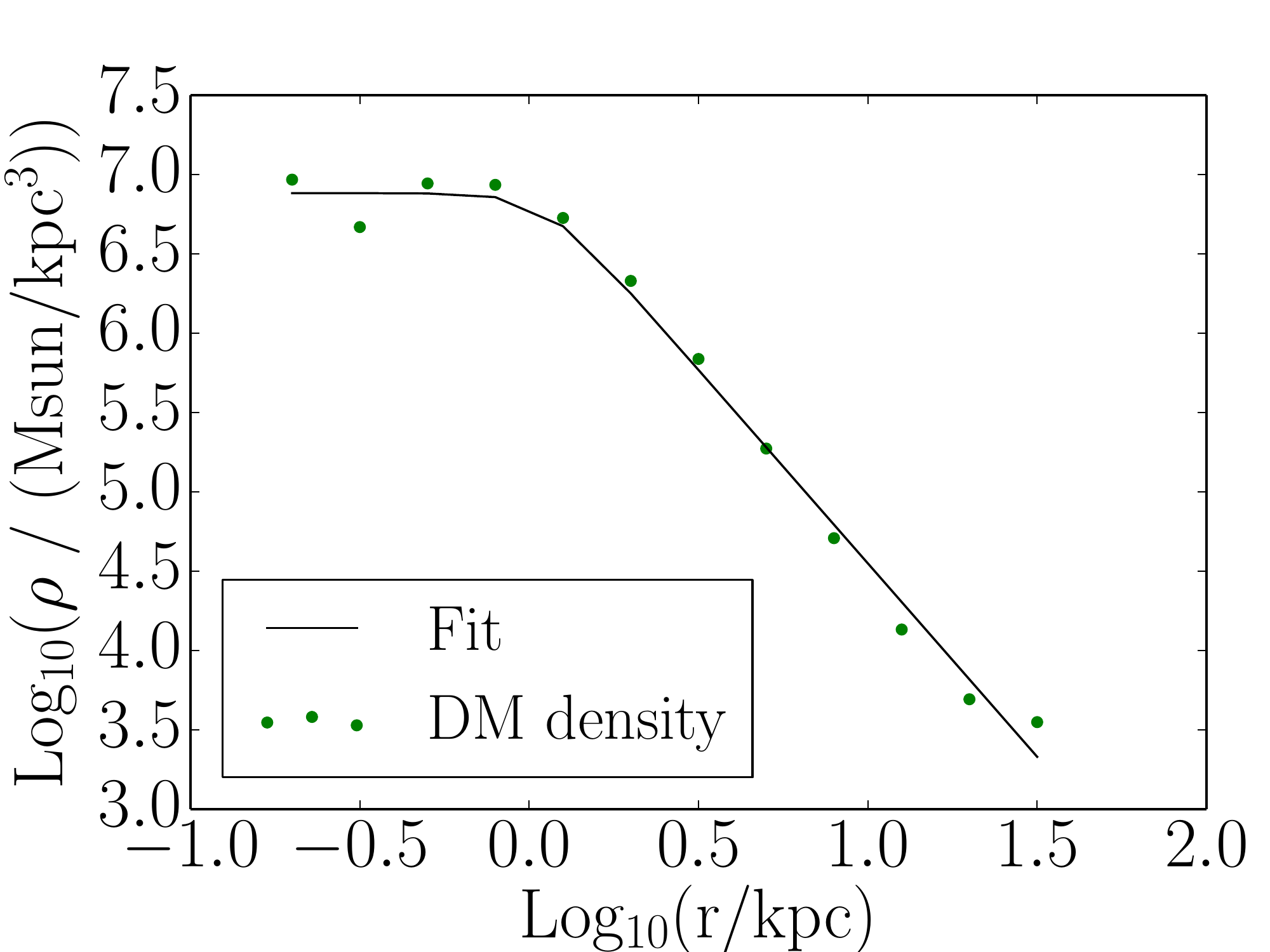}
 \end{subfigure}
   \begin{subfigure}[t]{0.32\linewidth}
    \includegraphics[width=1.\linewidth]{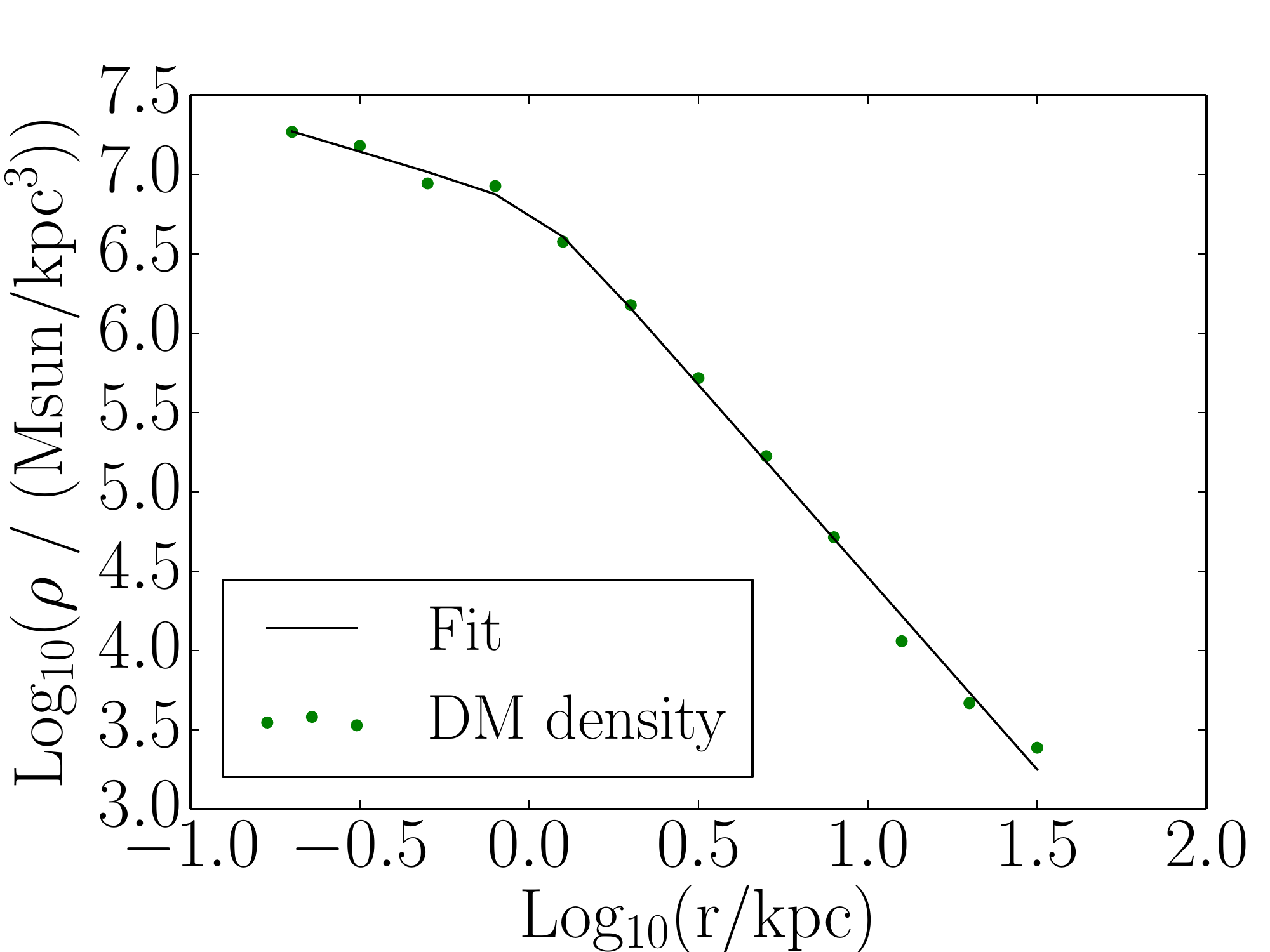}
   \end{subfigure}
 \caption{The three panels show typical DM density profiles for satellites from the DM-only run (left) and the hydrodynamical run,
 where the right panel shows a profile with a steeper inner slope with respect to the cored DM profile flattened by SN feedback in the central
 panel. We fitted the profiles with an $\alpha-\beta-\gamma$ function (equation \ref{eq:NFW}).}
 \label{fig:SatellitesFit}
 \end{figure*}

In this last section we make a concise analysis of the subhalo population of the host halo. The scientific goals
are twofold: on the one hand, we compare the subhalos found in the hydrodynamical run with their alter-egos from the DM-only run and see
if there are non-neglible effects that distinguish these populations. On the other hand, we look at the impact of star formation and its associated
feedback and see if similar effects that are observed in the case of the host halo also affect the subhalos. Because the galaxy of 
Halo B presents a lot of MW-like properties, we wandered
which results the simulation produces actually inside its halo.  
We only show here the subhalo results for Halo B, but we verified that, qualitatively, they do not vary
for the other halos. \\
First, we looked with AHF for the overdensities contained inside $R_{50}$ that is 301 kpc (288.5 kpc) for Halo B (Halo B-DM) and identified 308 (469)
subhalos. For each subhalo we then built the spherical density profile, starting from the center defined as the highest density point that is closest to the position 
indicated by AHF, and going out to 2.5 times $R_{\textrm{sub},50}$. In the hydrodynamical run, we include the stars in our calculation.
We then define $R_{\text sub}$ as the radius where the slope of the density profile (centered on the subhalo) becomes flat and reaches the local background density of the host. 
Summing up 
all the mass inside $R_{\text sub}$, we obtain $M_{\text sub}$.\\
Next, we analyse the subhalo's central DM density. 
Having defined the subhalo's edge, we construct the DM density profiles for the 25 most massive subhalos in the hydrodynamical and the DM-only simulation
going out to $R_{\text sub}$. While the resolution of the simulation is high enough to give meaningful results for these objects, the central profiles
of smaller clumps could be altered by a possible lack of resolution and therefore we limit the analysis to this sample. 
We fitted the subhalo's DM density profile with an $\alpha$-$\beta$-$\gamma$ profile (Equation \ref{eq:NFW}) and subdivided the subhalos in several populations, 
of which typical examples can be found in Figure \ref{fig:SatellitesFit}: subhalos from the DM-only run, subhalos from the hydrodynamical run that
possess stars (17/25), and subhalos from the hydrodynamical run
that did not produce stars (8/25). We resumed the results of the fit in the left panel of Figure
\ref{fig:SatelitesLumDist}. Most appealing is the comparison of the value of $\gamma$ in the different samples: while the satellites of the DM-only run tend
to have an inner profile with a slope close to -1 (as one would expect), the central profile of subhalos that formed stars in the hydrodynamical run is cored.
At the same time, the subhalos in the hydrodynamical run that do not have a star formation history have steeper inner density profiles. This result coherently
enlarges
the observations we made earlier for the host halos. Even in coarser non-isolated simulations like these realisations here, 
SN feedback has a strong impact, in a systematic and consistent way, on the central DM densities.\\

\begin{figure*}
 \centering
 \begin{subfigure}[t]{0.33\linewidth}
    \includegraphics[width=1.\linewidth]{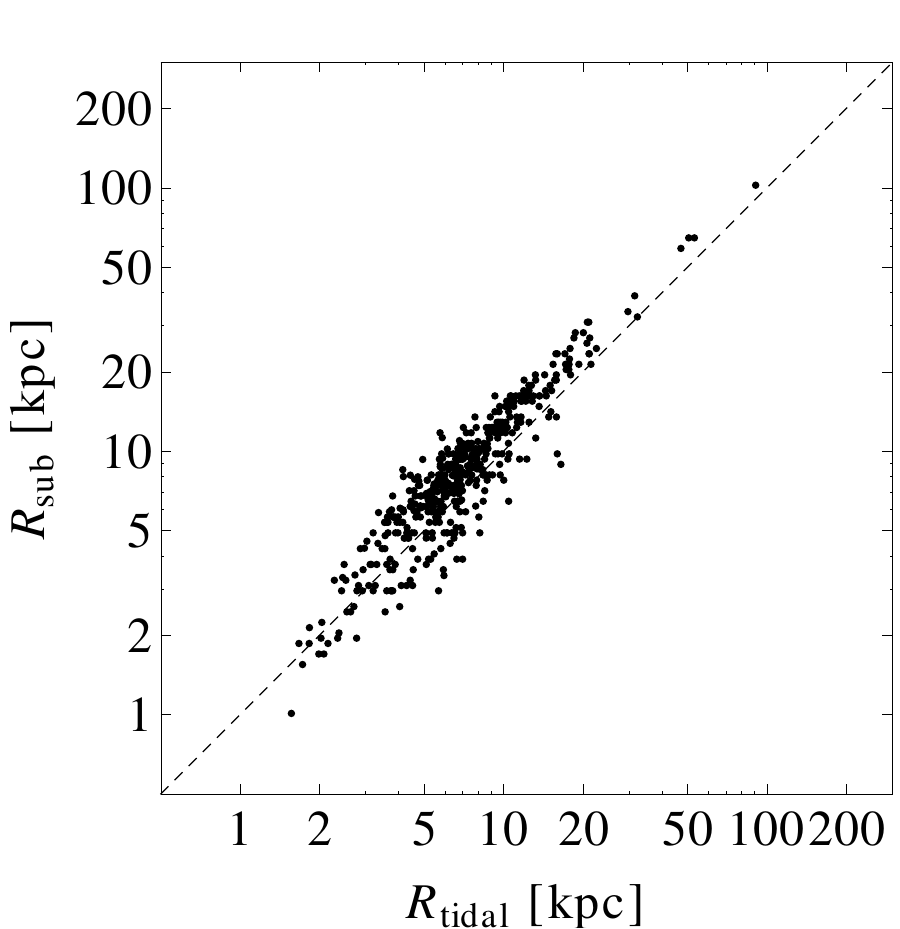}
 \end{subfigure}
 \begin{subfigure}[t]{0.33\linewidth}
    \includegraphics[width=1.\linewidth]{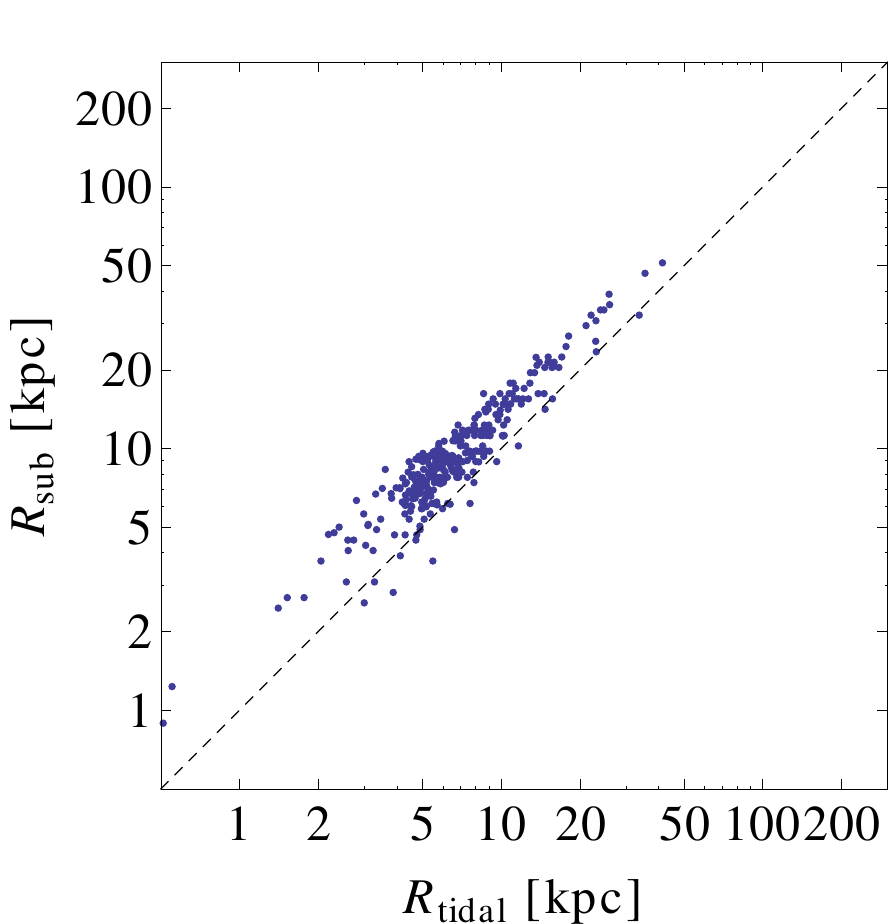}
 \end{subfigure}
 \begin{subfigure}[t]{0.33\linewidth}
  \includegraphics[width=1.\linewidth]{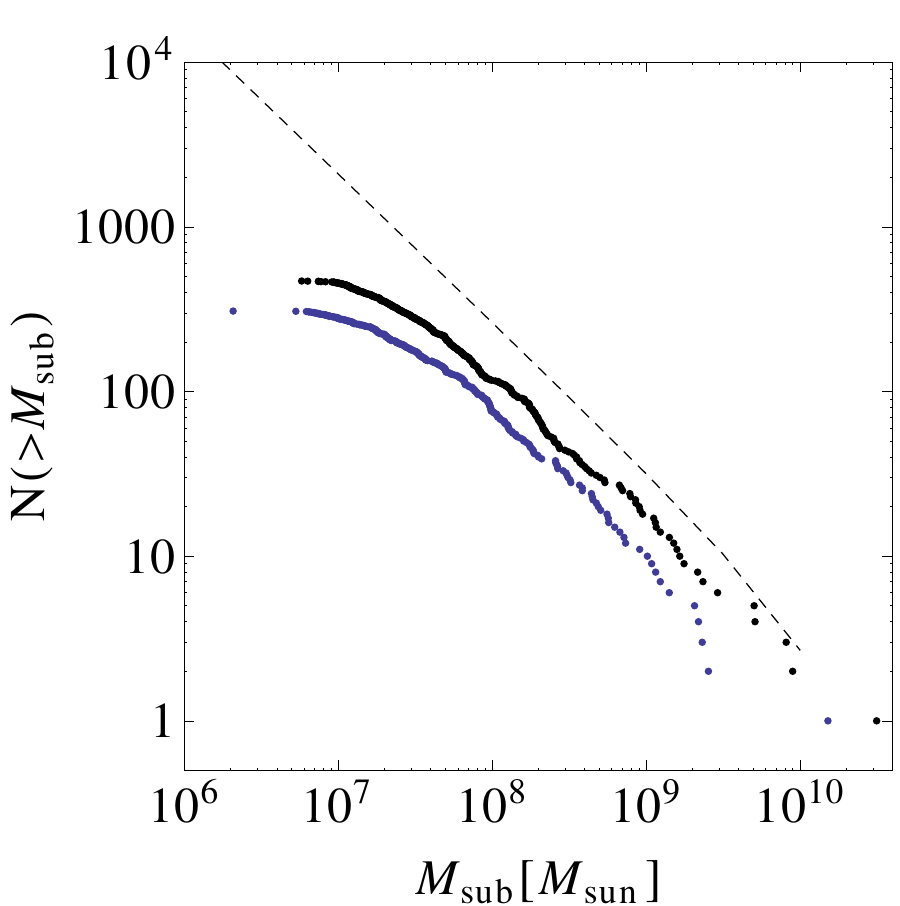}
 \end{subfigure}

 \caption{Left and central panel: the $R_{\text sub}$-$R_{\text tidal}$ relation for the subhalos the DM-only run (black points) and the hydrodynamical run (blue points), in which we included 
 star particles to calculate $R_{\text tidal}$. Right panel: the integrated clump spectrum for the the DM-only run (black) and the hydro run (blue). 
 The dashed curve shows $\int_{M_{\text sub}}^{max(M_{\textrm{sub}})} (\frac{dN}{dM})^{-1.9}dM $.}
 \label{fig:rsubrtidal}
 \end{figure*}

\begin{figure*}
 \centering
 \begin{subfigure}[t]{0.32\linewidth}
  \includegraphics[width=1.\linewidth]{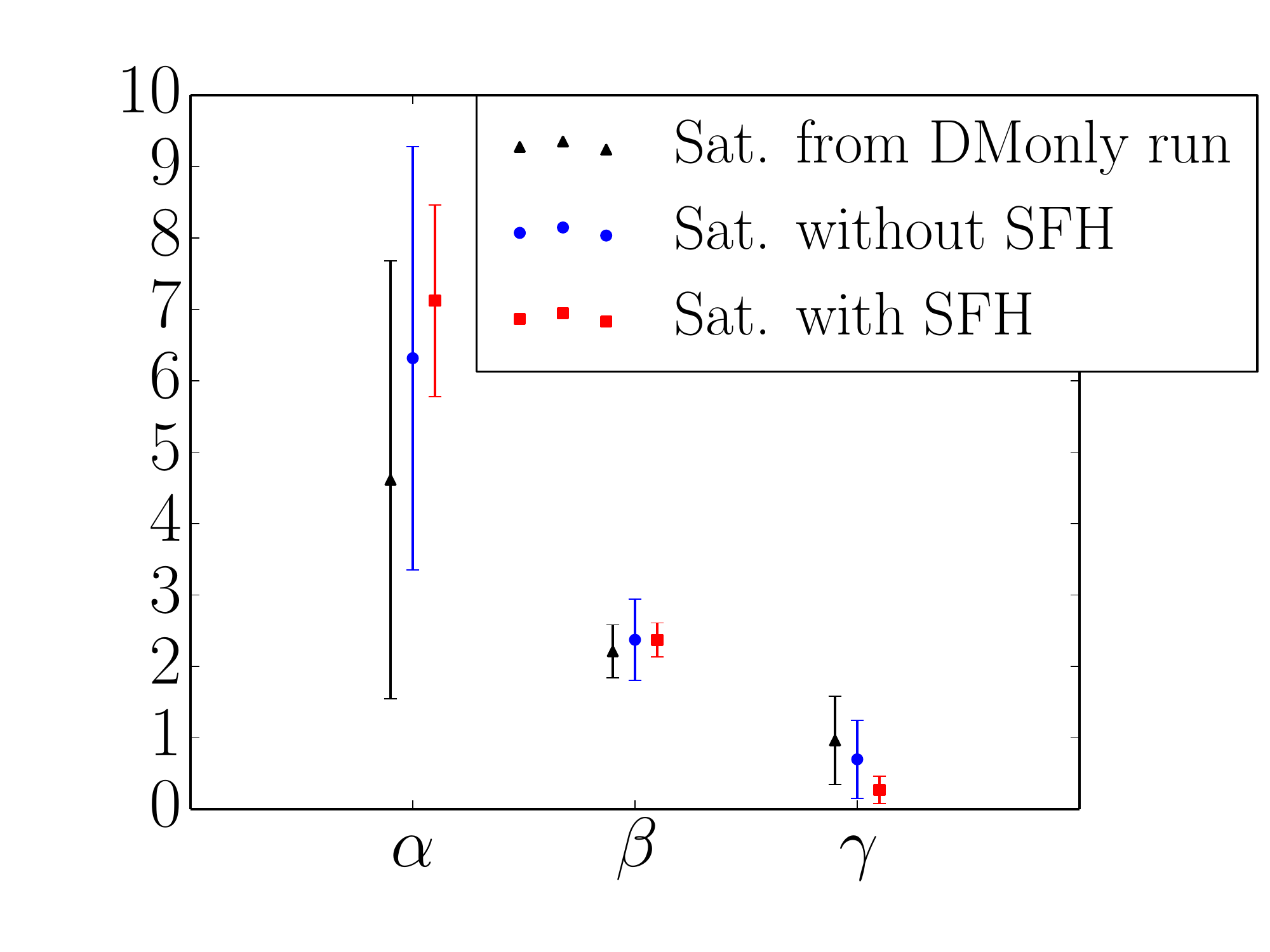}
 \end{subfigure}
 \begin{subfigure}[t]{0.32\linewidth}
  \includegraphics[width=1.\linewidth]{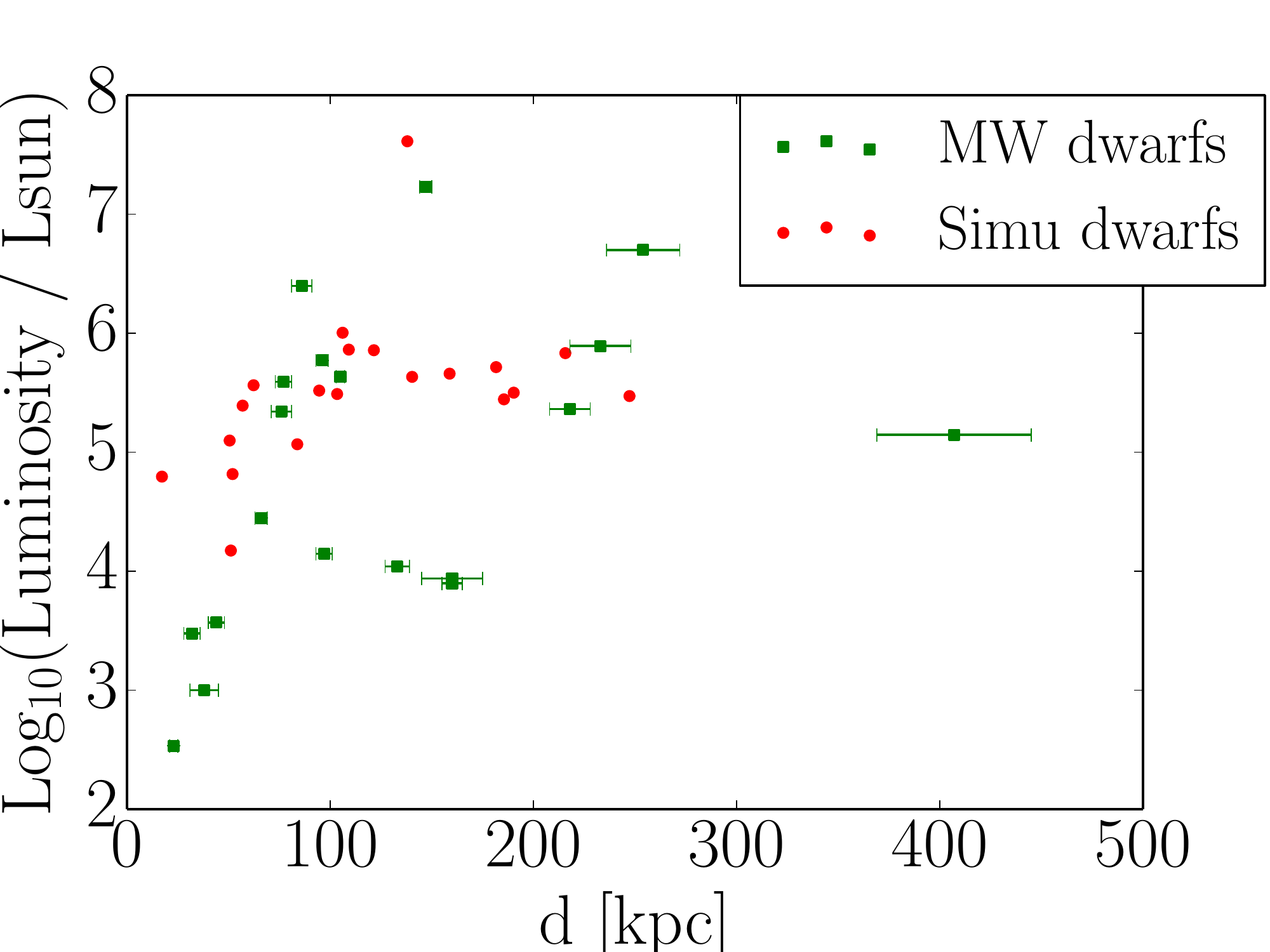}
 \end{subfigure}
   \begin{subfigure}[t]{0.32\linewidth}
    \includegraphics[width=1.\linewidth]{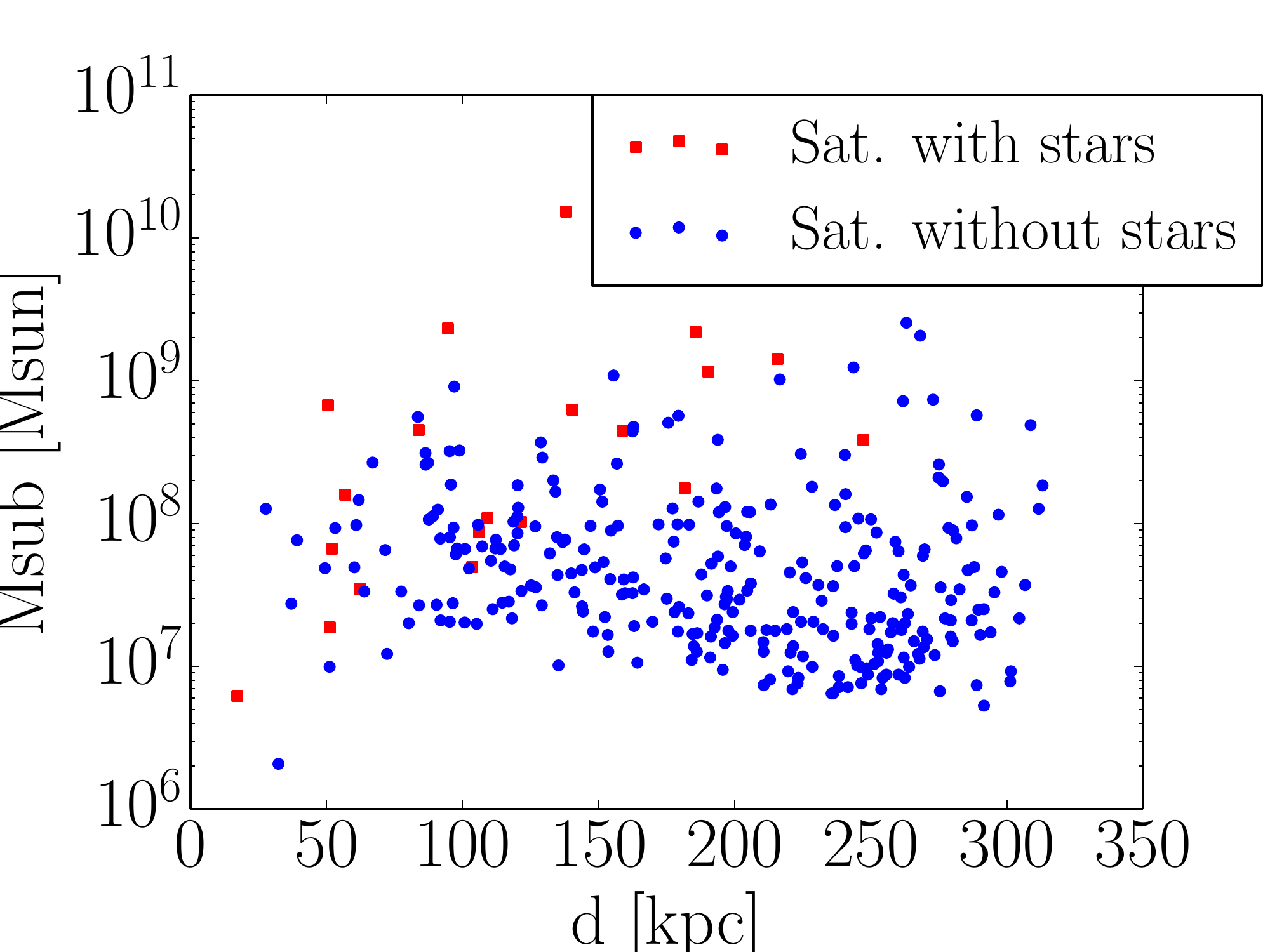}
   \end{subfigure}
 \caption{Left panel: Results of the fit (one $\sigma$ error bars) with equation \ref{eq:NFW} to
   the DM density profiles of the different satellite populations.\\
 Central panel: the V-band luminosities (for the satellites that were able to retain gas and eventually to produce stars) against the distance to the galactic center.
 For comparison, we also show the data from the Milky Way satellites \citep{2010MNRAS.406.1220W} (pre-SDSS/classical MW dSphs and post-SDSS MW dSphs).\\
 Right panel: for Halo B, $M_{\text sub}$ shown against the distance to the galactic center, blue points being DM-only satellites and red points satellites that formed
 stars (both populations from hydrodynamical run!).}
 \label{fig:SatelitesLumDist}
 \end{figure*}

In Figure \ref{fig:rsubrtidal} we compare the calculated subhalo radii with the tidal radii that depend on the subhalo's distance to the galactic center
and on $M_{\text sub}$. The tidal radius is the theoretical limit for a particle bound to a satellite that is orbiting in the gravitational field
of a host halo.
Knowing that the point-mass approximation for our systems is no longer valid because the satellites orbit within the body of the host system \citep{2008gady.book.....B},
we calculate it like~\citep{2008MNRAS.391.1685S}
\begin{equation}
 R_{\text tidal}=(\frac{M_{\textrm{sub}}}{(2-\textrm{d ln}M/\textrm{d ln}r)M(<d)})^{1/3}\cdot d
\end{equation}
where d is the distance to the galactic center and $M(<d)$ the mass contained inside a sphere of radius d. Within some scatter, we find a reasonable 
agreement between $R_{\text sub}$ and the tidal radius and validate the consistency of our method to calculate $R_{\text sub}$.\\
In figure \ref{fig:rsubrtidal}, we plot the the cumulative subhalo spectrum N($>M_{\text sub}$) for the DM-only run (black points) and the hydrodynamical
run (blue points). The dashed curve shows the integrated mass function 
\begin{equation}
N(>M_{\text sub})=\int_{M_{\text sub}}^{max(M_{\textrm{sub}})}\frac{\textrm{d}N}{\textrm{d}M}\textrm{d}M 
\end{equation}
where $\frac{\textrm{d}N}{\textrm{d}M}\propto M^n$ with $n=-1.9$ \citep{2008MNRAS.391.1685S,2012MNRAS.425.2169G}. 
We remark that both curves follow the tendency curve. They depart from a
slope of -1.9 due to the lack of resolution around $M_{\text sub} < 5*10^7 \mo$.
It is interesting that there is a clear shift in the integrated clump spectrum. As the overall curves do show the same behaviour and because they stay parallel, we conclude that
we witness a considerable statistical mass loss of the subhalos over the whole mass range in the hydrodynamical run compared to the DM-only version. 
This feature seems to be in agreement with \citet{2014MNRAS.444.1518V} who found a mass reduction for low mass satellites.\\
\citet{2013ApJ...765...22B} identified two mechanisms that could explain the mass loss of satellites which may be applicable to our simulation:
on the one hand, if the satellites are able to form stars, supernova feedback can flatten the DM cusp, so that DM particles with elongated orbits
are stripped more easily (in contradiction with \citet{2013MNRAS.433.3539G} where the effective model is not capable of transmitting a
sufficient amount of energy to the DM of the satellites in order to match the observed central densities of MW spheroidal dwarfs). On the other hand, 
tidal stripping due to the presence of the galactic baryonic disk is enhanced.
In agreement with \citet{2012ApJ...761...71Z}, our results seem to strengthen these interpretations as solution attempts 
for the $\Lambda$CDM missing satellite problem \citep{1999ApJ...522...82K} and 
the connected too-big-to-fail problem \citep{2011MNRAS.415L..40B}. Namely, if we recap what has been shown, all subhalos in the simulations are prone to statistical
mass loss, and only some of the most massive satellites are able to form stars. \\

 Analysing the hydro-run of Halo B,  we show in Figure \ref{fig:SatelitesLumDist} the satellite V-band luminosities of the simulation against the MW satellites 
 \citep{2010MNRAS.406.1220W}.
 To be consistent and for simplicity, we used the same procedure to derive the stellar luminosities that we used for the host galaxy by taking into account 
 the stellar ages and metallicities (for redshift 0!). We only counted the stellar luminosities for the dwarf galaxies that possess five or more star particles
 to make sure that we neglect subhalos that have one or several star particles at the border of the subhalo. Interestingly, the simulation subhalos populate
 a similar luminosity range compared to the discovered MW dwarfs. In addition, the most luminous simulation satellite is not alarmingly brighter than the brightest
 MW dwarf. However, these results are in tension with a recent study of satellites in the highly resolved ELVIS suite of $\Lambda$CDM simulations, that
 compare the local group galaxies (of the MW) with the simulated halos that are in a similar configuration \citep{2014MNRAS.444..222G}.
 In these DM-only simulations, they find that the number of subhalos that are theoretically massive enough to form stars at some point in their history largely exceeds 
 the observed amount of Local Group dwarf galaxies. However, in the recent work of \citet{2014arXiv1404.3724S}, it has been stated that zoom simulations
 that focus on a Local Group-like structure can reproduce the right order of magnitude of faint galaxies. This result 
 points toward the obligation to include the modelling of
 baryonic physics and makes the failure of $\Lambda$CDM less probable.\\
 In Figure \ref{fig:SatelitesLumDist}, we show for every subhalo of Halo B the virial mass $M_{\text sub}$ as a function of their distance to the galactic center, blue points denoting
 satellites without stars (i.e., DM-only) and red points for satellites that have formed stars. No clear correlation between mass or position of a satellite and the presence of 
stars within a subhalo can be determined. Higher mass subhalos seem to be able to form stars in some instances, though what makes these halos special relative 
to their DM-only counterparts is unknown.
On the other hand, low mass subhalos in the simulations generally seem to be unable to produce stars. 
 
\section{Summary and perspectives}\label{sec:summary}
In this paper, we presented three cosmological hydrodynamical simulations zoomed on Milky Way sized halos using the 
adaptive mesh refinement code RAMSES.
Adopting widely-used subgrid prescriptions to model star formation and stellar feedback and keeping them fixed during the on-going simulations, 
we have obtained and analysed the results of the three halos. We compared them to recent observational key results for simulations. 
In summary:
\begin{itemize}
\item We built realistic luminosity maps of the galactic stars taking into account their age and metallicity. From them, we derived brightness profiles in several
bands and computed their disk and bulge scales.
\item We constructed the circular velocity profiles and compared them with observational data from the Milky Way. Furthermore, connecting the stellar mass
to the rotational velocity at a specific radius after the velocity curve reached its maximum, we confirmed that the simulations lie within recent observational data of the Tully-Fisher relation.
 \item We verified that star formation in the simulation follows a Kennicutt-Schmidt law at redshift 0.
\item Regarding the total stellar mass, we confirm the outcome from \citet{Roskar:2013pia}, stressing that supernova feedback is not strong enough to diminish the gas content of the halo in order 
to prevent an (although postponed) excessive star formation. Recent simulations using a feedback scheme where outblowing winds are decoupled from the flow 
(e.g. \citet{2013MNRAS.436.3031V,2014MNRAS.437.1750M})
are able to expel a bigger quantity of the galactic gas out of the baryonic disk or even out of the halo and consequently, the gas reservoir 
used to form stars is smaller. Other attempts trying to ensure a more physically motivated star formation (through molecular based star formation) combined with additional stellar feedback 
\citep{2014arXiv1404.2613A,2013arXiv1311.2073H}  
also seem to produce good results with regard to the stellar-to-halo mass relation. 
\item Even if the stellar-to-halo mass ratio of the simulations presented in this paper is too high, we showed that if one applies 
standard techniques of stellar mass deduction from observational data, it is possible to reduce the gap between our simulations and observational results.
Because of the extensive consequences, this line of investigation definitely deserve more attention to ensure a more thorough treatment. 
\item We studied the DM density profile over time and observed a combination of a flattening of the inner DM density profile with a contraction 
of the DM density at the outer parts of the galactic disk. This effect, which is caused by stellar feedback, can produce DM cores of $\sim$ 5 kpc. 
We remark that the production of DM cores for this halo size has not been observed before in other simulations, in particular not 
in the aforementioned simulations using the decoupled winds feedback 
because their inner DM density has a cuspy profile. A comparison of their result with an analytical adiabatic contraction model would be interesting. 
Furthermore, we showed that the observed DM feature highly affects the local DM density at the solar radius.
\item We concluded with an analysis of the halo satellites. We showed that in the star forming subhalos, stellar feedback 
also modifies their inner DM densities and  lowers the inner slope
with comparison to satellites that do not form stars.
Moreover, the halos in the lower MW mass range seem to give a good matching between the Milky Way spheroidal dwarfs
and the brightest simulation satellites.
Using a self-designed tool to define the edge of the subhalo, we verified that it maps correctly
the theoretically defined tidal radius. In all the simulations, we found less substructures in the hydrodynamical run compared to the DM-only run, a general subhalo mass 
loss over the whole mass spectrum and, in particular, a mass reduction of the most massive subhalos. Therefore, we could confirm earlier results of 
isolated simulations. The subhalo mass and its distance to the galactic center seem to be uncorrelated to the capacity of forming stars, even though
low mass subhalos don't have a SFH in the simulation.
\end{itemize}
As we explicitly indicated before, Halo B mimics many observational properties of the Milky Way, even though its virial mass is in the lower part of the estimated
MW mass range and the tuned SN feedback used here leads to a less regular spiral structure.
Those quite realistic features predestine this simulation to be an interesting framework for DM detection studies 
\citep{2010JCAP...02..012L,2012PhRvD..86f3524N}. Indeed, we are going to take advantage of our better
 resolved simulations and use them as consistent frameworks for DM detection phenomenology.
These aspects are beyond the scope of the present paper and will be published in subsequent articles with
improved astroparticle calculations.

\section*{Acknowledgements}
We thank  
Valentin Perret for fruitful discussions and Jean-Charles Lambert for providing us with quick and efficient support when technical problems occurred.\\
We also want to thank the anonymous referee for the report. His concise and helpful remarks undeniably improved the final version.\\
The work is supported by the Fonds National de la Recherche, Luxembourg (Project Code: 4945372).\\
It was granted access to the HPC resources of Aix-Marseille Universit\'e financed by the project Equip@Meso (ANR-10-EQPX-29-01) 
of the program "Investissements d'Avenir" supervised by the Agence Nationale pour la Recherche.
The simulations were also performed on the Centre Informatique
National de l'Enseignement Superieur under GENCI allocation 2013.\\
This work has been carried out thanks to the support of the OCEVU Labex (ANR-11-LABX-0060) and the A*MIDEX project (ANR-11-IDEX-0001-02) funded by the "Investissements d'Avenir" French government program managed by the ANR.

\nocite{}
\bibliography{PaperSimuJournal}

\label{lastpage}
\end{document}